\documentclass[iop]{emulateapj}
\usepackage{color}

\usepackage{graphicx}
\usepackage{mathptmx}
\usepackage{soul, xcolor}
\setstcolor{red}
\usepackage[breaklinks=true, colorlinks=true, linkcolor=black, urlcolor=blue, citecolor=black]{hyperref}
\definecolor{myblue}{RGB}{0, 0, 0}
\definecolor{mygreen}{RGB}{0, 160, 0}

\def\Mstel{M_\ast}
\def\Mgas{M_{\rm gas}}

\def\Mhalo{M_{\rm halo}}

\def\resp{respectively}
\def\bfr{\color{black}}
\def\bfb{\color{black}}

\def\bfp{\color{black}}

\def\sfr{{\rm SFR}}

\gdef\imacs{{\it IMACS}}
\gdef\z5fract2{{\it z5fract2}}

\def\etal{\hbox{et al.}}

\gdef\ltsima{$\scriptscriptstyle \; \buildrel < \over \sim \;$}
\gdef\simlt{\lower.3ex\hbox{\ltsima}}

\gdef\gtsima{$\scriptscriptstyle \; \buildrel > \over \sim \;$}
\gdef\simgt{\lower.3ex\hbox{\gtsima}}

\gdef\about{\raise.3ex\hbox{$\scriptscriptstyle \sim $}}

\def\gs{\mathrel{\raise0.35ex\hbox{$\scriptstyle >$}\kern-0.6em 
\lower0.40ex\hbox{{$\scriptstyle \sim$}}}}
\def\ls{\mathrel{\raise0.35ex\hbox{$\scriptstyle <$}\kern-0.6em 
\lower0.40ex\hbox{{$\scriptstyle \sim$}}}}

\def\etal{\hbox{et al.}}

\def\Msun{\rm{\hbox{$\,$M$_{\odot}$}}}				
\def\ang{\hbox{$\,$\AA}}

\def\24m{\hbox{24\,$\micron$}$\,$}
\def\10-18{\hbox{$\times~10^{-18}$}}

\def\Tobs{\hbox{\emph{T$_{\rm obs}$}}}

\def\T0{{$t_0$}}

\begin{document}




\slugcomment{Accepted to ApJ, 23 October 2018}

\shortauthors{Dressler \etal\ }
\shorttitle{Late Bloomer Galaxies: Growing Up in Cosmic Autumn}

\title{Late Bloomer Galaxies: Growing Up in Cosmic Autumn
\footnote{T\lowercase{his paper includes data gathered with the 6.5 meter \uppercase{M}agellan \uppercase{T}elescopes 
located at \uppercase{L}as \uppercase{C}ampanas \uppercase{O}bservatory, \uppercase{C}hile.}}}

\author{
Alan Dressler\altaffilmark{1,*},
Daniel D.\ Kelson\altaffilmark{1},
and Louis E.\ Abramson\altaffilmark{1,2}
}

\altaffiltext{1}{The Observatories of the Carnegie Institution for Science, 813 Santa Barbara Street, Pasadena, CA 91101, USA}
\altaffiltext{*}{\href{mailto:dressler@carnegiescience.edu}{dressler@carnegiescience.edu}}
\altaffiltext{2}{Princeton University, Peyton Hall, 4 Ivy Lane, Princeton, NJ 08544, USA}

\begin{abstract}

Late bloomers are massive ($\Mstel >10^{10}\,\Msun$) galaxies at $z<1$ that formed the majority of their stars within $\sim$2 
Gyr of the epoch of observation.  Our improved methodology for deriving star formation histories (SFHs) of galaxies  at redshifts 
$0.45 < z < 0.75$ from the Carnegie-\emph{Spitzer}-\emph{IMACS} Survey includes confidence intervals that robustly distinguish late 
bloomers from ``old'' galaxies. We use simulated SFHs to test for ``false positives" and contamination from old galaxies to demonstrate 
that the late bloomer population is not an artifact of our template modeling technique. We  show that late bloomers account for 
$\sim$20\%  of $z\sim0.6$ galaxies with masses of the modern Milky Way, with a moderate dependence on mass.   We take advantage of a 1\% 
overlap of our sample with  HST (\emph{CANDELS}) imaging to construct a ``gold standard" catalog of 74 galaxies with high-confidence SFHs, 
SEDs, basic data, and HST images to facilitate comparison with future studies by others.  This small subset  suggests that galaxies  with both 
old and young SFHs {\bfb cover the full range of morphology and environment} (excluding rich groups or clusters), albeit with a mild but 
suggestive correlation with local environment.  We begin the  investigation of whether late bloomers of sufficient mass and frequency are 
produced in current-generation $\Lambda$CDM-based semi-analytic models of galaxy formation.  In terms of halo growth, we find a 
late-assembling halo fraction within a factor-of-two of our late bloomer fraction. However, sufficiently delaying star formation in such halos may be 
a challenge for the baryon component  of such models.

\end{abstract}

\keywords{
galaxies: evolution ---
galaxies: star formation ---
galaxies: stellar content
}

\section{Introduction: Star Formation Histories, Individual and Collective}

The global history of star formation is constructed by measuring the star formation rate (SFR) of galaxies over cosmic time 
in a sufficiently large volume to contain a representative sample of starforming galaxies.  The evolution  of this quantity, the 
\emph{star formation rate density}, (SFRD; Lanzetta, Wolfe, \& Turnshek 1995, Lilly \etal\ 1996,  Pei \& Fall 1995, Madau 
\& Dickinson 2014, hereinafter MD14), is well defined over most of cosmic history, $0<z<10$ (Oesch \etal\ 2014). Modulo 
corrections for starlight absorbed by dust grains and potential incompleteness from 
galaxies fainter than---and SFRs less than---the observation limit, the SFRD is an accurate description of the efficiency with which 
galaxies have grown since a time when their stellar masses were $\sim$1\% of what they are today.  The cumulative buildup of 
stellar mass, recovered from the mass function of galaxies over the same range of  epochs, is in good agreement with the integral 
of the SFRD (Dickinson \etal\ 2003; MD14).  It could be argued, then, that the story of {\it global} stellar mass production is now 
reasonably well understood, or at least well-characterized.  

``Cosmic noon'' is a good name for when the luminosity of the universe peaked (as is ``cosmic dawn"---first light).  However---as 
suggested by the title of this paper---the growth of galaxies relates better to cosmic \emph{seasons} rather than hours of the 
day; i.e., a cosmic spring ($3<z<6$), summer ($1<z<3$), and autumn ($z<1$). with cosmic winter yet to come.

In their comprehensive review of the history of cosmic star formation, MD14 parameterized the rise and fall of the SFRD as a 
polynomial (see MD14 Figure 9, Equation 15), a form that provides a good fit to the data but offers no insight into the physical 
processes that control the evolution of the global SFR, let alone the SFR histories of the individual galaxies from which it is composed.  
Taking this latter step requires a further constraint on the behavior of SFRs with epoch. The \emph{star formation main sequence} 
(SFMS; e.g., Noeske \etal\ 2007; Whitaker \etal\ 2012)---a now thoroughly studied correlation between stellar mass  and SFR at 
$z\lesssim6$---has been the favored method for constructing SFHs that, in aggregate, reproduces the SFRD (e.g., Peng \etal\ 2010, 
Speagle \etal\ 2014, or Tomczak \etal\ 2016).  From the near-unity slope of the SFMS the implication has been drawn that galaxies  
grow in direct proportion to their mass, modulo the rising zero point of the SFMS before $z\sim2$---faster growth---and a rapid decline 
after.  If the considerable scatter of the SFMS can be taken as a series of random perturbations on otherwise smooth growth, each galaxy 
can be fit by a set of conformal growth curves, identical up to a mass scaling (e.g., Peng \etal\ 2010, Leitner \etal\ 2012; Behroozi \etal\ 
2013).  However, since the SFMS bends from this unity slope at its high mass end---in effect, a general slowing of the stellar mass growth after 
the peak in the SFRD---some sort of ``quenching" mechanism is required to explain the declining SFRs of evolved, massive galaxies,  and its 
collective manifestation in the decline of the SFRD after $z\sim1.5$.

In a previous series of papers, Oemler \etal\ (2013, O13), Gladders \etal\ (2013, G13) Abramson \etal\ (2015), Abramson \etal\ 2016, 
A16), and Dressler \etal\ 2016, D16), we have  described a different approach that evolved from O13's identification of a fraction of
massive galaxies, $\log\Mstel\gtrsim10.6$, increasing in redshift up to $z\sim$1, that require rising SFRs around the epoch of observation 
(\Tobs).  This is a notable departure from what had long been inferred from studies of low-redshift galaxies: almost every present-day galaxy 
can be fit by a ``tau-model" of exponential decline (Tinsely 1972).  O13 found a fraction of these so-called ``young'' galaxies of $\sim$20\% 
by $z\sim0.8$.  Such galaxies had been found in previous studies (e.g., Cowie 1996, Noeske \etal\ 2007),  however,  the larger sample of O13 
showed that their prior characterization as starbursts was not tenable (see Figure 4 of O13).   Although late-rising SFHs are observed for 
many present-epoch \emph{dwarf} galaxies (e.g., Gallagher \etal\ 1984), the identification of rising SFRs for a substantial fraction of 
massive galaxies ($\Mstel > 10^{10}$\Msun) at redshifts $z>0.3$ was new information for understanding the evolution of common galaxies 
(see, for example, Kelson 2014).

Responding to the inadequacy of tau-model SFHs for this population, Gladders \etal\ (2013; G13) explored the idea that individual SFHs 
might be better parameterized by a two-timescale \emph{lognormal} form. This idea that came from realizing that the SFRD itself is well 
described as a single lognormal, with timescales of $T_{0}\approx 5.2$ Gyr (associated with the midpoint of mass buildup), and a 
characteristic duration of 5.7 Gyr.\footnote[3]{FWHM =  $2\,\exp(T_{0}-\tau^{2})\,\sinh(\sqrt{2\,\ln(2)}\,\tau)$, where 
$\exp(\tau)\approx 1.9$ Gyr.}

Abramson \etal\ (2015) and A16 explored the implications of this parametric SFH model in terms of the galaxy stellar mass function. 
These studies found a very good match even up to $z\sim8$, remarkable because the parameters of the G13 model relied only on data 
from galaxies at redshifts $z<1$.  A16 expanded this to include the slope, evolution, and scatter of the SFMS and other aspects of the SFMS 
``grow and quench" picture, finding again that the G13 lognormal SFHs model accounted equally well for observations of principal ensemble 
behaviors.  The paper concluded that identifying a uniquely ``good'' description of galaxy evolution required new observational constraints.

Towards this end, D16 investigated individual galaxy spectral energy distributions---SEDs---and their implied SFHs in order to test 
the efficacy of the two approaches, which to that point were evaluated mostly in terms of distribution functions and scatter-plots/scaling-laws.  
D16 used spectrophotometric data from the Carnegie-\emph{Spitzer}-\imacs\ study (CSI, Kelson \etal\ 2014, K14) that combined  
broad-band photometry and \imacs\ (Dressler \etal\ 2011) prism observations for $\sim$20,000 galaxies in the XMM field of the \emph{Spitzer 
SWIRE} survey (Lonsdale \etal\ 2003).  These data were used to construct SEDs analyzed in terms of SFHs.  The innovative methodology of K14 
was to model the SED as the sum of 6  epochs of (constant level) star formation, the first from redshift $z=5$ to 1 Gyr before the epoch 
of observation, \Tobs, followed by five 200 Myr periods over that final Gyr.   D16 defined a quantity \emph{z5fract} as the fraction of the total 
stellar mass generated before the final Gyr of the observed galaxy.  Using \emph{z5fract} as a proxy for the galaxy's mean age brought 
attention to a population of late-growing galaxies with a fraction of old stellar mass---i.e., that formed in the first SFH bin---below 50\%.  
This population amounted to about 20\% of the sample, reminiscent of the fraction of  young galaxies found in studies cited above and 
the G13 model analysis.  D16 was a step beyond the earlier work, though, because the identification of individual SFHs, while crude, 
offered for the first time the possibility of discriminating between the ``grow and quench" and ``a diversity of lognormals" pictures.

In this paper, we improve and refine the SFH analysis of D16, focusing on the reality of what we called ``late-bloomers," galaxies 
at $z\sim0.6$ where the majority of its stellar mass appears to have formed \emph{later than} $z\sim1$. {\bfb (See Chauke \etal\ 
2018 for a recent complementary study at {\bfp $z=0.6$--1}.)} This includes a critical look at the possibility that our earlier analysis simply failed 
to detect large populations of old stars because of insufficient  sensitivity or the hiding of these stars by dust.  By exploiting the highest 
signal-to-noise ($S/N$) data of the CSI XMM field and making full use of the thousands of duplicate measurements collected in the survey, 
we arrive at a improved SFH analysis, particularly with respect to the amount of old stellar mass in what appear to be genuinely young 
galaxies.  Further, we add confidence intervals to the SFHs to accurately characterize their uncertainties, and perform a rigorous analysis 
of simulated SEDs to assess the impact of $S/N$ on SFH derivations and to determine the level at which late bloomers might be contaminated 
by misidentified older objects.

With measures of confidence in our SFH fits of CSI data, we catalog a ``gold sample" of 74 galaxies with high confidence SFHs 
and HST (CANDELS) imaging.  We provide coordinates and other basic data, including RGB images, SEDs, and derived SFHs 
to enable other researchers to observe and analyze these galaxies and compare their results with ours.

Indeed, new datasets and methods now make this kind of analysis not only possible, but robust (see, e.g., Pacifici \etal\ 2012, 2016; Iyer \& 
Gawiser 2017; Chauke \etal\ 2018). Extant and future facilities can support inferences regarding the full diversity and character of individual 
galaxy growth and so directly confront theoretical evolutionary models in their native domain.  This work joins the above cohort of 
complementary studies in establishing the first more-than-tentative footholds in this new regime.

The paper is organized as follows: Section \ref{sec:data} reviews the methodology of D16 in the context of the fidelity of the SEDs.
There, we define ``late bloomer'' (Section \ref{sec:lbdef}), and describe improvements to the analysis made through a purposeful 
attempt to falsify our claim that such galaxies are common and span a wide range in stellar mass.  Such careful scrutiny and skepticism are 
justified by the challenge late bloomers may present to conventional wisdom about the growth of stellar mass.  Section \ref{sec:LBfract} 
presents our best assessment of the global late bloomer fraction at $z\sim0.65$. Section \ref{sec:catalog} presents the catalog of 74 galaxies 
with HST imaging, secure SEDs,  and high-confidence SFHs, intended to encourage tests of our results with other techniques and analyses.  
Section \ref{sec:lbchar} describes basic properties of late bloomers using larger sample of $\sim$7600 galaxies of {\bfp $\Mstel>10^{10}$\,\Msun} 
with high-quality SEDs and well-constrained SFHs to explore the late bloomer fraction trends with stellar mass, redshift, and environment.  
The discussion in Section \ref{sec:context} focuses on the compatibility of late bloomers with a $\Lambda$CDM semi-analytic model of 
stellar mass growth, including the implications of this work for abundance matching and scaling laws as tools in studying {\bfb galaxy} evolution.  
Section \ref{sec:summary} distills the major conclusions of the paper and describes possible {\bfb next-steps---including ongoing efforts
to obtain higher resolution \imacs\ spectra---to significantly improve SFH constraints and better distinguish late bloomers from truly old 
galaxies.} 
 
\section{Improvements in Deriving SFHs from CSI Spectrophotometric Data}
\label{sec:data}

The results of D16 on  SFH diversity were not anticipated in designing the CSI program.  Its primary goal was to improve 
on previous measurements of the fraction of passive galaxies as a function of redshift and stellar mass, in connection with 
the question of which processes might lead star forming galaxies to a temporary or permanent  cessation of star formation.  
Because passive populations are most luminous in the near-IR, the choice of a sample from the \emph{SWIRE} project, 
magnitude-limited by \emph{Spitzer} 3.6 $\micron$ imaging, yielded improved sensitivity to passive populations at 
redshifts up to $z\sim1.5$ and thus a better measure of the evolution of their fraction of the total galaxy population.  

The goal in CSI of measuring redshifts to 2\% or better---important both for SED analysis and for distinguishing galaxy 
clustering---led to a better method of extracting SFHs from the combination of \imacs\ prism spectra and broad-band 
photometry.  This necessarily included an accurate assessment of star formation in the final Gyr before the epoch of 
observation (\Tobs), as distinct from earlier star formation. This in turn lead to the parameterization of the SFH as a single 
early epoch of star formation beginning at $z=5$ followed by 5 epochs of 200 Myr duration in the final Gyr.   Slicing  cosmic 
history in this way well addressed the question of whether galaxies were passive or active and provided, for passive galaxies, 
potential evidence of relatively recent star formation.  Such information that might inform the mechanism  of what is commonly 
called ``quenching" of star formation.

The fraction of stellar mass a galaxy produced in the Gyr before \Tobs\ was a matter of special interest because of O13's study 
of field galaxies (in the \imacs\ Cluster Building Survey) that inferred  \emph{rising} SFHs for a growing fraction of galaxies 
at redshifts $0.2<z<0.8$.   Rising SFHs were implied by specific star formation rates (sSFRs) that exceeded what is allowed for 
\emph{constant} star formation systems---the limiting case for exponentially declining ``tau-model" (Tinsley 1972). 
The CSI data moved this work beyond a comparison of sSFR distributions at different epochs to the actual identification of 
galaxies with rising SFHs after $z\approx1$.  

In fact, such rising SFHs are implied by characteristic SFRs that are {\it twice\/} the lifetime average of galaxies (e.g., Kelson 
2014, see also G13).  The importance of such results is easy missed.  Even modest fractions of galaxies with late-rising SFHs imply 
a break in mass rank ordering, making traditional techniques of abundance matching an exercise that is dubious, at best.  The 
implication that the relative positions of galaxies within scaling relations like the SFMS do not stay fixed vitiates {\it any\/} ability 
to connect progenitors and descendants over cosmic time using scaling relation data and stellar masses alone.


\begin{figure}[t]
\centerline{
\includegraphics[width=3.7in, angle=0]{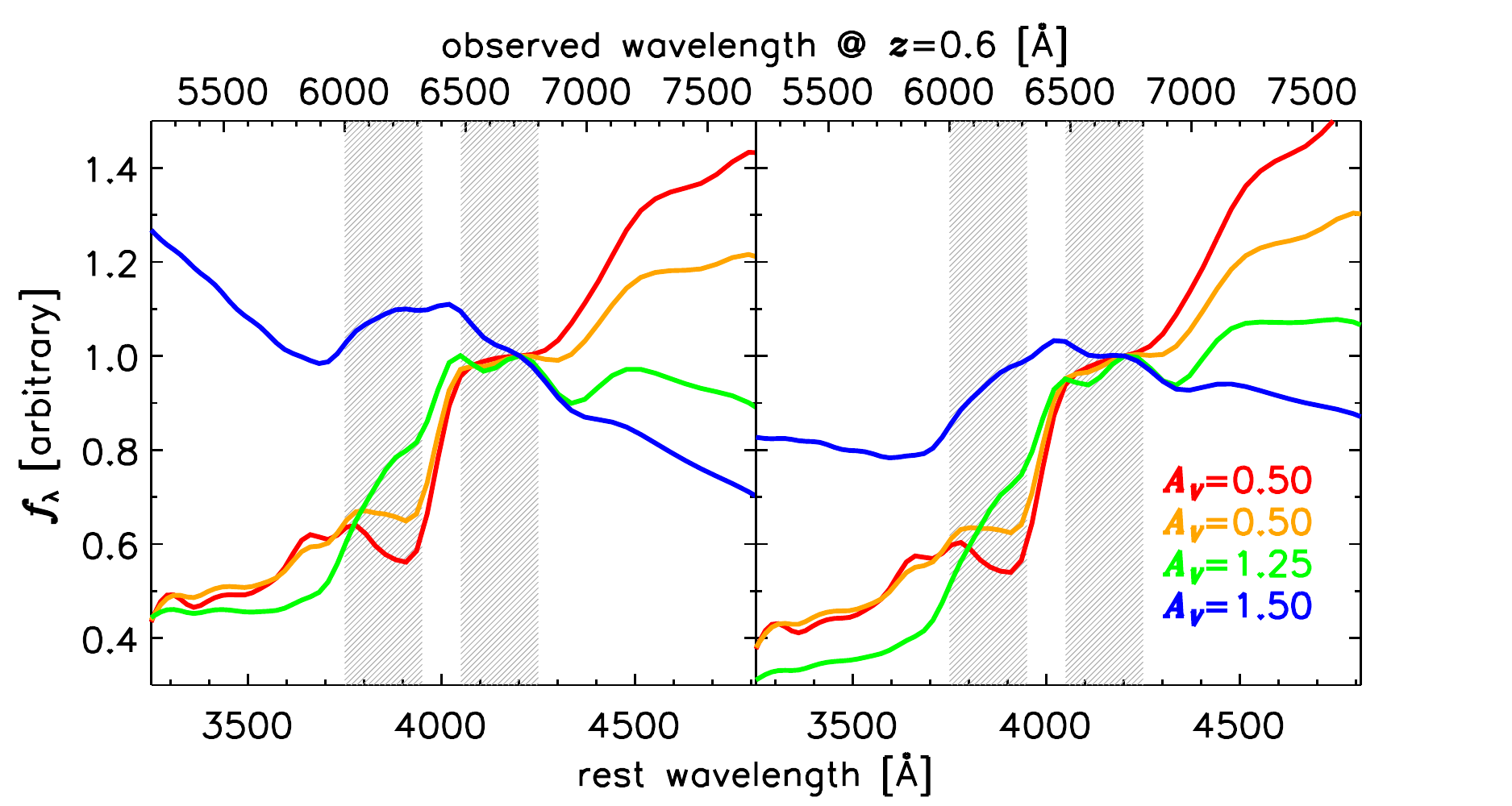}
}
\caption{Four template SEDs for {\bfr solar metallicity} stellar populations with a range of ages. {\bfr (left) SEDs of population
unattenuated by any dust; (right) the same SEDs but attenuated by levels of dust given in the figure.}
The blue line is a young population, constant SFR from
200 Myr prior to \Tobs; green line, from 1 Gyr to 200 Myr; orange line, from 2 to 1 Gyr; red line, from $z=5$ to 2 Gyr before \Tobs.
Populations 1 Gyr or younger are easily distinguished from older populations, while 1--2 Gyr, though less distinct, has a significantly
different SED than that of constant star formation from $z=5$ to $z=1$ (or star formation that ended earlier).  Adding a 1--2 Gyr epoch of
star formation into the CSI SFH modeling enables the recovery of a longer timescale of late star formation, as is expected given our
knowledge of the SFHs of the Milky Way and its neighbors inferred \emph{directly} through color-magnitude diagrams.  Each template is
normalized at rest-frame 4200\,\AA; vertical dashed lines show bandpasses measuring the D4000 break at $z=0.6$. {\bfr Even when the
young populations are highly reddened, SEDs have structure, around the Balmer and 4000\AA\ breaks for example, that provide strong
leverage on the relative fractions of old, intermediate, and young stellar mass.}
}
\label{fig:stelTemps}
\end{figure}

After D16, we recognized the advantage of pushing further back in time---from 1 Gyr before \Tobs\ to 2 Gyr---effectively separating the old 
from young stellar populations at $z\approx1$ (given the sample $\langle z\rangle\sim0.6$).  Figure \ref{fig:stelTemps} shows that 
the stellar template of a 1--2 Gyr old population, essentially F stars, can be distinguished from the template of the $\sim$5 Gyr 
of old stars that preceded it.\footnote{For a galaxy at $z\sim0.6$, the difference shows up as the depth of the D4000 break 
($6000<\lambda_{\rm obs}<6300$\ang ) and the slope of the continuum redward of $\lambda_{\rm obs}\sim7000$\ang,  
as seen in Figure \ref{fig:stelTemps}.}  By adding this epoch of intermediate-age star formation, our analysis has purchase on star 
formation after $z\approx1$ that occurred prior to the easily recognizable 0--1 Gyr population. Furthermore, our new model SFHs 
are less likely to incorrectly ascribe star formation that is relatively recent to the oldest stellar population.  We discuss the 1--2 Gyr 
population further in Section \ref{sec:2gyr}.

\begin{figure*}

\centerline{
\includegraphics[width=4.5in, angle=0.0]{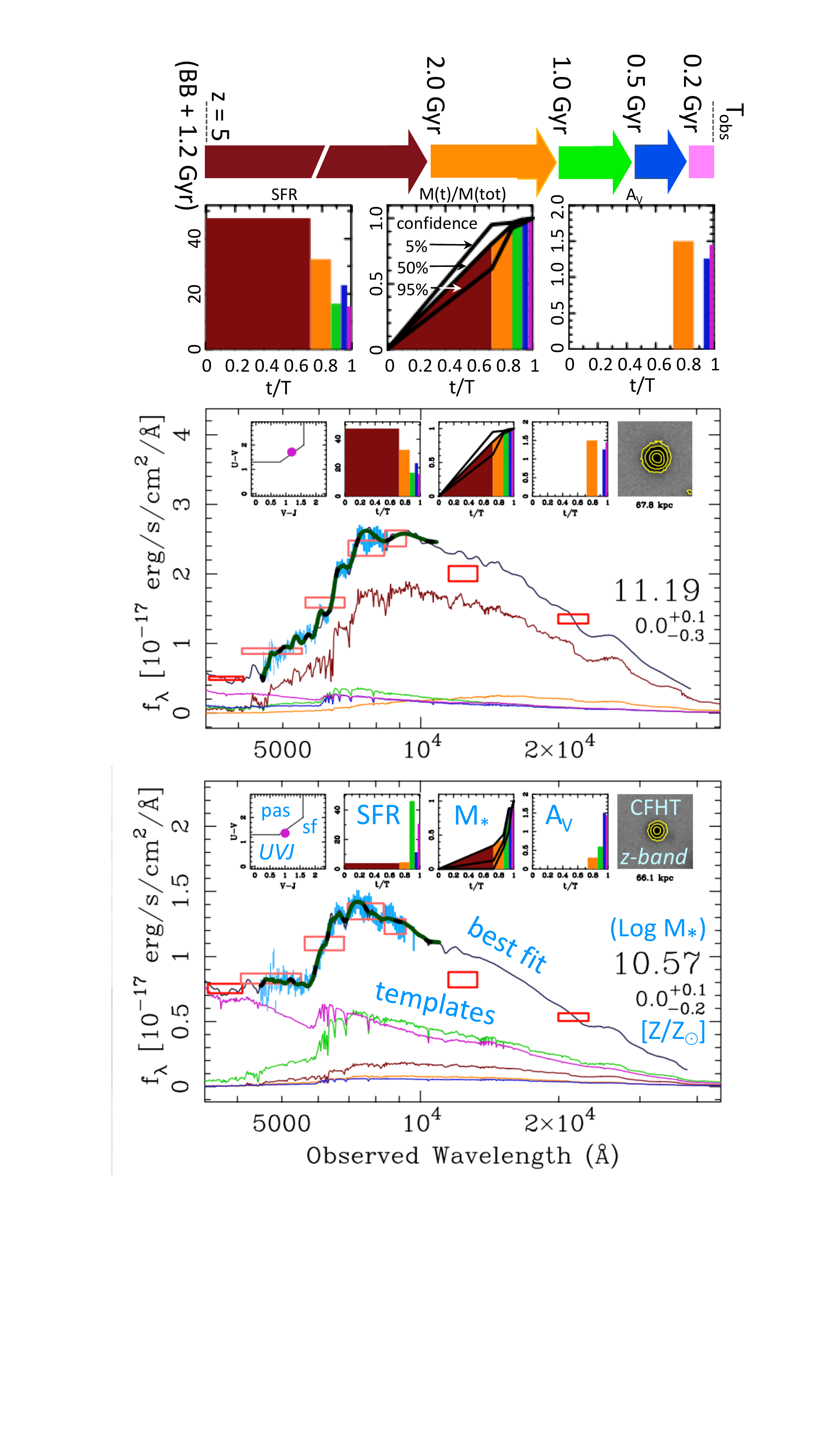}
}

\caption{Two example SEDs from CSI.  At the top of the figure are the 5 ``age bins" (in order of decreasing age, the brown, orange, 
green, blue, and magenta arrows) used to parameterize the SFH.  Each SED includes: (1) the data---broad band photometry (8 red boxes; 
{\it Spitzer} 3.6\,$\mu$m fluxes were used for selection only, not in SED fitting) and the \imacs\ prism spectrum (blue trace); (2) the 
most-likely SED fit (black line); (3) stellar templates for constant star formation in the five color-coded age intervals; (4) $\log\Mstel$; 
{\bfp (5) the metal abundance of the fit model.  The thin black line is the maximum likelihood fit---the sum of the templates, but at the 
lower spectral resolution of the prism data.  The width of the heavier line through the prism data covers the 5\%--95\% uncertainty of the 
fit model.}  The five boxes above the SED show: (1) \emph{UVJ} diagram (passive vs active SF); (2) the mean SFR  over each age interval 
(SFH); (3) {\bfb fractional} stellar mass growth; (4) extinction ($A_V$ mag) for each age interval; (5) isophotal $z$-band image from 
CFHT Legacy Survey. The three black lines in the center box show the confidence interval derived for stellar mass growth---5\%, 
maximum likelihood, 95\%. The top SED shows a well-constrained SFH; nevertheless, the range of  60\%--95\% contribution from the old 
population (born 2 Gyr or more before \Tobs) covers both a dominant old galaxy and one that has formed stars more-or-less constantly up to 
\Tobs.  The bottom SED has a confidence interval that appears wider, but an old contribution of less than 50\% (built up in $\sim$5 Gyr 
compared to 2 Gyr) identifies this galaxy---with high confidence---as a \emph{late bloomer}.  
}
\label{fig:schematic}
\end{figure*}

Figure \ref{fig:schematic} (top) shows our new scheme for parsing SFHs into five age intervals: {\bfp (1) constant SFR from 
200 Myr prior to \Tobs; (2) 500 to 200 Myr; (3) 1 Gyr to 500 Myr; (4) 2 to 1 Gyr; (4) from $z=5$ to 2 Gyr before \Tobs.} In addition 
to  better time resolution of SFHs back to $z\approx1$, these choices represent an improvement over the four equal-length time 
bins because they are attentive to the natural timescales of stellar populations as expressed in stellar spectra.   

Figure \ref{fig:schematic} also explains how the SEDs are presented in Section \ref{sec:catalog}.  The constraining data are photometry in 8 
broad bands (\emph{ugrizJK$_s$}---the red boxes) and an \imacs\ prism spectrum covering (rest-frame) 3000--4500\ang\ (the blue trace). 
The  solid black line shows the best fit to the photometry and spectrum produced through the modeling.\footnote{The methodology used 
to derive the average SFR in each bin from these data is described in detail in K14 and D16.} The five components to the model SED are 
shown as template spectra at the derived flux level for each age bin where star formation has been detected. The templates are color-coded 
to match the SFRs and integrated stellar mass, as shown in the enlarged left and center boxes above the upper SED.  The modeling also 
solves for  $A_V$ extinction (right-center box, following Calzetti \etal\ 2000) and metal abundance (not shown) for each of the stellar 
populations.  

The two example CSI SEDs shown in Figure \ref{fig:schematic} from the statistical sample (see Sec \ref{sec:LBfract}) span the wide diversity 
of SFHs found by D16.  The upper example is a $z=0.62$ galaxy dominated by an old stellar population, the kind of galaxy that has long been 
recognized in this  subject.  It is noteworthy, however, that this very massive galaxy is not ``quenched," but shows a continuing  history of star 
formation that, in the best fitting model, adds $\sim$25\% in stellar mass since $z\approx1$.  The lower example is a $z=0.586$ galaxy 
dominated by star formation after $z\approx1$.  Active star formation is well detected in both galaxies, beginning 500 Myr before \Tobs, 
and accompanied by 1.3--1.5 magnitudes of extinction.


However, the ``best fit" model of the SED  is not the whole story.  The most important modification of our approach since D16 is 
the extraction of confidence intervals for the SFHs, which previously had been recorded only as the fit of maximum-likelihood (ML).  
Adding the bounding values that specify the 5\% and 95\% confidence fits provides an essential metric of SFH reliability.  Focusing on 
the center box (the stellar mass growth), we see that the SFH of the top SED constrains the population of stars older than \Tobs-- 2 Gyr  
to between 60\% to 95\% of the stellar mass.  In other words, our confidence interval runs from a galaxy that is almost completely 
old to one that has essentially ``constant star formation" from $z=5$ to \Tobs.  The 5\%--95\% confidence interval for the bottom example 
runs from a galaxy with zero old stellar population to one with as much as $\sim$35\%.   Either way, this galaxy is found to be very 
young, despite its Milky Way (MW)-like mass (i.e., $\log\Mstel\in[10.5,10.8]$).

We use these SFH confidence intervals in the following discussion to define a class of massive galaxies that has not been recognized: 
{\bfb massive} $z\sim0.6$ galaxies whose {\bfb stars} formed mainly within 2 Gyr following $z\sim1$, instead of the preceding  
$\sim$5 Gyr.   We call these ``late bloomer" galaxies. 

\subsection{Why Add a 1--2 Gyr Stellar Template?}
\label{sec:2gyr}

Figure \ref{fig:stelTemps} shows that stellar populations of age $\ls$1 Gyr can be unambiguously separated from the light  of much older stars.  
D16 used this as a conservative, reliable way to distinguish young from old populations.  However, from the vantage point of galaxies 
observed $\sim$5 Gyr earlier than today, with their lower fraction of very old stars---all less than 7 Gyr old---a 2 Gyr separation of old 
and young populations is feasible given sufficient $S/N$, as Figure \ref{fig:stelTemps} also shows.

There are good reasons for adding 1--2 Gyr-old stars to the ``young" category.   First, separating ``young" from ``old" populations at 1 Gyr 
made it likely that a significant fraction of recent star formation, 1--2 Gyr before \Tobs, was erroneously credited to a population that 
was, on average, \emph{much} older.  This is mitigated by defining old as ``stars forming earlier than 2 Gyr before \Tobs." Second, the 
identification of late bloomers based on populations less than 1 Gyr from \Tobs\ implied an almost bursty history, while it is more 
sensible to expect this late epoch of star formation to have lasted several Gyr {\bfb (e.g., Chauke \etal\ 2018)}.  Even if the timescale 
for mass growth is as short as $\sim$1 Gyr, this is far from the conventional situation of incremental mass growth in a burst, since the mass 
growth has been sufficient to surpass the mass of old stars, those born prior to 1 Gyr, or now, 2 Gyr.  In {\bfb the appendix} 
we show, through the analysis of simulated SFHs, that detections of 1--2 Gyr-old populations are generally reliable, 
and thus an improvement in identifying late bloomers.

\subsection{The Re-Definition of a ``Late Bloomer''}
\label{sec:lbdef}

Based on the addition of the 1--2 Gyr star formation template, we thus revise the definition of a late bloomer used in D16 based on 
\emph{z5fract}---the fraction of mass formed earlier than 1 Gyr before \Tobs---to one based on \z5fract2, the fraction of mass 
formed {\it 2} Gyr or more prior.   Table 1 (see Section \ref{sec:catalog}) includes \z5fract2\ and recalculated \emph{z5fract} values 
based on our new SFHs, with the modifications described above.  

Formally, we define a \emph{late bloomer} as a galaxy in which $\le$50\% of the stellar mass formed in the epoch $z=5$ to 2 Gyr 
before \Tobs\ at the 95\% confidence level; i.e.,
\begin{equation}
	\z5fract2_{95\%}<0.5
\label{eq:lbdef}
\end{equation}
 
\noindent{Through this definition---which our simulations show leads to the best compromise between sample purity and completeness (see
 {\bfb the appendix})---mass growth in the final 2 Gyr for late bloomers exceeds that of the previous 5 Gyr.   Obviously, this implies rising 
SFRs (i.e., accelerating mass growth). Moreover, though, it implies SFRs are rising {\it faster than linearly} at these epochs:  if $\sfr\propto t$, 
$\Mstel(t)\propto t^2$, for which it so happens that \z5fract2 $\simeq\,0.5$ at $z=0.65$.\footnote{Constant SFRs imply ${\it z5fract2}=0.7$.} 
This means that our late bloomer definition is conservative---it excludes galaxies that have substantial (though not super-linear) increases in 
SFR over the final 2 Gyr.  As such, our abundance estimates for late bloomer-like systems should be higher than what we quote in Section 
\ref{sec:LBfract}. }

Although the galaxies on the other side of  this divide are dominated by star formation before $z\sim1$, our data strongly suggest that 
not all of these followed similar SFHs that were at some point quenched on a $\ls$1 Gyr timescale.  Rather, we see these older galaxies 
as including those whose star formation rose and and fell rapidly in the first few Gyr of cosmic history, and those that rose slowly with 
more-or-less continuous star formation until $z\sim0.6$---if not all the way to the present epoch.  For the purposes of the following discussion, 
we define old galaxies as those with $\z5fract2\ \geq0.85$,  and constant-star-formation galaxies as forming 50\%--85\% of their mass 
before $\Tobs-2$\,Gyr ($0.5 < \z5fract2 < 0.85$).  The latter interval begins with SFRs that are rising in the final 2 Gyr (not super-linearly) 
and ends with SFRs that peaked before $z\approx1$ and are slowly declining by $z\sim0.6$.

\subsection{Spectral Characterization of Late Bloomers}
\label{sec:specChar}

Outfitted with an improved toolkit for turning SEDs into SFHs, we focused on the major issue of the relative proportion of 
young to old stellar populations in our sample galaxies.   

Much of the leverage for deriving a SFH from a SED rests in the low-spectral-resolution ($R\sim30$) prism spectrum, which is 
particularly sensitive to young and intermediate-age populations, and the 4 broad-band photometric indices \emph{i, z, J, K} that 
measure the contribution from the red-giant-branch stars of old stellar populations. (The {\it Spitzer} 3.6\,\micron\ detection 
band is omitted in the SED fitting to avoid selection bias.) Galaxies with star formation during the 2 Gyr before \Tobs\ (since 
$z\approx1$, for our sample)  show unambiguous evidence through the Balmer break, primarily from A stars, but also from F stars.  
This signature is readily distinguishable from the contribution from older stars to this spectral region, which exhibits a strong 
D4000 break and other prominent features---the G-band, Ca H\&K lines and CH complexes.  (This spectral region is called the 
``break region" in the discussion to follow.)  Massive galaxies can have substantial contributions from both old and young stars, 
so the spectral resolution $R\sim30$ of the prism observations is an important advantage over broad-band photometry SEDs in 
distinguishing the contributions of both young and old populations.  

The old stellar population (i.e., older than 2 Gyr at \Tobs) dominates the flux further into the red and infrared,  so good broad-band 
photometry beyond 6000\ang\ is generally sufficient for assessing their contribution. However, the added constraint of the break 
region may be essential if the older population is viewed through a dusty disk, which can further diminish or eliminate its signature 
at rest-frame $\lambda\sim4000$\ang.   Since testing the reality of late bloomers depends critically on whether or not an old 
stellar population can be detected, we tested the dust attenuation issue through simulations of mock spectra that included substantial 
``hidden" old stellar mass.  These tests are summarized in the appendix {\bfb and included in our full sample result uncertainties (Section \ref{sec:LBfract}). The upshot is that the impact of such hidden mass is unlikely to be dramatic, such that the observed late 
bloomer abundance is accurate to roughly $\pm0.05$ in absolute terms over a wide range of $S/N$, $A_{V}$, and $\sfr(\Tobs)$ mixes.} 

\subsection{Role of Duplicate Observations}
\label{sec:dupes}

{\bfp CSI has a $\sim$20\% repeat observation fraction to
facilitate empirical estimates of measurement errors, which can be nonlinear functions of the
observations. In general, these duplicates can unfortunately not be used to test the repeatability of binary 
classifications such as being a late bloomer. Such groupings are like a biased coin toss, 
with the bias set by the selection method's purity. In our case, simulations suggest (and the 
assessments below reveal) a $P\simeq70\%$ purity, which is not close enough to unity to guarantee 
repeat classifications ($\propto P^{n_{\rm obs}}$) agree. Indeed, at $P=0.7$, we expect to repeat-confirm
only 58\% of the initial late bloomer candidates.\footnote{\bfp Counterintuitively, due to the 
commensurately higher initial false-positive fraction, $P=70\%$ produces the same fraction of 
repeat classifications as a paltry $P=30\%$ (though many more will be intrinsically incorrect). 
The purpose of Section \ref{sec:catalog} is to provide targets for {\it new} observations 
with potentially greater selection purity/discriminatory power than our own.}

The proper use of repeat observations at moderate purity is instead to characterize formal
measurement errors in a continuous---not binary---property. This is analogous to repeated 
flux measurements in noisy images: the scatter between them reflects the true underlying noise level. 
In the context of late bloomers, the analogy is repeated assessments of the amount of old stellar 
mass in a galaxy. The scatter between estimates reflects the noise floor on this estimate. Our
classifications should be reliable if this floor is $\lesssim$50\%. Fortunately, this is the case.

Averaged over total stellar masses $10\le \log\Mstel^{\rm tot} \le 11$, the RMS scatter in the 
{\it differences} of maximum likelihood old stellar mass fractions across duplicates is 
$74\% \pm 2\%$. Assuming Gaussianity, this is $\sqrt{2}\times$ the formal error on a 
single measurement, implying an acceptable $52\% \pm 1\%$ 1\,$\sigma$ noise floor on 
\z5fract2. If these estimates are restricted to galaxies whose first observation implied 
identically zero old mass, the noise floor---closer to a real upper limit on \z5fract2 in these 
cases---drops to $46\% \pm 2\%$.

As discussed above and in the appendix, however, we do not select late bloomers using the 
maximum likelihood old mass fractions, but their estimated 95\% upper limits. For galaxies with 
zero maximum likelihood old mass, we find the mean of these limits in the second observation to be
$75\% \pm 3\%$ ($10\le \log\Mstel^{\rm tot} \le 11$). For Gaussian noise, this is $1.960\times$ 
the standard deviation, implying just a $38\%\pm 1\%$ \z5fract2 1\,$\sigma$ noise floor. 
Including all galaxies raises this to $48\% \pm 1\%$.

In sum, both duplicate-based approaches support the conclusion that our LB selection method 
is accurate at the 1\,$\sigma$/68\% level. Reassuringly, this is fully consistent with the 
$\sim$70\% purity we infer from simulations (see the appendix).

Comparing duplicate observations and SFHs also helped us optimize our SED fitting, modifying 
the procedure described in K14 by narrowing dust and metallicity constraints for the oldest population. 
This allowed finer parameter grids, which improved redshift estimation accuracy. Because of the 
similarity in shape and location of the Balmer break in young populations and the 4000\,\AA\ break 
in old populations, better redshifts led to higher fidelity SED-derived SFHs.}

\section{The Late Bloomer Fraction of CSI Galaxies at \lowercase{\it z}\,$=0.6$}
\label{sec:LBfract}

In this section we quantify the fraction of CSI galaxies that were late bloomers $\sim$6 Gyr ago. This measurement simply entails
summing the weights of the relevant galaxies, described in K14. Every galaxy's weight is the inverse of the completeness estimated 
for its magnitude, color, and local source density, divided by the number of times a galaxy was observed.  The CSI late bloomer 
fraction (LBF) is therefore the sum of the weights for galaxies with $\z5fract2_{95\%}<0.5$ over the sum of the weights of the
galaxies in the full catalog.   {\bfp In order to ensure fidelity in late bloomer fraction estimates, we reduced the size of the full 
XMM-SWIRE CSI catalog of 50,000 high quality redshift measurements to approximately 22,000 
systems with high-quality observations and stellar masses above $10^{10}\,\Msun$.}

{\bfb To assess the LBF's sensitivity to sample depth, $S/N$, redshift uncertainty, and spectrophotometric quality variations, 
we used seven different versions of the CSI dataset based on different cuts in these dimensions. We fold the RMS scatter 
(a few percent)  between LBFs from these different selections into the error bars in our plots. These different versions of 
the catalogs are valid statistical samples with their own completeness estimates as functions of magnitude, color, and source
density, with sizes ranging from 8,000 to 12,000 galaxies with $\Mstel>10^{10}\,\Msun$ at $0.45<z<0.75$. For the measurements 
of LBF evolution (Section \ref{sec:when}), this sample is extended over $0.25<z<0.75$ and is 25\%--30\% larger, depending on 
the selection cuts.}

\begin{figure}[t]
\label{fig:lbfs}
\centerline{\includegraphics[width=0.85\hsize]{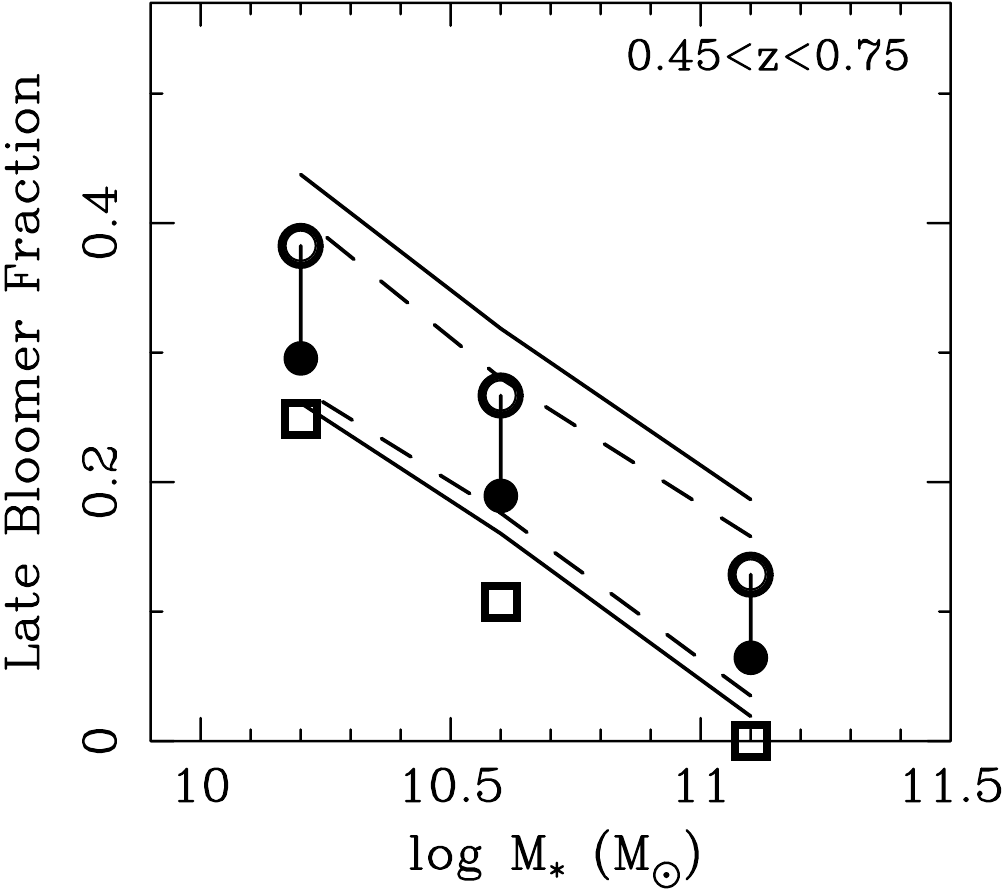}}

\caption{Late bloomer fractions at $0.45<z<0.75$ in three different stellar mass bins. A substantial fraction of galaxies 
at these epochs had recently experienced runaway growth.  {\it Open circles}: Raw CSI measurements; {\it Filled circles}: 
Reasonable lower bounds on true LBFs based on simulations in the appendix; {\it Open squares}: Pessimistic lower bounds 
assuming the raw measurement at $\log\Mstel\sim11$ represents {\it only} false positives.  {\bfp The upper dashed line 
trace 1-$\sigma$ upper bounds on the observed LBFs.  The upper solid line includes an additional $\pm 0.05$ uncertainty 
in those upper bounds due to potential uncertainties in the correction factors to LBFs, as derived from scatter between 
different simulations of CSI data.  While the filled circles trace reasonable lower bounds on the LBFs, using plausible
corrections to the measurements, the lower dashed line is the 1-$\sigma$ lower bound on these corrected LBFs from the 
formal errors alone.  The lower solid line includes an additional uncertainty of $\pm 0.05$, derived from scatter 
between different simulations of CSI data.}}

\end{figure}

Figure \ref{fig:lbfs} shows the measured LBFs as functions of stellar mass at $\langle z\rangle=0.6$ (open circles). Simulations
of CSI data described in the appendix allow us to quantify the systematic bias in these measurements due to contamination by 
false positives (galaxies that grew less than half their mass in the last 2 Gyr).  According to these simulations, this bias is approximately 
zero if late bloomers are selected using the 95\% upper limits on \z5fract2---the definition we adopt (Equation \ref{eq:lbdef})---with 
an additional uncertainty of $\pm 0.05$ due to model-to-model variations, uncertainties in the mix of quiescent and star-forming galaxies, 
and uncertainties in the underlying mix of SFRs at \Tobs.

The simulations also let us construct plausible lower limits to the LBFs, with systematic contamination at levels of $+0.05$ for 
quiescent galaxies and $+0.10$ for star forming galaxies. These plausible lower bounds are shown in Figure \ref{fig:lbfs}
using the filled circles. 

A third, empirical, and more pessimistic approach is to assume that all late bloomers in the stellar mass bin $\log
M>10.8$ are false positives, such that the LBF in that mass bin is identical to the contamination rate (at least for populations
with the same mix of quiescent and star forming galaxies). By scaling the high mass LBF by the ratio of the quiescent fraction in
each bin to that at high mass, we can empirically correct each bin for the potential contamination by false positives. These
results are shown by the open squares.

Formal uncertainties in the raw measurements are estimated through bootstrapping. Systematic uncertainties in those measurements
are estimated using the RMS variation in LBF derived from multiple variations of the CSI catalog tailored in different ways, such
as varying $S/N$ restrictions, prism spectral quality, how well the spectra have colors that match the photometry,
etc. The dashed lines show their quadrature sum above and below the raw measurements and plausible lower bounds. The additional
uncertainty of $\pm 0.05$ due to model-to-model variations, uncertainties in the true mix of quiescent and star-forming, and
uncertainties in the underlying mix of ongoing star formation rates, added in quadrature, is thus shown by the solid lines.

Taking these measurements at face value, several striking conclusions are readily apparent:
\begin{itemize}
	\item Many galaxies at least doubled their stellar mass between $z=1$ and $z=0.6$.
	\item About $20\%$ of MW-mass galaxies did this, and $\sim$30\% of galaxies at half the MW's mass.
	\item More massive galaxies do this less, but, even at $\Mstel > M^*$, the LBF remains $\sim$5\%--10\%.
\end{itemize}

These relatively simple observations strongly confront the basic picture, commonly held, that most galaxies in this mass range
generally grew early, with much slower rates of growth after cosmic summer. They strongly contradict paradigms in which 
galaxies are thought to simply grow along the SFMS and quench {\it en masse\/}: Even modest numbers of galaxies---to say 
nothing of 20\%---with sustained, rapid, late growth, as shown above, have strong consequences for analyses of galaxy evolution 
that rely on the preservation of mass (and so abundance) rank ordering.

Given the strong implications for these measurements, we devoted great effort to simulating CSI data with the aim of understanding
how to make late bloomers ``go away." That is, we tried to identify what fraction of CSI late bloomer identifications
could be due to noise and, for example, dust effects, which could systematically hide old stellar mass and artificially make intrinsically 
non-late bloomers appear as late bloomers. The simulations allowed us to derive plausible corrections to the CSI measurements, 
but did not in any way indicate that these were substantially biased by errors in the SED-inferred SFHs.

The upshot of this exercise is that the $z\sim0.6$ LBF is highly unlikely to be zero for galaxies with stellar masses {\bfb at least} up 
to $\log\Mstel\sim10.8$.   Our most conservative assessment is that the LBFs of galaxies with half-to-all the MW's mass is at least 
10\%--20\%.  Stated most plainly, roughly one-in-five of today's MW-mass galaxies went through an extreme growth spurt between 
$z=1$ and $z=0.6$, having evolved only lackadaisically before then, and perhaps since. This is an odd conclusion from the standpoint 
of SFMS-integration  or abundance matching, but, try as we may, we cannot avoid it.

The simulations only provided one scenario to make our LBFs a complete procedural artifact: forcing half the
stars to have been formed in the first 1--2 Gyr, to be hidden or partially obscured underneath the stars that formed subsequently.
But even even in these cases, the inferred SFHs from the SED fitting represent the formation histories of the final $\sim$5 Gyr.
The implication is that such galaxies would have had lacunal histories: initial early ``bursts,'' rising again only much later to
rapidly grow the final quarter of their stellar mass in the 2 Gyr prior to $z\sim 0.6$. To contrive that all late bloomers---again, 20\%
of MW-mass systems---grew that way seems at least as challenging as accepting them as real late bloomers in 
the sense of Gladders et al.\ (2013), Kelson (2014), or Kelson, Benson, \& Abramson (2016; hereafter KBA16).

\section{A Catalog of Galaxies with HST Imaging and Secure SFHs}
\label{sec:catalog}

This section provides a catalog of galaxies with high-confidence SFHs to serve as a reference sample for comparison with 
and by other studies.  The 5 square degrees of the XMM-SWIRE field containing the sample in this paper overlaps by 
$\sim$1\% with CANDELS HST imaging (Koekemoer \etal\ 2011).  We chose to focus on this area for our catalog since 
high-resolution images of the calibrator galaxies should be indispensable in evaluating SFHs in terms of galaxy structure and 
morphology, the presence of near neighbors, distribution of star formation, and notable features (for example, tidal tails) that 
could be identified.  Images also set the stage for later observations and studies motivated by interesting SFHs.

{\bfp To assemble a representative set of high-quality SFHs for the catalog, we started with the statistical sample from 
Section \ref{sec:LBfract} restricted to the redshift range $0.45<z<0.75$. (For the median redshift of $z\approx0.6$, 
1 and 2 Gyr before \Tobs\ are $z\approx0.80$  and $z\approx1.0$, respectively.)  We further reduced the sample by} 
cutting at (1) $S/N>15\,{\rm pix^{-1}}$ (for the prism spectrum continuum), (2) $\Delta z<1.5\%(1+z)$ $(\pm0.024$ 
at $\langle z\rangle\sim0.6$), and (3) more restrictive quality criteria on the spectral flux calibrations. Thus, out of ~230 
CSI galaxies in the redshift range of interest, these restrictions yield a sample of 128 high-confidence CSI sources with joint 
CANDELS F606, F814, and F160W ACS/WFC3 imaging with which to make RGB images. We focus on 74 of these galaxies 
(with 6 duplicates) as a ``gold standard'' sample below, for which confidence in their early mass fractions is high, based on 
the analysis of duplicates, and the simulations described in the appendix.


\begin{figure*}[t]

\centerline{
\includegraphics[width=0.98\textwidth, trim = 0cm 0cm 0cm 0cm, angle=0]{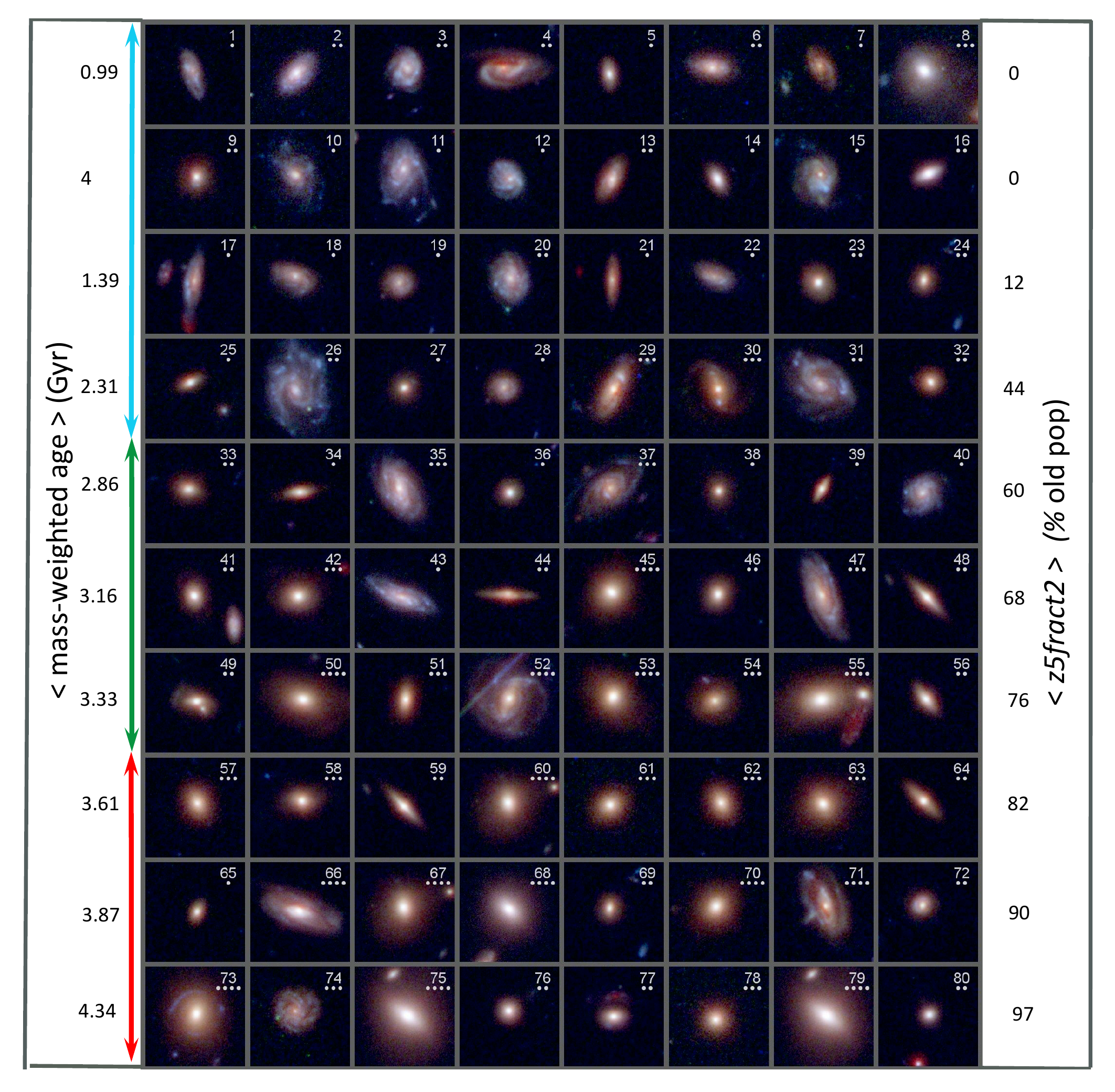}
}

\caption{\emph{CANDELS} HST images for high-confidence SFHs from this study.  80 observations of 74 galaxies with 6 repeat SED 
measurements (see Figure \ref{fig:mosSEDs} and Table 1). Images (identical for repeats) are arranged by increasing \z5fract2, 
as is indicated by the mean value of \z5fract2\ (right column) for each row, expressed as the percentage of old stellar population 
(rather than a fraction).  Images are \emph{RBG}, created from F606, F814, F160 images from WFC3; each box has a physical 
scale of $\sim$35 kpc on a side. The number in the upper right corner of each image is the catalog number in Table 1.  The dots below 
the catalog number give the measured stellar mass for each observation binned by factors-of-two: $\log\Mstel = 10.0$--10.3 (one dot), 
10.3--10.6 (two dots), 10.6--10.9 (three dots), 10.9--11.2 (4 dots).  {\bfp The \emph{mass-weighted age in Gyr (not to be confused with 
commonly used \emph{light-weighted} age}from the vantage point of \Tobs, averaged over each row, is given in the left column.}  
Three ranges of SFH fits are indicated: (blue arrow) late bloomers---95\%  confidence $<$50\% old population ; (green arrow) old, approximately constant star formation---95\% confidence $>$50\% old population;  (red arrow) old---95\% confidence $>$70\% 
old population.  The duplicate IDs---19/28, 24/69, 48/59, 57/62, 60/67, 75/79---show that SFHs place both members of the pair in 
the same category---late bloomer, constant star forming, old galaxies.}

\label{fig:mosaic}
\end{figure*}


\begin{figure*}[t]

\centerline{
\includegraphics[width=8.0in, angle=0]{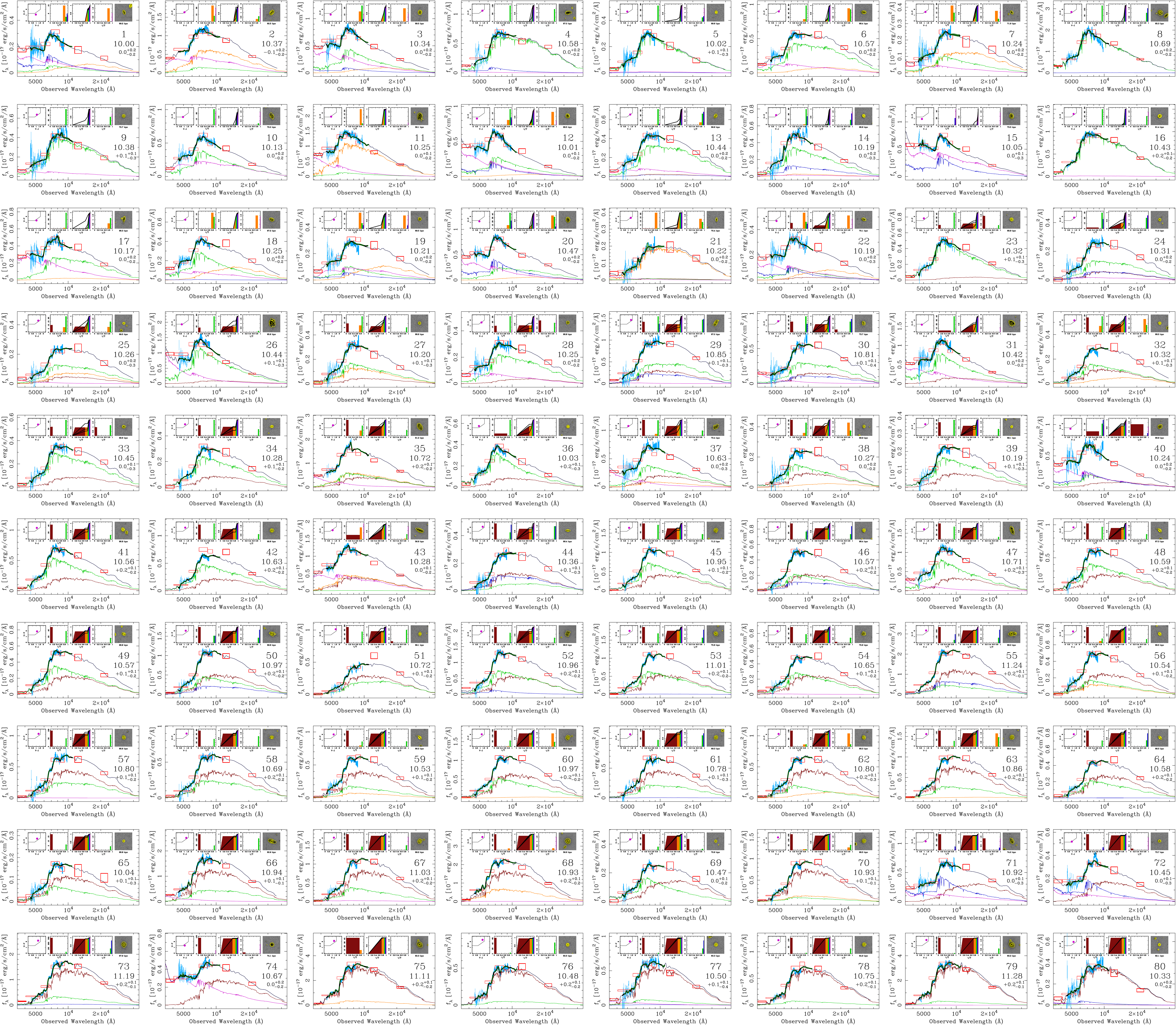}
}

\caption{SEDs for Galaxies with HST images and high-confidence SFHs from this study.  Documentation on how to read these 
SED plots is given in Figure 2.  The SEDs are arranged to match the images in Figure 4 and are include the image/catalog number 
for convenience.} 

\label{fig:mosSEDs}
\end{figure*}


\begin{figure*}[t]

\centerline{
\includegraphics[width=6.5in, angle=0.0]{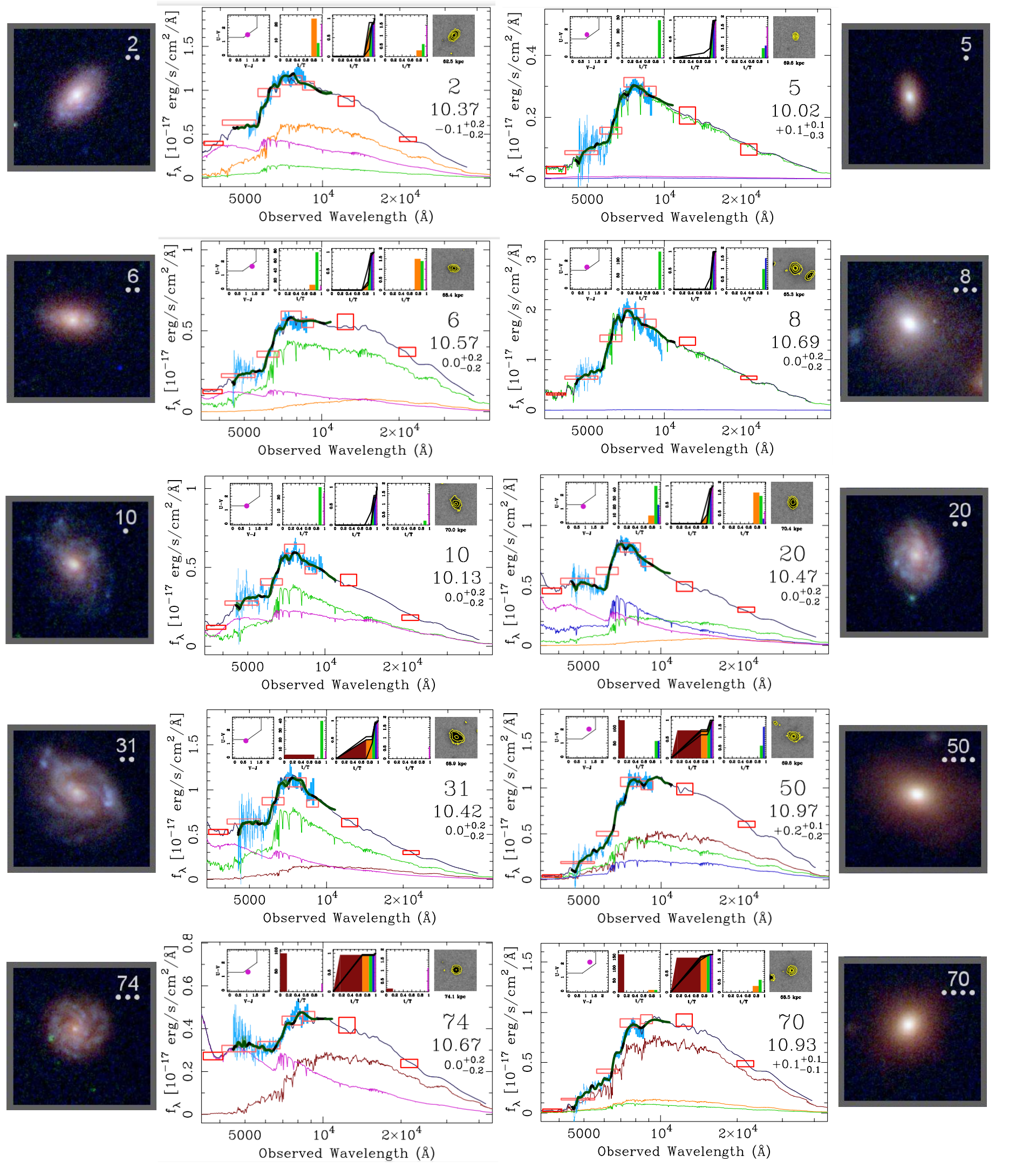}
}

\caption{Images and SEDs for 10 galaxies with high-confidence SFHs from this study, showing the range of
histories and morphologies of galaxies, from old galaxies to late bloomers.  The top three rows are for late 
bloomers: each galaxy has a contribution from an old population (earlier than \Tobs--2 Gyr) of less than
50\%, at the 95\% confidence level.  IDs 5 and 8 appear to have ``early type" morphologies and a single stellar 
population of age 0.5--1.0 Gyr that well fits the SED, as discussed in the text.  {\bfp It is worth noting that, in the
UVJ diagram (top left box), three of the six late bloomers are clearly situated in the ``passive" area, showing that this 
diagnostic is indicative of the most recent epoch of star formation rather than a more general indicator of long term 
star formation or passivity.  The galaxies in the bottom two rows have old populations of more than 50\% at the 95\% 
confidence level. The CSI SFHs for IDs 31 and 50 are more-or-less constant in time (a diagonal line in the mass growth 
plot) while 74 and 70 have are dominated by old stars ($>$70\%) with contributions of $<$10\% of stars born within 
2 Gyr of \Tobs.  With respect to morphology,} IDs 50 and 70 are early types, while ID 31 is clearly a spiral disk galaxy.  ID 74 
shows a spiral pattern, but this is apparently due to a small fraction of stellar mass in a starforming disk, in an otherwise 
spheroidal, old galaxy.} 
\label{fig:imSEDs}
\end{figure*}

Figure \ref{fig:mosaic} is a mosaic of the 80 HST images of the 74 catalog objects.\footnote{IDs 19/28, 24/69, 48/59, 57/62, 60/67, 
75/79 are repeats (Tables \ref{tab:inCANDELS0}, \ref{tab:inCANDELS1}).}  Figure 5 shows the corresponding SEDs for each 
observation as described below. Table \ref{tab:inCANDELS0} identifies these objects by RA and DEC and provides basic data for 
each of the 80 observations: $i$-band magnitude; prism spectrum $S/N$; $z$ (1$\sigma$ errors); {\bfp $\log\Mstel$\,(5\%--95\% 
confidence interval); fractional mass-growth history (1$\sigma$ errors) at 2 Gyr (i.e., \z5fract2), 1 Gyr (i.e., \emph{z5fract}), 500 Myr, and 200 Myr before \Tobs; and \emph{n2}, local galaxy number density in Mpc$^{-3}$ derived from the full CSI catalog ($r=2$ comoving 
Mpc aperture)}.

Figure \ref{fig:mosaic} is arranged by increasing \z5fract2: late bloomers are on top with the oldest galaxies at the bottom.  {\bfp The 
sample is not strictly speaking a random draw} from the 128 possibilities because it is biased to more-certain SFHs and to late bloomers, 
which make up 40\% of the galaxies in this ``gold sample," compared to 20\%--30\% of observed galaxies overall (see Section 
\ref{sec:LBfract}, Figure \ref{fig:lbfs}.  The numbers to the right of each row give its $\langle\z5fract2\rangle$ as a \emph{percentage}, 
with mean mass-weighted-ages on the left. All quantities are derived from the SFHs/SEDs of Figure \ref{fig:mosSEDs}.  

{\bfb As the appendix} explains, based on simulations, the \z5fract2 values for the \emph{late bloomers} in Figure \ref{fig:mosaic} 
can be treated as lower limits: the 0\% in the first two rows reflects non-detections of a population that may be as high as 60\%. {\bfb However, the 
simulations also suggest that the late bloomer sample in these figures/tables is about 75\% pure, and that a majority of the  25\% contaminants that 
formally breach the $\z5fract2<0.5$ bound have, in reality, $0.5 < \z5fract2 < 0.6$.} These are still very young galaxies, considering that this is the 
fraction of {\bfb mass} produced in the first $\sim$5 Gyr compared {\bfb that made in just the} 2 Gyr before \Tobs. {\bfp In any event, future 
investigations using better data with higher resultant purity ($P\gg0.75$) in their SFH classifications should concur on about $3/4$ of our late bloomer 
classifications (see Section \ref{sec:dupes}).\footnote{SFH reconstructions of similar quality to ours should agree at the $\sim$60\% level.}  We encourage such
new observations but are open to sharing our own data for use in novel, improved analyses.}

\subsection{Morphology and Galaxy Age}

Figure 4 is divided into three sections: 4 rows of late bloomers (the blue range), 3 rows of more-or-less constant star formation 
(the green range), and 3 rows of the oldest galaxies (the red range). {\bfp The main bias of the 74 galaxy catalog with respect to a random 
draw of the full CSI \emph{SWIRE} catalog is only an over-representation of late bloomers at the $\sim30$\% level. Therefore, we 
can use this subsample to make two remarkable assertions about galaxy evolution.}  The \emph{CANDELS} images in 
Figure \ref{fig:mosaic} tell us that 

 \begin{enumerate}
	\item Massive galaxies, $\Mstel > 10^{10}$\Msun, come in a range of ages, including those that formed all their stars early 
	in the universe, those that formed them over a long time, \emph{but also late bloomers}, which formed most of 
	their stars after the universe was already 6 Gyr old.  There is a trend with mass in the sense that less massive galaxies are 
	generally younger, but as we 	have found many times in studies of galaxy evolution, all types are represented at all masses, with 
	differences only in the mean;
 
	\item Galaxies of all ages come in all morphologies.  There is a trend in Figure \ref{fig:mosaic} that younger galaxies are 
	generally more disky and the oldest galaxies are more likely to be spheroidal, but again, for each class, 
	all morphological types are represented: only the mean type changes.
\end{enumerate}

The first point above is made even in the first row of galaxies, for which {\it no} old population is detected: three have {\bfb masses}
close to 10$^{10}$\Msun\ (IDs 1, 5, 7),  and one galaxy is more massive than the Milky Way (8).  Of the 7 different galaxies in 
the bottom row---almost entirely old stellar populations---three have masses of $\sim$2 $\times$ 10$^{10}$\Msun\ while 2 
exceed 10$^{11}$\Msun.  In a similar vein, disk galaxies with large, presumably massive bulges are found all over the mosaic, 
for example, IDs 2, 8, and 23 among the late bloomers and 53, 60, and 75 among many of the old galaxies.  Likewise, small bulge 
galaxies are probably expected for late bloomers---for example, 1, 15, and 22---but they are found all the way down the age 
sequence, for example, 43, 52, 66, 74.  

The point is that galaxies encompass SFHs that are very fast, very slow, or very late in a way that is correlated to mass and 
morphology, but not in a strong way.  Simple models that predict the appearances of galaxies, old to young, tell
only part of the story.

\subsection{Ten Examples of Late Bloomers and Their Elders: Images and SFHs}

Figure \ref{fig:imSEDs} provides images and SEDs for 10 galaxies: 6 late bloomers and 4 old, 2 of which are have long-term,
continuing star formation.  Along with the two SEDs of Figure 2 and its discussion in Section \ref{sec:data}, this figure
will help the reader interpret the compendium of 80 SEDs for the 74 cataloged galaxies in Figure \ref{fig:mosSEDs}.

Among the 6 selected late bloomers, we see a variety of SFHs, but all developed after $z\approx1$, and for each the envelope 
of ``allowed" SFHs is very narrow.  By definition, none have a detected old stellar mass fraction of $>$ 50\% as an upper limit, 
but in these 6 examples, even within the 95\% confidence interval, there is \emph{no} evidence for a stellar population that formed 
earlier than \Tobs\ -- 2 Gyr (the brown rectangle is not present).  However, we reemphasize that this does not preclude the presence 
of a sizeable old population: our sensitivity to $<$10$^{10}$\Msun\ is borderline, and less than 3 $\times$10$^9$\Msun\ in old 
stars is very unlikely to be detected, as the discussion in {\bfb the appendix} explains. All 6 late bloomers show considerable 
star formation in the 1--2 Gyr and 0.5--1.0 Gyr populations (orange and green) or both, suggesting that these are galaxies with a 
several-Gyr history of star formation, as opposed to a few-hundred Myr burst.  One late bloomer is passive at \Tobs\ and---of the five 
that are starforming---SFRs have mainly declined in the last 0.5 Gyr.  All are dusty, with measurable, sometimes large extinction at all 
ages $<$2 Gyr.  Four of the 6 are disk galaxies {\bfb (for the four rows of late bloomers in Figure \ref{fig:mosaic}, 17 of 24 are)}, 
and all of these have small-to-moderate bulges.

The most striking feature of this group of late bloomers comes from IDs 5 and 8, both massive galaxies that would, by appearance, 
normally be considered early types. \emph{These appear to be entirely composed of A stars} (the green 0.5--1.0 Gyr population). 
Only a single young stellar template is required for a very good SED fit.  This suggests a remarkable SFH, to be sure, but perhaps more 
remarkable is the \emph{amount} of stellar mass involved: $\sim$10$^{11}$\Msun\ of gas was turned into stars in approximately 1 Gyr. 
(An all-green SFH solution is likely consistent with star formation over $\sim$2 Gyr, whose mean age is that of the green age bin, but this 
is hardly less surprising.) As we saw in the statistical sample discussed in Section \ref{sec:LBfract}, if these massive, spheroidal galaxies 
with a sudden stellar buildup after $z\approx1$ are not common, neither are they rare. Based on this small sample, 3 others---a total 
of 5 of the {\bfb 24} late-bloomers in Figure \ref{fig:mosaic}---appear to have the same morphology and whiter ``color'' that distinguish 
them as a different kind of system from the redder images of spheroidal galaxies identified as truly old, visual evidence that these 
are a different kind of system.  

In summary, not only is it remarkable that there are massive galaxies that make most of their stars around $z\sim0.8$, but 
moreover, some of these have the the ``early" morphology associated with old, massive galaxies, and certainly not associated 
with a large amount of young stars.  ``Late bloomers" already land in the ``unexpected" category, we think, but arriving in many 
morphological types adds to a puzzle, which must be solved---if understanding galaxy evolution is the goal.

For IDs 50 and 74, the two examples of more-or-less-constant star formation, the presence of multiple epochs of star formation 
is the common feature of this type.\footnote{Our data and methods probably cannot determine if the absence of one or two of the four late time intervals that make up the 2 Gyr before \Tobs\ is real---see, e.g., the missing 1--2 Gyr bin of IDs 31 and 50, or 
the 200--500 Myr bin of 31. Either object could have had a continuous star formation history.} Of the two, one is very actively 
forming stars but the other is not---it appears to be an old galaxy (and old looking, too) that had a MWs mass of gas dumped on it in 
the last Gyr or so, which it promptly turned into stars. ``Accretion" seems more than an understatement, but neither does this 
resemble a major merger, given the large gas fraction required.

The two very old galaxies in the bottom row are more than 95\% old, but each shows signs of a small amount of late star formation---a 
late ``frosting," and a heavy one at that.  For galaxy 74 the star formation is ongoing and a prominent spiral pattern is seen, but it seems 
likely that the bulk of the stellar mass is spheroidally distributed, with a thin disk hosting the brief return to life of this galaxy.  Galaxy 70 
is an example of what we would likely call, looking back from the present epoch, a completely old galaxy, suggesting that late episodes 
of significant star formation 5--6 Gyr ago were common for these as well (see also, e.g., Treu \etal\ 2005).  Neither of these galaxies 
will have been passive for all of its life up to ``today"---i.e., they will have left the quiescent region of the {\it UVJ} diagram after initially 
entering it---something that deserves more investigation with larger samples that include morphologies.

\section{Discussion}
In this section, we begin with a phenomenological description of late bloomers in the broader context of what is known about the evolution 
of MW-like galaxies.  We then ask how late bloomers fit into the dark matter halo/$\Lambda$CDM picture via a comparison to a 
semi-analytic model. This is only a short introduction to what could be a complex and broad-reaching challenge:  What aspects of the widely 
accepted picture of halo and stellar mass growth---and the agents of abatement that lead to a ``cosmic winter" of galaxy building---are actually 
physically illuminating.

\subsection{Late Bloomers: What, When, Where, How, and Why?}
\label{sec:lbchar}

For newly recognized phenomena, these simple interrogatives define a traditional first step towards understanding.  In this section, we review 
the state of the subject of late bloomer galaxies as we now see it.

\subsubsection{``What?''}

The ``what" of late bloomers rests on the reliable determination of SFHs that are not part of the canon---i.e., do not correspond
to those generated by integrating scaling laws such as the SFMS. The direct implication is that a significant fraction of 
\emph{presently} MW-mass (and probably greater) galaxies experienced most of their star formation in the ``cosmic autumn"
instead of ``cosmic spring" {\bfb or ``summer,''} something that has not been recognized from studies of present-epoch galaxies (though 
cf.\ Mar\'{i}nez-Garc\'{i}a et al.\ 2017).  From a theoretical perspective, late bloomers {\it might} be outside expectations if their 
growth in stellar mass departed substantially from the growth of their dark matter halos, as judged from, e.g., $\Lambda$CDM 
$N$-body simulations.  In this context, late bloomers could be galaxies whose SFHs are delayed with respect to their dark matter 
halo growth. Section \ref{sec:context} discusses the above questions in more detail.


\begin{figure}[t]
\centerline{
\includegraphics[width=\linewidth, angle=0, trim = 1cm 0.5cm 1cm 0.5cm]{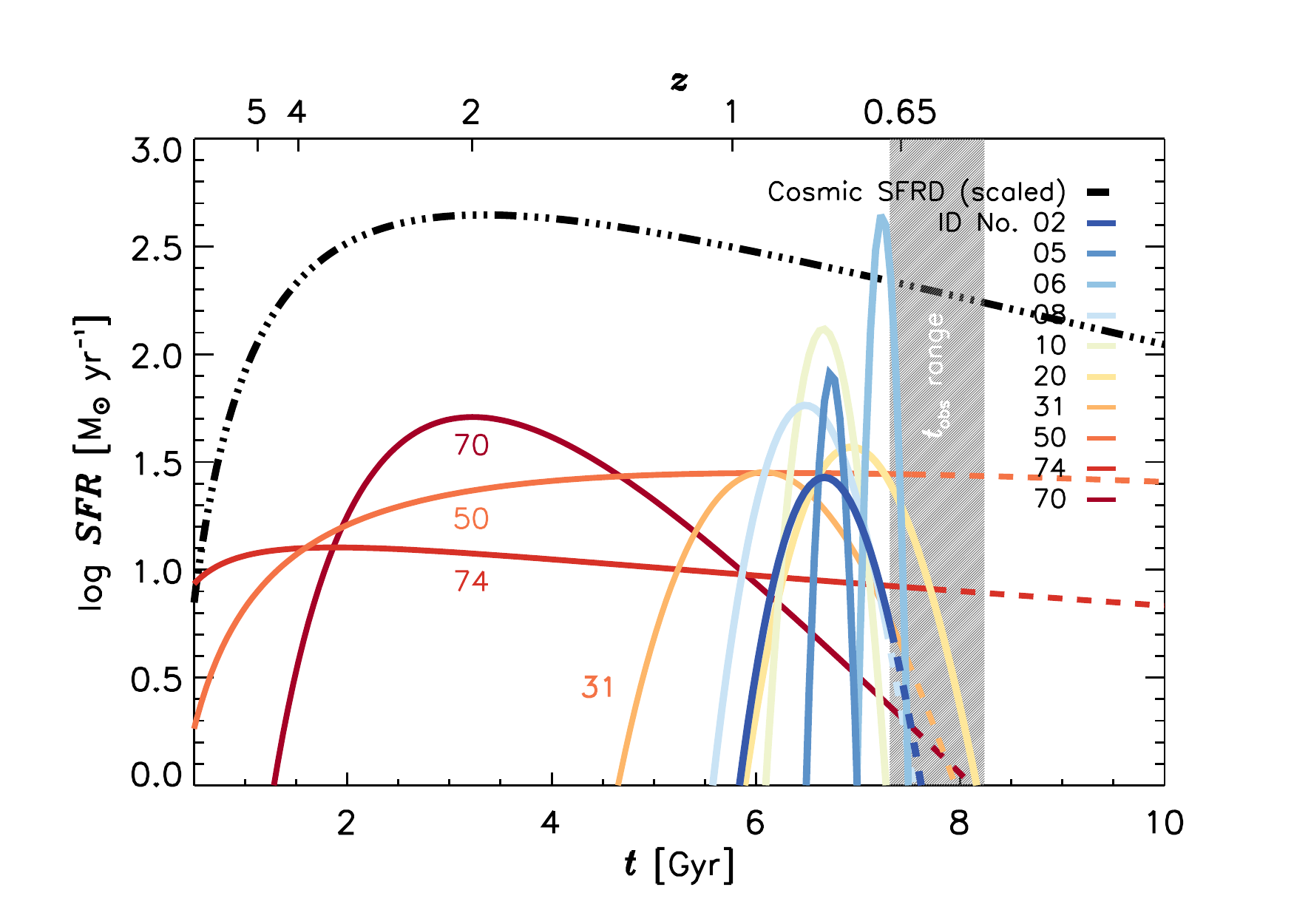}
}
\caption{Lognormal SFH fits for the 10 galaxies in Figure \ref{fig:imSEDs}.  The dot-dashed black line is the cosmic SFR density.  
The six late bloomers (blue-to-green lines) have the form that is unique to a two-parameter SFH model (like the lognormal).  
They peak at late times, after $z=1$, but form their considerable stellar masses in 1--2 Gyr,  in stark constrast to the SFRD  
and the SFHs of old galaxies. {\bfp Two of the 4 old galaxies in Figure 6, IDs 70 and 74, peak early---near or before the 
SFRD peaks.  ID 74 appears in its lognormal parameterization to have a very extended history of star formation, but this is probably 
an artifact of significant star formation at \Tobs; such bimodal SFHs are poorly described by a lognormal.  ID 50 is an example of
a galaxy whose SFH is approximately constant.   ID 31 appears to rise well after the SFRD peak, but its star formation history is  
broad and might have been called a late bloomer if observed 1-2 Gyr earlier.} 
} 
\label{fig:lognormals}
\end{figure}

\subsubsection{``When?''}
\label{sec:when}

The ``when" of late bloomers is best approached by finding a way of characterizing their exceptional SFHs.  Our parameterization of SFHs as 
lognormals is such an approach.  While lognormal SFHs may be an approximation and subject to alteration by  late processes like galaxy mergers 
and baryon accretion, the success of this model in reproducing the ``bulk properties" of galaxies demonstrates its suitability as a starting point 
(G13, A16). 

In Figure \ref{fig:lognormals} we show fits to lognormal SFHs for the 10 galaxies of Figure \ref{fig:imSEDs}, obtained by solving for the $(T_0,\tau)$ 
pairs that best reproduces each systems' total mass at the end of the five CSI SFH bins.  The two old galaxies, 74 and 70, show small $T_0$ but long 
$\tau$, not far from a traditional exponential model (Tinsley 1972). Our measurements are upper limits---$T_0$ could be earlier and $\tau$ 
shorter---since our ability to age-date loses sensitivity before the peak in the SFRD at $z\sim2$.  We are on firmer ground in the case of 
more-or-less-constant star formation, ID 50, whose timescales are within our effective time horizon.  

{\bfp The SFHs that are best captured by a lognormal parameterization} are the 6 late bloomers (and perhaps one marginal case, ID 31): these 
are well constrained by the best-measured star formation rates, $\ls$2 Gyr old populations, and the minimal to less-than-equal contribution 
from an old population. Thus, {\bfp at these stellar masses,} the \emph{long} $T_0$, short $\tau$ values for late bloomers are the distinguishing 
features of this type {\bfp (see Figure \ref{fig:mosaic})}, a combination of values that may or may not be consistent with {\bfp a} conformal 
{\bfp galaxy} growth model.\footnote{\bfp Mathematically, late bloomers can also come from galaxies with long $T_{0}$ and long $\tau$; i.e., 
delayed but monotonically rising SFHs. However, we believe the objects illustrated in Figure \ref{fig:lognormals} are more representative 
of this phenomenon at the approximate mass of the Milky Way.  This is because, for example, constraints from the evolution of the galaxy 
stellar mass function suggest such systems do not maintain the necessary super-linearly rising SFRs for many Gyr after $z\sim0.6$ 
(e.g., Moustakas et al.\ 2013).}

\begin{figure}[t]
\centerline{
\includegraphics[width = 0.45\textwidth]{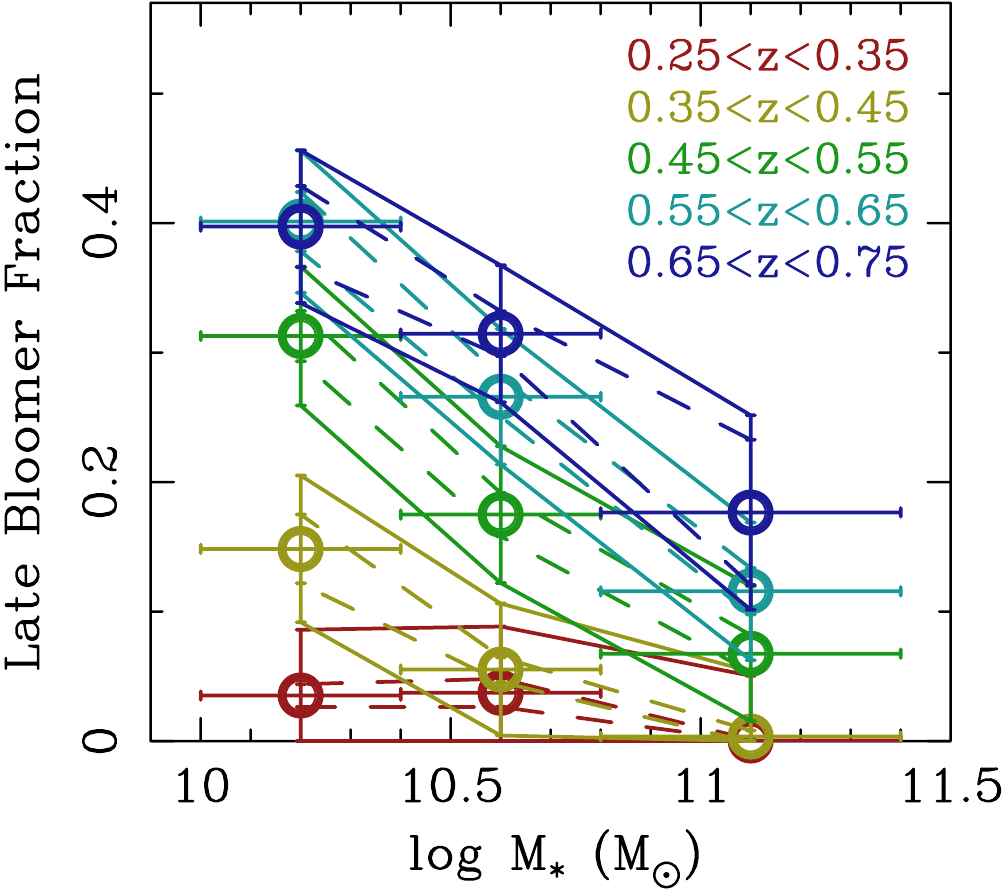}
}
\caption{The dependence of the late bloomer fraction (LBF) on mass and redshift.  The LBF declines with increasing mass but also with 
decreasing redshift: later than $z\sim0.3$, late bloomers effectively disappear at $\log\Mstel>10$. 
} 
\label{fig:lbfMassTime}
\end{figure}

We know, from the adequacy of simple declining exponential models to fit the the SFHs of local galaxies, that late bloomers are essentially 
extinct today, at least for galaxies more massive than 10$^{10}$\Msun.  By this we mean that, if measurements are made for $z\approx0$ 
galaxies to focus on star formation that occurred since 2 Gyr before the present epoch ($z\approx0.16$), there should be no late bloomers; i.e., 
no galaxies with a substantial fraction of the stellar mass formed during that period.  The decline of the late bloomer population from 
$\sim$20\% at $z\sim0.8$  to near zero today was first documented in O13 and confirmed in the D16 study.  Figure \ref{fig:lbfMassTime} 
shows this result from the CSI sample for this study. At  $z=0.7,  0.6, 0.5, 0.4, 0.3$, the LBF noticeably declines from $\sim$30\% to 
$<$5\% at the MW's \emph{current} mass of $\log\Mstel\sim10.7$.

{\bfb Note that the LBF at $\log\Mstel\sim10.2$ and $z\sim0.7$ is about twice that at $z\sim0.4$ (2 Gyr later) for $\log\Mstel\sim10.6$ . 
Hence, a meaningful fraction of late bloomers will go on to double {\it once more} over the course of their lives, growing 0.6\,dex in 4\,Gyr.}

It is worth distinguishing this signal from the very different question of whether we could \emph{recognize} that a present-epoch galaxy 
was a late bloomer at $z\sim0.6$ (i.e., was a descendant of one of the CSI systems). That answer is {\bfp probably \emph{no}: with present 
techniques based on integrated stellar populations, it would be very difficult to recognize that a $z\sim0$ galaxy had its major star formation 6 Gyr ago rather than 8+ Gyr ago.}   For this reason{\bfb---and the near-zero $z\approx0$ LBF---}we should not be surprised that there 
has been no hint of late bloomer SFHs from studies of present-epoch galaxies.\footnote{However, galaxies near enough to produce HR 
diagrams---e.g., the PHAT study of M31's disk (Dalcanton \etal\ 2012)---may offer such an opportunity.}

Are there late bloomers at higher redshift, $z\gs1.0$, for example?  G13's original lognormal realization suggests that short $\tau$ SFHs 
spread from the earliest times (starting with old galaxies like IDs 74 and 70) through the SFRD peak, all the way to $z\sim0.3$, but 
disappear at lower redshifts (consistent with Figure \ref{fig:lbfMassTime}.  In principle, CSI data at $z\gs1$ should be able test this, but 
beyond $z>1.2$, our data will not be able to answer this question: The break region redshifts out of the optical, and in any case we 
imagine that the shorter age of the universe would make it very difficult to distinguish long $\tau$ from short $\tau$  for galaxies with $T_0$ 
approximately that of the peak in cosmic star formation. The takeaway is that massive {\bfb late bloomers probably} cover 
the whole of cosmic history up to a few Gyr ago, even if, {\bfb operationally, they are hard to 
distinguish as a separate class of objects at $z\gtrsim1$.}

\subsubsection{Where?}
\label{sec:where}

\begin{figure}[t]
\centerline{\includegraphics[width=1.00\hsize]{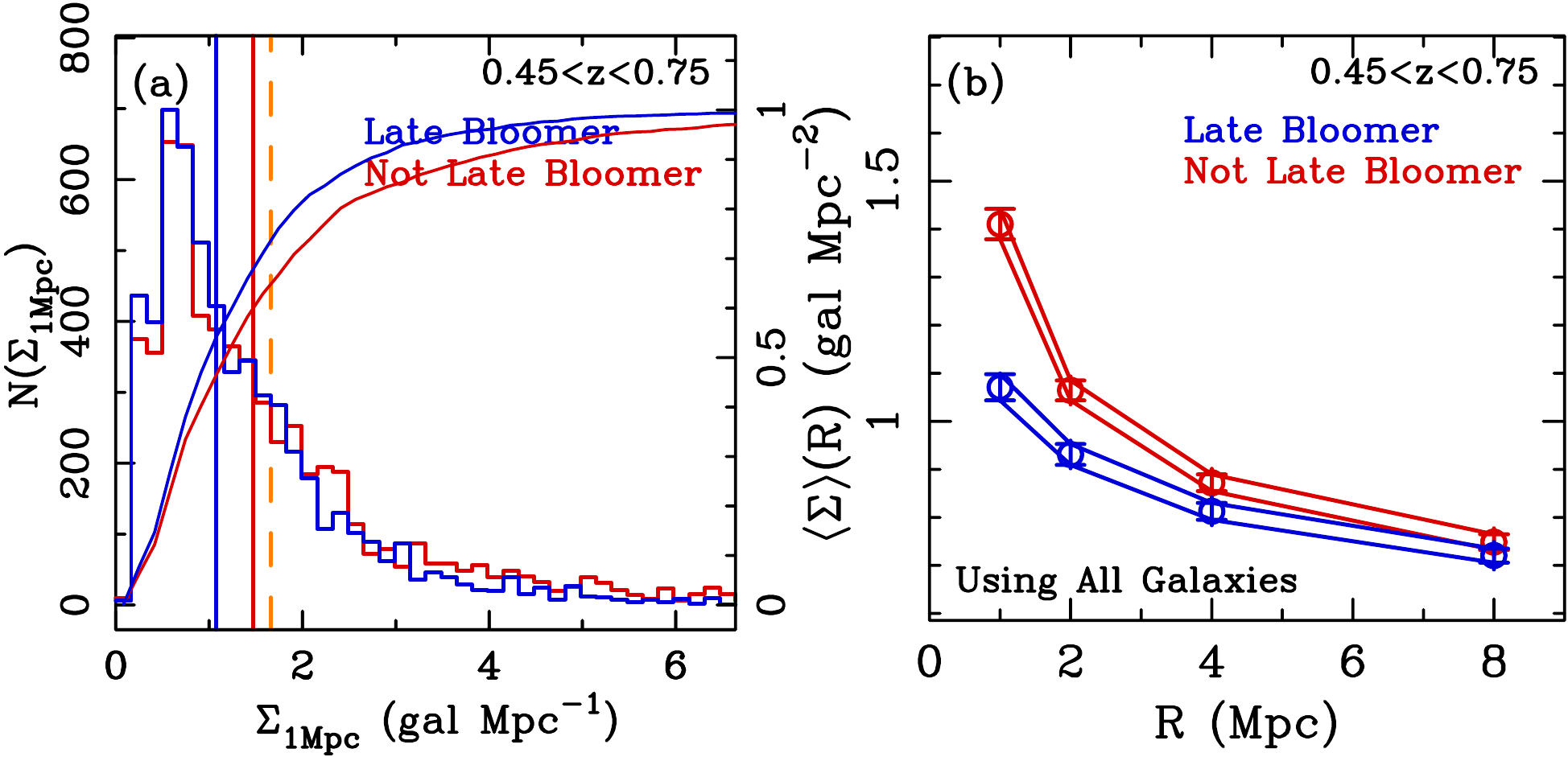}}
\caption{(a) (thick lines) Distributions of local projected galaxy densities within projected radii of 1 comoving Mpc and 
$\Delta z\pm 0.02(1+z)$ around late bloomers and non late bloomers. Cumulative distributions are shown using the
thin lines. Orange vertical line marks the 75th percentile of the full distribution of galaxy densities. Mean local densities 
for late bloomers and non-late bloomers are shown by the blue and red vertical lines, respectively.  (b) Mean local projected 
galaxy densities around late bloomers and non-late bloomers as a function of project scale over
which densities are measured.
\label{fig:where1}}
\end{figure}

Figure \ref{fig:where1}, left, shows the distribution of galaxy overdensities around secure late bloomers (LBs) and non-late bloomers (NLBs) 
for the full CSI sample, measured in 1 Mpc apertures.\footnote{Note: the $\Delta z = \pm 0.02\,(1+z)$ redshift range in which projected galaxy
densities are computed has a comoving length of $\sim$300 Mpc, much larger than the transverse size of the volumes these density
measures probe.} From this perspective the ``environments'' of LBs appear similar to those of NLBs, with mean densities differing by only 
$\sim$25\%. Figure \ref{fig:where1}, right, shows the correlation of mean local galaxy density with the scales over which they are measured; 
NLBs live in regions with slightly higher density on scales at least to 8 projected comoving Mpc.

\begin{figure}[h!]
\centerline{\includegraphics[width=1.00\hsize]{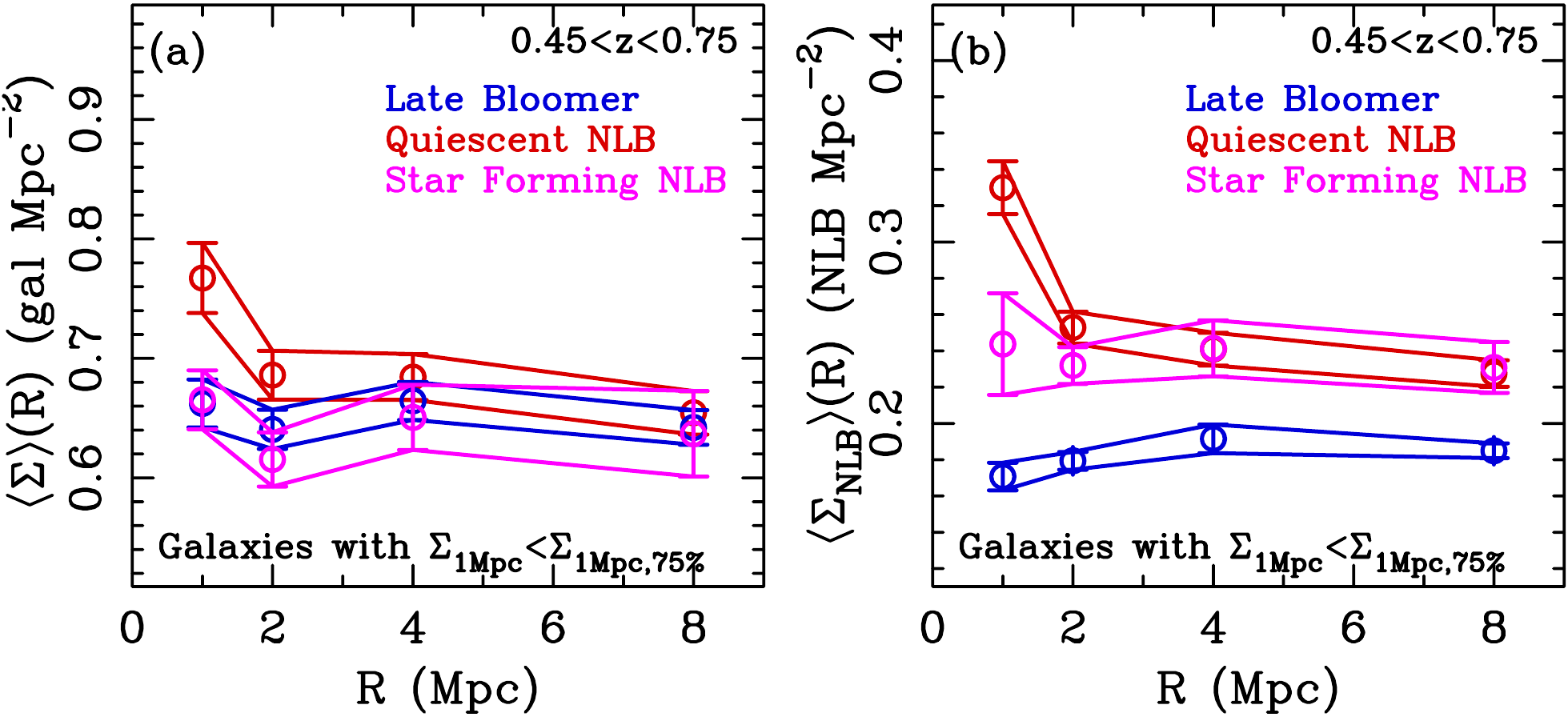}}
\caption{(a) Mean local projected galaxy densities around those late bloomers and non-late bloomers with $\Sigma_{\rm 1\,Mpc}$ 
less than the 75th percentile (orange line in Figure \ref{fig:where1}) as a function of project scale over which densities are measured. 
Late bloomers and non-late bloomers live in neighborhoods with similar numbers of neighbors.  (b) Measuring densities of non-late 
bloomer neighbors, we see that late bloomers prefer not to livearound non-late bloomers, and that non-late bloomers prefer to live 
around each other. This homophily of galaxy SFHs either reflects long-term exposure to similar environmental forces, or different 
sets of initial conditions that set ensembles on paths to beingearly or late forming.
\label{fig:where2}}
\end{figure}

Some part of this signal is due to the fact that the galaxies least likely to be late bloomers---elliptical galaxies and other
passively evolving systems---are known to live in the densest regions (Dressler 1980) and are the most strongly clustered
(Davis \& Geller 1976, Loveday \etal\ 1995). We can remove this signal by eliminating all galaxies in the upper quartile in
projected galaxy density ($\Sigma_{\rm 1\,Mpc}$, orange dashed line). {\bfp Figure \ref{fig:where2} presents the results.}

The remaining LBs now appear to inhabit almost exactly the same environments as ordinary, starforming galaxies (see blue and violet
lines in Figure \ref{fig:where2}{\it a}). One might {\bfb therefore} conclude that, {\it in the general field\/}, LBs do not live in particularly interesting
places. However, this is not the case: When one measures environments by counting {\bfb only} the NLB galaxies---the
systems generally thought of as ``normal,'' with SFHs that began their decline at $z\gtrsim1$---a very different signal emerges.

Figure \ref{fig:where2}{\it b} shows that {\it late bloomers avoid non-late bloomers\/}. Although, projected on the sky, LBs do not
live in regions specially marked as over- or under-dense in all galaxies, late-growing galaxies nevertheless {\it do live in
special places\/}: ones with markedly fewer NLBs (starforming or not).

Two implications stem from this finding. One is simply (again) that LBs are a real phenomenon, not some noise-selected subset of the
normal galaxy population.

The second is more profound: galaxy SFHs trace {\it environmental histories\/}. {\bfp Moreover, since {\it z5fract2} depends on the SFR
averaged over the first $>$4\,Gyr-wide bin, {\it they do so over long timescales}.} Though they may have just as many neighbors, late
bloomers do not {\it grow up\/} in non-late bloomer neighborhoods; those neighborhoods apparently do not foster, {\bfp and have never 
fostered,} LB behavior {\bfb (see also O13)}.

In sociology, ``homophily'' is the tendency of individuals to form relationships with other, similar individuals. In astronomy, ``red galaxies cluster'' is an example of homophily. Here, however, we are encountering
homophily not as a function of {\it present\/} attributes, but {\it historical\/} ones. The signal in Figure \ref{fig:where2}{\it b} suggests
that galaxy {\it childhood\/} and/or {\it inheritance\/} matters: Either star formation behavior and performance over Hubble timescales 
reflects (1) {\it prolonged} childhood exposure to similar environmental {\bfp factors}, or (2) {\bfp accumulated} biases towards early or late 
growth inherited from initial conditions (e.g., KBA2016). {\bfp We cannot yet say, specifically, what these factors/biases are, but they would
seem to be} poorly encoded by the overdensities inhabited at \Tobs\ (e.g., Mo \& White 1996). Halo mass at \Tobs\ must also be a poor proxy 
under either scenario, outside of the most clustered halos hosting the most quiescent NLBs. Only once you have the SFHs can you tease out 
these key facts.

\subsubsection{``How?"}
\label{sec:how}

``How" can be approached in many ways, but we think the heart of the question is this: How can major star formation in 
$\sim$one-fifth of massive galaxies be postponed by billions of years compared to their peers? 

The obvious appeal is to major mergers, but two arguments push back on this possibility. First, as established by a number of studies 
using a variety of techniques (e.g., Bell \etal\ 2006; Williams \etal\ 2011; Man \etal\ 2016) the major merger fraction at these epochs 
is only about 6\%, less than $1/3$ of the CSI LBF estimate.  Second, such mergers would have to be extremely gas rich/entail systems 
with quite low $\Mstel/\Mhalo$.  Recall: our SEDs and SFH inferences are sensitive to the mass present before the late bloomer ``episode.'' 
Hence, to explain MW-mass late bloomers in which little or no {\bfb old stellar} mass is detected (not atypical; see Figure \ref{fig:mosaic}), 
two $\sim10^{12}\,\Msun$ halos that are mostly gaseous would have to merge. Given that effectively all results from abundance matching 
show such haloes having, on average, the {\it highest} stellar mass fractions (e.g., Moster et al.\ 2013; Behroozi et al.\ 2013a), this scenario 
seems highly unlikely.   Assuming an 0.2 dex scatter in $\Mstel/\Mhalo$ (Behroozi \etal\ 2013b), and that the MW's 
$\Mstel\sim10^{10.7}\,\Msun$ is representative, our $\Mstel\sim10^{10}\,\Msun$ detection threshold corresponds to a 
more-than-3\,$\sigma$ low-side outlier for a given halo mass, far too rare to account for the late bloomers.

While it is true that dark matter halos show a range of collapse times---the result of a spread of initial densities at a fixed mass---a 
doubling of collapse times for a fraction of $\sim$20\% of $\sim$10$^{12}$\Msun\ halos may not comport with current 
$\Lambda$CDM simulations (see Section \ref{sec:context}). If not, baryonic physics is the only plausible agent.  Indeed, the 
importance of ``feedback'' in modeling galaxy evolution has grown rapidly in this decade, particularly as a feature needed by 
the simulations to retard or stop the growth of massive galaxies (to match the observed mass function) when plenty of gas 
remains to fuel star formation.   By expelling gas into a galaxy's halo, winds driven by vigorous star formation and/or 
supernovae, or powered by AGN outbursts fed by gas \emph{inflow}, could suppress star formation---temporarily 
or permanently.  

The problem of invoking  feedback to explain late bloomers is that such influences are least expected here.  Late bloomers have 
formed a smaller fraction of stars for their halo mass, so feedback from star formation is minimized, and galaxies that have had little
stellar mass growth are not good candidates for large supermassive black holes.  This does raise an interesting and answerable 
question: what is the incidence of Seyfert nuclei in late bloomers compared to ``normal" populations?  Again, the most 
remarkable feature of this phenomenon is that, with $\ls$10$^{10}$\,\Msun\ of mass in old stars, how do these systems 
retain $10^{10}\ls M_{\rm gas}\ls10^{11}$\Msun\ of gas to $z\sim1$, some 6 Gyr after the big bang, and then begin to form stars 
at furious rates of 10--100\,\Msun\,yr$^{-1}$, long after their elders passed through that phase.  \emph{How}, indeed?

\subsubsection{``Why?''}

At the heart of the ``why" of late bloomers, we think, is the fundamental question of what processes shape the SFHs of galaxies---all 
galaxies.  The reigning paradigm has been ``grow and quench,"  the idea that stellar mass grows along global scaling laws, such as the 
SFMS, until some feedback mechanism sharply curtails star formation, or ends it altogether. The picture is widely accepted, even in the 
absence of a {\bfb uniquely} successful model for the quenching mechanism.  Furthermore, ensemble properties of galaxies---for example, mass 
functions and fractions of active versus passive galaxies---provide very weak constraints, easily satisfied by very different models 
(A16; KBA16).  {\bfb Sufficient numbers of i}ndividual examples of  galaxies in the act of quenching are not identified, and models and 
predictions of what observations might be discriminating are not offered. 

Quenching, to be meaningful, is by definition a short timescale process that requires an \emph{event} that alters the course a galaxy 
would otherwise take.  There is scant evidence for this at low redshift.  Most SDSS ``green valley" galaxies are \emph{not} recently-quenched  
galaxies from the ``blue cloud,'' as had been suggested (e.g. Faber \etal\ 2007): expected UV color evolution and spectral features (strong 
Balmer absorption without star formation) is seen in only a few percent of the population (Schawinski \etal\ 2014;  Dressler \& Abramson 
2015; Rowlands \etal\ 2018).   In fact, green valley galaxies are evolving slowly towards the red  sequence, as star formation slowly 
ebbs.  Acknowledgement of this fact has led to the introduction of an oxymoronic  ``slow quenching" to describe what is certainly 
better characterized as galaxy \emph{evolution} over a Hubble time. 

Perhaps, as originally suggested, massive galaxies at high redshift truly quench---rapidly---but a ``smoking gun" requires the 
identification of a mechanism and its observation.  An appropriate alternative is one that invokes Hubble-timescale processes---which, 
by definition, become rapid at high-$z$---to shape the SFHs of all galaxies, like the lognormal model we have developed (G13, D16, 
A16).  As with quenching, the success of this model has been largely judged by its ability to reproduce the mean properties of galaxy 
populations through cosmic time.  However, as we have shifted our focus to the SFHs of individual galaxies (D16 and this paper), we 
believe that this richer data set moves the discussion from ``why do galaxies quench?" to ``why do galaxies follow a Hubble-time-scale 
form?" (See also Pacifici \etal\ 2016.) For both rapid and slow forming galaxies we see a common theme: galaxies grow as long as their 
gas fractions are high and fall as stellar mass overtakes available gas for further star formation.  However, the critical physics that 
translates this observation into a lognormal or similar SFH  remains elusive.

We suggest that late bloomers are key to understanding what shapes star formation histories because they are simply non-existent
within grow and quench scaling law-driven pictures, but they are clearly real. The ``why'' of late bloomers is, then, their role in properly and 
fully describing the histories of star formation for all galaxies.  Any model that does not produce late bloomers must be incomplete, 
but models in that category should view late bloomers as an opportunity to learn about and/or tune the physical inputs to their star formation 
prescriptions.  What is clear is that late bloomers are a new, potentially strong constraint on future simulations and theory of galaxy growth.

\begin{figure}[t]
\includegraphics[width = \linewidth, trim = 0.5cm 0cm 0cm 0cm]{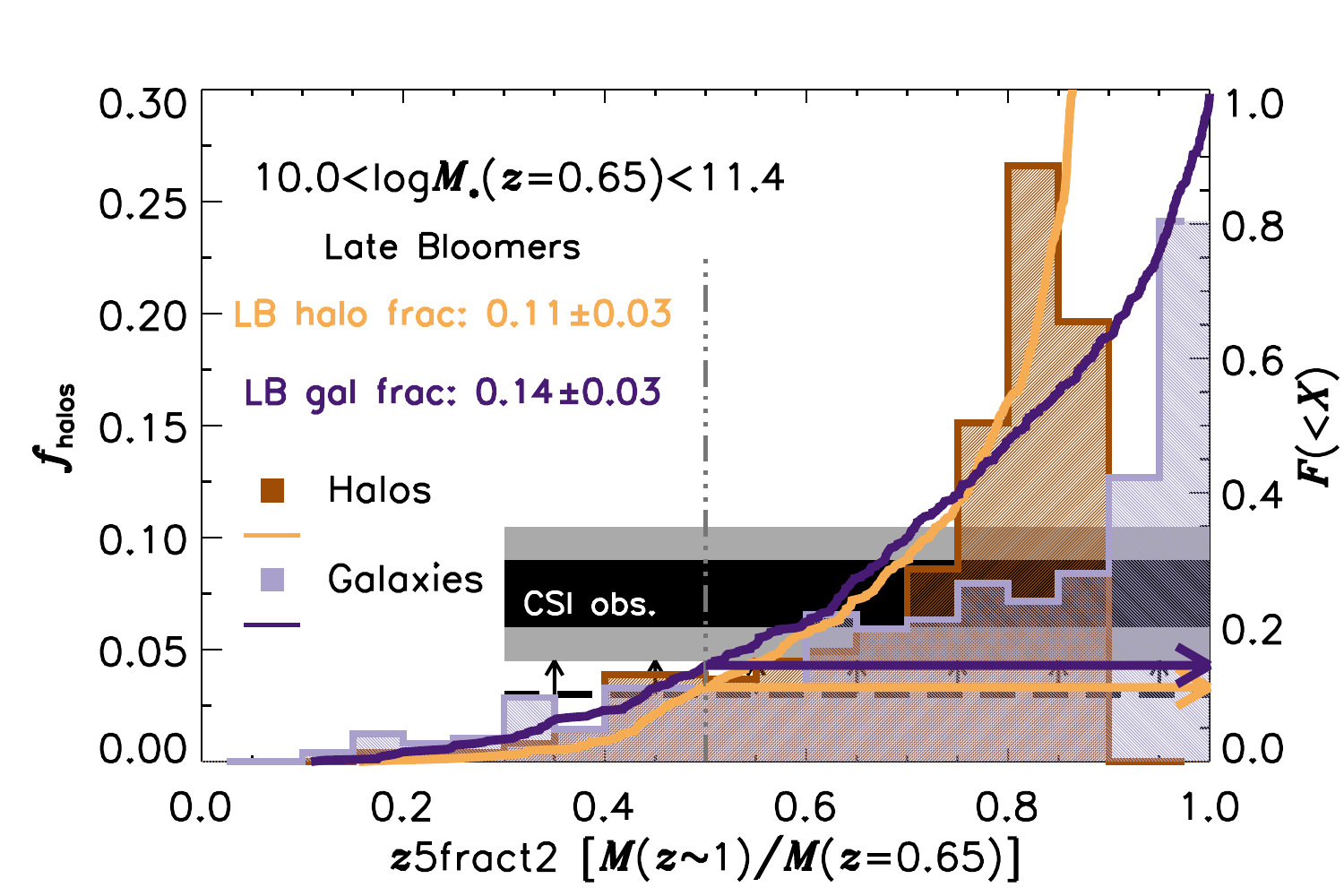}
\caption{Theoretical halo (orange) and galaxy (purple) \z5fract2 values (left axis) and LBFs (right axis) based on the 
{\tt GALACTICUS} semi-analytic model. {\bfb Orange and purple horizontal arrows highlight the halo and galaxy
model LBFs, \resp, while the black/gray solid horizontal bands show the CSI empirical results (see Figure \ref{fig:lbfs}).} Only model halos harboring galaxies above the CSI completeness limit in the empirical 
analysis---$\log\Mstel\geq10$---are considered. Binomial uncertainties show the 95\% model confidence interval. {\tt GALACTICUS}' halo LBF is consistent
with the empirical lower-limit: its galaxy LBF with the best observational estimate at $\sim$1\,$\sigma$.} 
\label{fig:samResults}
\end{figure}

\subsection{Late Bloomers, Dark Matter Halos, and the Grow and Quench Paradigm}
\label{sec:context}

In this section, we consider the late bloomers we have found at $\sim$20\% abundance in our $z\sim0.6$ sample in the 
context of a semi-analytic model of galaxy evolution in a $\Lambda$CDM universe.  Three issues that stand out are: (1) Are a significant 
fraction of $\sim$10$^{12}$\Msun\ dark matter halos still assembling at $z\le1$?  (2)  Are theoretical prescriptions used to model 
baryon evolution able to delay stellar mass growth with respect to halos?  (3) What implications do the late bloomers have for established 
methods of inferring galaxy SFHs from global scaling laws (and quenching prescriptions)? Below, we address these questions in order. 
Our results are based on the well-tested {\tt GALACTICUS} semi-analytic model by Benson 2012. They are robust to resolution at least 
over minimum halo masses of $10^{9}$--$10^{10}\,\Msun$; the {\tt .xml} input file from which they were derived is appended as 
an ancillary data file.

\subsubsection{``Late'' Halo Growth at Milky Way Scales} 
An obvious question to ask when trying to understand late bloomer galaxies in the $\Lambda$CDM framework is whether a 
similar number of halos also doubled in mass so rapidly at these redshifts. If so, at least the diversity in halo growth trajectories 
\emph{could} encompass that in galaxy SFHs.

To answer this question, we ran {\tt GALACTICUS} in a ``standard'' DM+baryons mode (revision 6169:394a64c6b493; see Knebe \etal\ 
2018), tracking 3000 halos. We 
then selected all 489 halos at $z=0.65$ harboring galaxies with $\Mstel\geq10^{10}\,\Msun$.  We identified their most massive progenitor 
2 Gyr earlier ($z=1.08$) and defined ${\it z5fract2}_{\rm halo}$ as the ratio of the progenitor to descendant masses. We repeated this 
calculation for the corresponding galaxies. Figure \ref{fig:samResults} shows the results.

\begin{figure}[t]
\includegraphics[width = \linewidth, trim = 0cm 0.5cm 0cm 0cm]{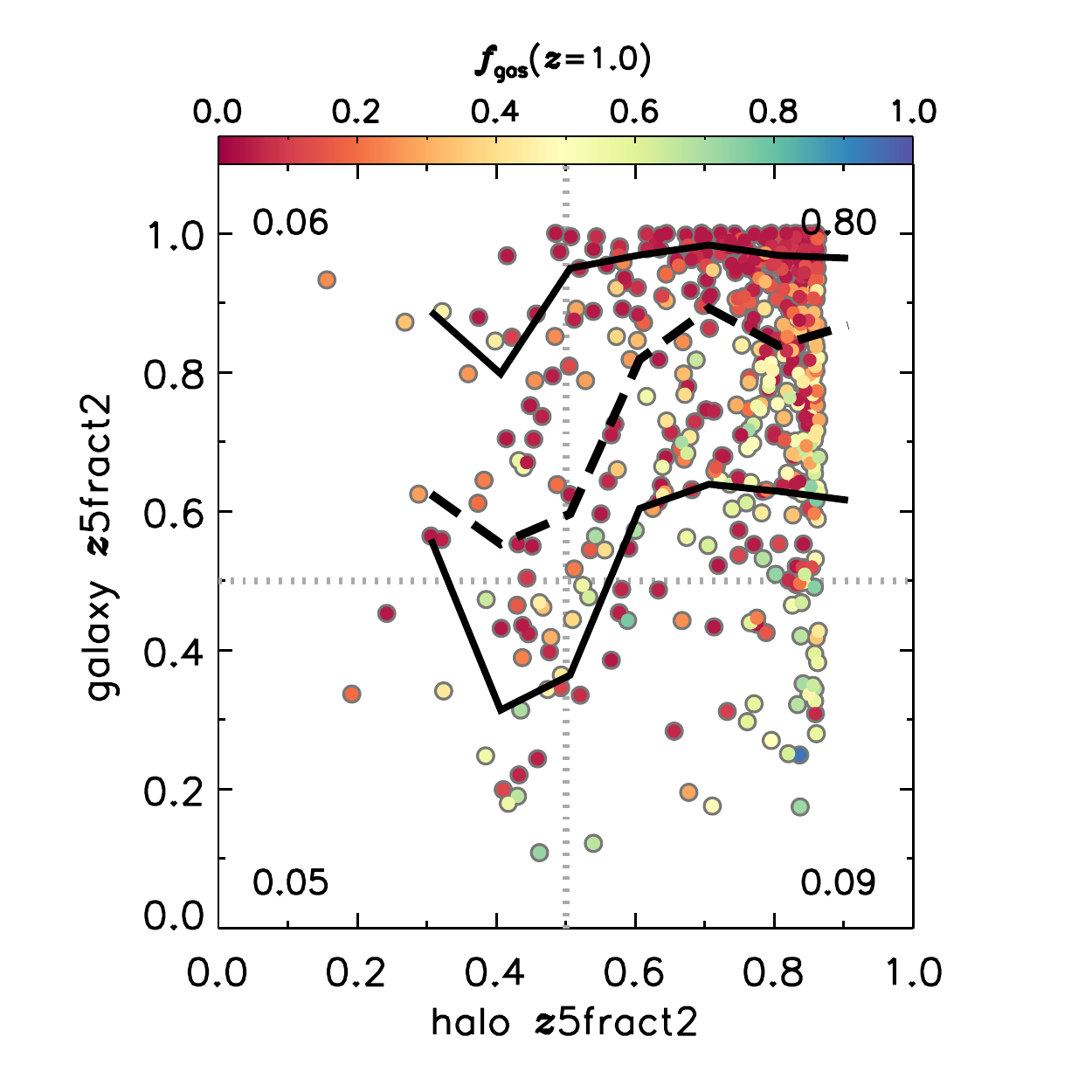}
\caption{Halo vs.\ galaxy \z5fract2\ from the {\tt GALACTICUS} SAM. Points are colored by gas 
fractions---$M_{\rm gas}/(M_{\rm gas} + \Mstel)$---at $\Tobs$ minus 2\,Gyr ($z\approx1$). 
{\bfb Dashed/solid black lines show the median/1\,$\sigma$ spread, \resp, in bins of
0.1\,dex in $\z5fract2_{\rm halo}$.} 
Late blooming {\it galaxies} tend to be gas rich, though late blooming {\it halos} can host
gas-poor galaxies {\bfb (perhaps suggesting these are not analogous to the actual halos of CSI
late bloomers)}. The fraction of objects is printed in each galaxy--halo (non-)LB quadrant.} 
\label{fig:haloGal}
\end{figure}

The cumulative distributions in that plot reveal that, while most halos only grew $\sim$20\% (in agreement with the traditional vision of 
massive galaxy growth), 8\%--15\% of those harboring CSI-detectable galaxies did indeed double in mass in the 2 Gyr preceding \Tobs\ 
(95\% confidence). These numbers rise to 11\%--17\% when examining the simulated galaxies themselves. These fractions are at most a 
factor of $\sim$2 away from the CSI observations, suggesting that at least the \emph{number} of halos is sufficient to account for the late 
bloomer phenomenon, and \emph{global baryonic prescriptions} are capable of producing them. 
As such, the existence of these objects was predictable from numerical modeling, though we are unaware of any works that drew attention 
to them. Certainly, there does not seem to be a compelling reason to rule-out late bloomers from a dark matter assembly perspective (see also Giocoli et al.\ 2012).

\subsubsection{Connecting Galaxy to Halo Growth} Though there may be sufficient late-blooming halos and galaxies in {\tt GALACTICUS}, 
the question remains as to the physical connection between these two entities; i.e., is halo assembly sufficient to account for the timing of galaxy 
growth, or is there something else at play (Section \ref{sec:how})? Figure \ref{fig:haloGal} plots \z5fract2 vs.\ ${\it z5fract2}_{\rm halo}$ to 
investigate this link.

While there is a weak trend in the expected sense---late-growing galaxies are more likely to occupy late-growing halos---the scatter evident 
in this diagram is extreme: galaxies might span 40\% in \z5fract2 at fixed halo assembly history (to the extent it is encoded by 
$\z5fract2_{\rm halo}$). As such, baryonic effects are playing a substantial role in late bloomer formation beyond what can be accounted for by 
the halo histories. This finding echoes results from Diemer et al.\ (2017; see  their Figure 9), though we note no late bloomers were found in 
that analysis of the ILLUSTRIS hydrodynamical simulation (Vogelsberger \etal\ 2014), suggesting (perhaps unsurprisingly) that assessing 
the above baryonic effects will no doubt be sensitive to the specifics of the modeling.

Nevertheless, a clue to the nature of such phenomena that is hopefully so macroscopic as to be insensitive to such details lies in the color 
coding in Figure \ref{fig:haloGal}. This shows galaxy gas fractions, $\Mgas/(\Mgas + \Mstel)$ (an observable, in principle), at 
$\Tobs-2$\,Gyr. Predictably, objects with the highest gas fractions---60\%--80\%---are much more likely to be late bloomers at 
fixed halo assembly history than galaxies with the lowest. {\bfp Indeed, examining the $\Mstel$--$\Mhalo$
relationship (not shown) reveals that non-late blooming galaxies in late blooming halos (top-left corner 
of Figure \ref{fig:haloGal}) live in abnormally massive halos for their stellar mass. This suggests 
they are centrals of assembling groups---i.e., the high environmental density tail we 
excluded when discussing LB environments in Section \ref{sec:where}. As such, they would typically be
passive, consistent with their below-average gas fractions in {\tt GALACTICUS}.} 
Though substantial scatter persists even here, this kind of statement 
would represent meaningful input by modelers as to how, where, and why observers might find late bloomers at other epochs or 
along other axes: if quantitative predictions for, e.g., the {\it mean and scatter} in gas temperatures or molecular fractions, {\it local} 
environments, bulge-to-disk ratios, or kinematics were made, these would be testable by future targeted observations or 
surveys. Other useful inputs include the {\bfb AGN} fraction among late bloomers, their metallicity distributions (stellar and gas-phase), and 
areas of parameter space that are {\it certainly} forbidden under $\Lambda$CDM halo assembly.  Correct \emph{predictions} in any of these veins would go a long way towards reassuring the community that a model captured something fundamental 
about galaxy evolution that qualitatively distinguished it from others with diverging answers.

\subsubsection{Implications for Scaling Law-Based Inferences}

Independent of their physical implications, we believe that the late bloomers demonstrate a central, mathematical fact that
the community must recognize if we are to gain a meaningful sense of the narrative of galaxy evolution. Simply, \emph{late bloomers
cannot be described by any model based on abundance matching or the integration of scatter-free scaling laws} (e.g., the SFMS).
These consequences follow from the fact that late bloomers break mass rank ordering; i.e., though they may occupy the same bin 
at \Tobs\ as equal-mass systems with constant (or even linearly increasing) SFHs, they were \emph{arbitrarily less massive} 2 Gyr in the 
past.  As such, they must have jumped over all galaxies in the intervening mass bins to reach the endpoint at which they were 
identified. If mass---halo or stellar---is taken as the controlling parameter for an object's growth rate---as it is in abundance matching 
or SFMS integration---this phenomenon obviously cannot occur.

The implications of this fact could not be more profound: if relative positions on scaling relations do not stay fixed, the above methods
become effectively useless for identifying---let alone {\it characterizing}---the progenitors or descendants of any galaxy, or 
even mass-limited sample. Of course, by definition, they are accurate {\it in the mean}. However, if, as we find, fully 1-in-5 systems are not 
only ``outliers'' here, but \emph{contradictory of the methods by which the mean is defined}, it is clear not only that the ``average galaxy''
is unrepresentative of important physics, but that approaching the problem from this vantage point {\it mathematically forbids even 
the recognition of this fact}, to say nothing of illuminating its causes (A16).  

This is not to claim that this issue has so far been unknown---Torrey et al. (2017), for example, perform a detailed investigation of 
the size and character of the effect of mass/abundance rank-order breaking based on numerical simulations. Studies of this kind 
provide important insights as to where and when galaxies "jump" each other, and statistical corrections to account for this phenomenon. 
We encourage further efforts in this vein, but the claim we are making here is that understanding its physical causes/making and testing 
potentially discriminating theoretical predictions with more global ramifications depends on actually identifying the galaxies that are doing 
it: late bloomers. Only in this way can we hope not only to learn the amount of SFH diversity, but understand why galaxies take the 
paths they do through that envelope.

Regardless, the implication from late bloomers on rank-order breaking suggests that further attempts to combine stellar and halo mass 
functions and SFR scaling laws (of any depth and redshift) will not be edifying.  Instead, attention must be paid to inferring appropriately 
complex SFHs from the SEDs that the above exercises would otherwise have required (Pacifici et al.\ 2016; KBA16; Iyer \& Gawiser 2017; and Abramson et al.\ 2017 provide steps  in this direction).

\section{Summary and Future Work}
\label{sec:summary}

The principle goal of this paper has been to make a compelling case for the reality of late bloomers,  massive galaxies that built the 
majority of their mass at a time when most galaxies were in notable decline.  Toward this end, the paper contains {\bfp five} major sections:

\begin{itemize}

\item{A description of changes made in our spectral fitting program to improve its sensitivity to late bloomers and to provide 
SFH confidence intervals to assess the likelihood {\bfb that a galaxy fits that classification};}

\item {\bfp The first robust measurement of the late bloomer fraction from individual 
		galaxy SEDs, showing 1-in-5 present-day MW mass systems (more or less 
		depending on epoch and mass) belong to this class. 
		We encourage new, higher-quality observations to verify this finding and are 
		open to sharing our data for novel/improved reanalyses.}

\item{A catalog of galaxies with high-confidence SFHs, images, and basic data that can be used as a standard sample for independent
studies by others employing different methods;}

\item{A discussion of the implications of late bloomers in the context of numerical models of structure growth that include prescriptions
for star formation, and for popular ideas about galaxy evolution, such as the SFMS, quenching, and abundance matching;}

\item{An appendix with a detailed description of the simulated SFHs to test the sensitivity of our methodology to detecting
old stellar populations as a function of $S/N$ and in the presence of dust extinction.}

\end{itemize}

Our conclusion is that late bloomers at redshifts $0.45 < z < 0.75$ are real and that they represent a significant minority population
of galaxies that grew to Milky Way mass and above by the present epoch.  The abundance of late bloomers declined rapidly beginning
at $z\sim0.3$ and became extinct (for massive galaxies) by the present epoch. Comparing with theoretical work on galaxy SFHs suggests
that a late-assembling population of dark matter halos available to host late bloomers is $\sim$10\%, consistent with our lower bound 
to the fraction of late bloomers, arrived at by assuming that all $\sim$10$^{11}$\,\Msun\ examples are false positives.  However, it
remains puzzling how such halos would have avoided star formation for $\sim$5 Gyr and reached $z\sim1$ with enough available gas to 
fuel the relatively rapid onset of star formation seen in late bloomers. 

This idea of a diverse SFHs, including this $\sim$20\% fraction that grew  late and rapidly, but also declined relatively soon thereafter, 
is at odds with the prevailing paradigm of an ordered, conformal set of SFHs quenched internally by a mass-related process or 
externally through environmental agents.  Because the paradigm employs the SFMS to infer SFHs and ``abundance matching" to 
relate the growth stellar mass with respect to dark matter halos, this picture has had considerable impact in the study of galaxy 
evolution---the existence of late bloomers will be important to understanding the limitations of this approach.  We have, therefore, 
devoted most of this paper to examining the basic methods that underlie the discovery of late bloomers, and used synthetic SFHs and 
simulations to demonstrate that our work is sound.  We believe we have made this case sufficiently well that it will be insufficient for 
others to simply dismiss late bloomers as the result of some unknown difficulty in spectral synthesis, but instead require observations 
and analyses purposed at testing our results and finding possible mistakes or errors.  To support this effort, we have provided a 74-galaxy 
sample with high-confidence SFHs with basic data and HST images.

Our work suggests several obvious possibilities for future observations that will clarify many of the issues raised in this paper.  First, 
our modeling of late bloomers is built on spectral templates that have higher resolution than the $\sim$30\,\AA\ prism spectroscopy 
used in the CSI study.  As such, we can predict with confidence what higher resolution spectra of late bloomers should look like, both for the 
confirmation of Balmer absorption lines from younger stars and from metal lines from older populations.  We are following up the catalog 
sample presented here with \imacs\ observations at $\sim$10\,\AA\ resolution and expect that others testing our results will likely obtain 
such improved data.

A second opportunity is investigate the nature of the stochastic component of SFHs as judged by the fraction of late bloomers that are
passive at \Tobs.  This unambiguous observation should {\bfb yield insights into the} duty cycles of episodes of vigorous star formation.
This will be especially interesting extended over the redshift range $0.3<z<1.0$ that our CSI data cover well for the kind of analysis we 
present here.

Finally, towards understanding where late bloomers fit in the context of most galaxies that accomplished the majority of their
star formation before $z=1$, it will be important to investigate such properties as AGN incidence, indicators of major mergers or accretion
events, and local environment that may hold clues to the late blooming of this remarkable population.

\section*{}

The authors thank the scientists and staff of the Las Campanas Observatories for their dedicated and effective support over the 
many nights of the CSI Survey.  We also thank A.~Benson for valuable assistance in running and understanding his numerical model.
We especially appreciated Simon Lilly's fair-minded consideration of late bloomers as examples of physics outside his paradigm of 
galaxy evolution---this paper was encouraged by his challenge to convince him of the reality of late bloomers.  Lastly, we are grateful 
to the referee for a careful reading and thoughtful report, which asked for additional information that both strengthens our results 
and should also help readers better understand our methodology and approach.


\appendix

\section{How Many Late Bloomers Are Imposters?}
\label{sec:app}

An important goal of this paper has been to compute the global fraction of galaxies that formed at least half their stars between 
$z=1$ and $z\sim 0.6$. In this appendix we show, quantiatively, how observational noise drives the best-fit solutions away from the
intrinsic growth trajectories of the galaxies.  We also show how to compensate for such problems without unnecessarily imposing 
priors on the early growth histories of galaxies, something which could bias the inferred \z5fract2 distribution.


\subsection{Why Simulations Are Needed To Inform Selection and Statistical Corrections}
\label{sec:whysims}

The principal side effect of observational noise is to diminish one's ability to detect old stars, especially when the SED is
dominated by young(er) stars. For example, if young stars produce 90\% of a galaxy's light, then, at $S/N\simlt 10$ at a handful of 
wavelengths, one has virtually no purchase on the old stellar component. This issue is the chief reason why parameterized SFHs are 
seen by many to be so useful: they bring along prescribed (and proscribed) amounts of stellar mass according to implicit preconceptions of 
galaxy growth. Obviously, the problems only get worse when young stars produce yet more light, or at earlier cosmic times, when 
younger galaxies are more common. Inferences about their properties from SEDs will become increasingly dependent  on input 
assumptions about prior growth.

In terms of $S/N$ and information content, we have in CSI independent flux measurements at ${\cal O}(10^2)$ wavelengths, with
most of the data discussed in this paper at $S/N\gg 10$.  Theoretically, we retain statiscially meaningful leverage on the old stellar 
mass in CSI with $\sqrt{N}\sim 10$, even when young stars make up $\sim 90\%$ of the light.

In practice, however, degeneracies arise between dust and the ages of older stars that may lie underneath younger stellar populations: the 
residual light from such brighter stars might be too noisy to disentangle the effects of reddening from age. Given the complexities of these 
degeneracies in the data, we must assess their effects through simulations.

More concretely, galaxies that did grow late may have SEDs that appear optimally fit by histories with earlier growth, and vice
versa.  One strategy, then, would be to identify the metrics by which one identifies as many late bloomers as possible, with the
greatest completeness.  However, such a strategy may be plagued by many false positives.  In particular, the SEDs of galaxies that 
grew most of their mass early but also had sufficient numbers of young or intermediate age stars, when subject to observational noise, 
could be falsely identified as late bloomers. A balance must be struck between the purity of the samples identified as late bloomers 
and its completeness.

We prefer a strategy that minimizes (or at least mitigates against) false positives,  even at the expense of inadequately capturing the 
census of true late bloomers.  Obtaining non-zero, but believable, lower bounds on late bloomer fractions already provides strong, negative 
implications for common approaches to galaxy evolution, for example, the use of mean growth trajectories and abundance matching. Clearly, 
the identification of a sample that is, say, 95\% complete but only 10\% pure is not very useful.  Ideally, its much mores sensible to construct 
samples that are simultaneously $>50\%$ complete and $>50\%$ pure\footnote{Erring on the side of purity is prudent anytime one believes
their target population is small; i.e., when there is a risk that {\it selected} samples will be dominated by false positives unless the
selection fidelity is good.}. Only through simulations of the data can one hope to verify such procedures.

When certain classes of noisy SEDs may have diminished fractional sensitivity to old stellar mass, one faces two choices for how to 
proceed:
\begin{enumerate}
\item Include strongly constrained early star formation histories as priors in SED fitting, at least for (relatively) early times; or
\item Impose minimal SFH priors and use the broad confidence limits that arise from the increase in noise while maintaining 
(empirical) ignorance about the underlying nature of (early) growth trajectories.
\end{enumerate}

The former has been an attractive approach in the community, with parameterized SFHs commonly in use to mitigate the larger
uncertainties in any weighing of old stellar mass. In order to use maximum likelihood estimates of \z5fract2, they must first
be derived using priors that adequately reflect the true underlying distribution of growth histories.  Unfortunately, inputing what 
one thinks is the true underlying distribution of growth histories is prone to lead \z5fract2\ distributions that resemble the 
starting assumptions.  Hence, the appropriate choice of strategy must be (2).

{\bfp
With such an approach, many noisy SEDs may not statistically require the presence of any old stellar mass, but simultaneously the best-fit
stellar population parameters ought still be accompanied by increasingly broad confidence intervals for \z5fract2. By definition, when the
fraction of old stellar mass is increased for such galaxies to levels at which goodness-of-fit metrics depart from optimality, these levels
of old mass can be considered as (nearly) ``maximum-old-mass'' models---as the \z5fract2\ values are
then pushed (nearly) as high as possible without statistically violating the constraints of the data.}

We desire simply a census of the number of galaxies in CSI that have no more than half their stellar mass in old stars. Thus, for our 
application, the true nature of early growth trajectories is unimportant, so long as we can place (strong) upper limits on the old mass 
present in the galaxies we identify.  For example, we are not particularly interested in distinguishing between galaxies that have 10\% 
of their mass in old stars from those that have 20\%---they are equally interesting to us cosmologically and astrophysically.  Similarly, by 
first trying only to find galaxies that grew {\it at least\/} half their stellar mass in the last 2 Gyr, one can relegate to a later date the more 
detailed
questions about which grew 70\% and which grew 85\%, why they did so, and maybe even how.

It is in this context that we now proceed to demonstrate that the best-fit \z5fract2 values do not serve our purposes for quantifying how 
many galaxies grew most of their stellar mass after $z=1$, and that the best way to select late bloomers is to use their 95\% upper 
confidence limits on \z5fract2.


\begin{figure*}[hbt]\figurenum{A1}
 \includegraphics[width = \hsize]{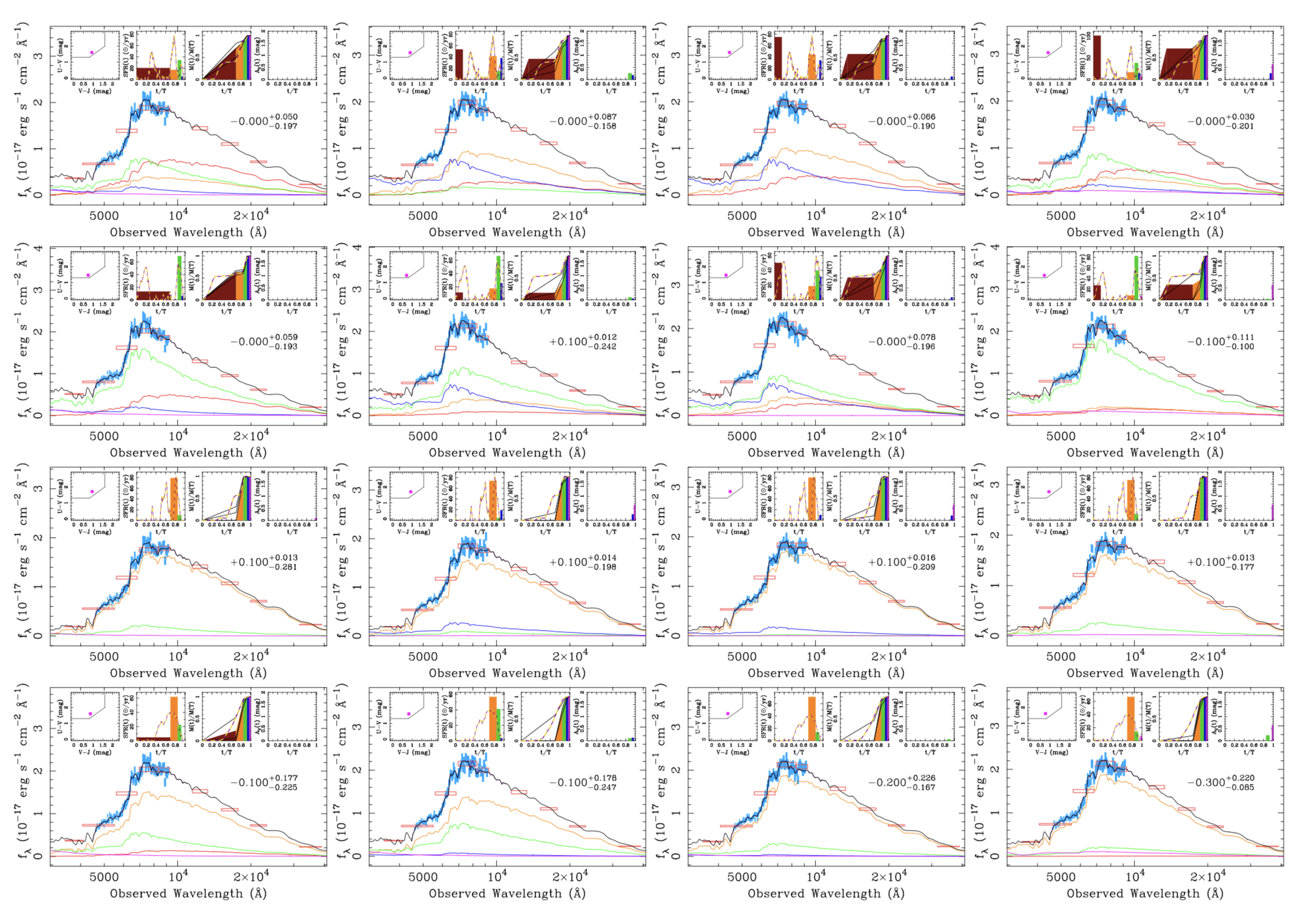}
\caption{Rows show example observations for different mock galaxies from Models A (top two rows) and D (bottom two rows)
at $S/N=20\,{\rm pix^{-1}}$.
{\bfr The bottom row is for a mock galaxy with half-solar metallicity.} Each simulation uses 100
mock galaxies observed 100 times to probe how inferred SFH confidence intervals circumscribe the true SFHs. At this
$S/N$, the best-fit SFH solutions (middle two insets) capture many of the features of the underlying true SFH, shown as
yellow dashed lines. That our SED fitting generally recovers these basic SFH features is the key result of these
simulations. {\bfr The metal abundances inferred from the SED fitting are listed inside each panel.}
\label{fig:simspec}}
\end{figure*}


\subsection{Simulations of CSI Data and their Implications}
\label{sec:sims}

The ideal set of mock SFHs is one that fills the entire possible {\it empirical} parameter space, with a broad diversity of early mass growth 
independent of later SFRs. By definition, the non-Markovian stochastic SFHs from Kelson (2014),\footnote{Specifically, we use those with 
$H=1$; see Kelson (2014); KBA16).} fit the bill perfectly---they are scale-free and have stochastic increments that 
are zero in the mean.  Furthermore, the $1/f$ noise in these SFHs provides a realistic match to the SFMS and its scatter, as well as a 
diversity of shapes similar to that seen in real galaxies and hydrodynamic simulations (see Kelson 2014; KBA16). Using these as a basis of our simulated 
CSI data provides a broad range of intrinsic \z5fract2 properties and a broad range of ongoing SFRs from quiescence through the vigorous 
rates of growth seen at intermediate redshift.

The first step was to turn these mock SFHs into SEDs using the Flexible Stellar Population Synthesis models (Conroy, Gunn, \& White 2009), 
fixed (solar) metallicity, $5\le S/N\le 80$ for the prism spectra, and variable amounts of dust reddening.  In addition to histories generated 
in a manner consistent with Kelson (2014), we also simulated mock galaxies in which arbitrary amounts of ancient stellar populations were 
added, allowing for different levels of dust reddening for the old stars compared to that covering the stellar populations that grew over a more 
extended period of time.  The point of this is to vary the amount of old mass that is potentially difficult to detect or even hidden under the noise, 
contriving situations that bracket what is empirically unknown to help us construct sensible bounds on the fractions of galaxies that grew late.

The simplest runs assume zero dust reddening (Model A), with others having more complexity, such as random screen reddening between 
$0\le A_V\le 1$ mag independent of mock galaxy properties (Model B), or random screen reddening up to a maximum $A_V$ set by the 
SFR at \Tobs\ normalized to a maximum of $A_V=2$ mag (Model C).  Another set solely used histories that have $\sfr=0$ in their final 
30 Myr timestep to help compare sensitivities to \z5fract2 in both star-forming and passive mock galaxies (Model D).

The first three simulations (Models A--C) are built on an unbiased random sampling of Kelson (2014) growth trajectories, and therefore
have intrinsically high LBFs of 46\%, and (apparent) quiescent fractions (QFs) of $\sim 15\%$---derived from inferred sSFR in the last 
200 Myr bin (classification by {\it UVJ} selection results in a QF of $\sim 17\%$).

Model D has a low intrinsic LBF of 7\% because it is comprised of stochastic SFHs with zero ongoing SFR; these histories have a long 
SFR coherence time. The ``observed'' 200 Myr sSFRs imply a QF of $\sim 87\%$ ({\it UVJ} classification implies a QF of 95\%).

We also ran a fifth model---Model E---identical to Model B except in that 25\% of the mass is added as very old stars hidden with $A_V=2$\,
mag.  This test probes the extent to which old stellar mass that only modestly impacts galaxy SEDs may bias measurements of the early mass 
fractions. It has a a similarly low QF to Model B, despite the significantly lower intrinsic LBF.

These simulations test sensitivity to the relative amounts of old stellar mass at a range of $S/N$ and broad levels of dust reddening.  In such 
models the old stellar populations may be more attenuated than is realistic, though as starforming disk galaxies are seen edge on, the effects 
of dust reddening on integrated colors can be significant (Patel et al. 2012).  Having greater amounts of dust mixed with the old stars than is 
present in the starforming disks, however, is an infrequent occurance, even though our stellar population fitting certainly can model such mixtures.

Figure \ref{fig:simspec} shows examples of such mock SED observations in a manner similar to the presentation of CSI data in Figure \ref{fig:mosSEDs}.
Qualitatively, late-time mass fractions are generally consistent with the true values from the input SFHs. Table \ref{tab:simlbf} summarizes 
CSI's ability to quantitatively select galaxies that grew more than half their stars in the 2 Gyr prior to observation from the five sets of simulations. For
each set we provide the ``observed'' quiescent fractions (again using the 200 Myr age bin from the best-fit SED), as well as the \emph{intrinsic} LBF 
of the set, followed by the \emph{measured} LBFs using selection of LBs by $\z5fract2_{ML}<0.5$ and $\z5fract2_{95\%}<0.5$.  Both reddening and 
noise change the ability to detect/identify late bloomers {\it quantitatively\/} but not {\it qualitatively\/}. Serious degradation in the ability to 
reliably identify late bloomers occurs once the $S/N\simlt 10$ per pixel, for reasons already elucidated.


\subsubsection{The Four Questions}

Having run hundreds of SFHs at many $S/N$, $A_{V}$, etc., according to the above schemes, we can now answer The Four Questions:
\begin{enumerate}
\item What metrics can be used to robustly measure the late bloomer fraction?
\item How accurate are the observed late bloomer fractions?
\item How pure is the late bloomer selection?
\item What are the chief contaminants in an observationally selected late bloomer sample?
\end{enumerate}

\paragraph{Metrics for Measurement of LBF}
Selecting late bloomers using $\z5fract2_{ML}$---the maximum likelihood \z5fract2 estimates---always over-estimates what (and which)
fraction of the population grew most of their mass in the last 2 Gyr. Too many galaxies that did not grow so late have best-fit SED parameters that 
identify too little early stellar mass due to noise in the data and the way it reduces the statistical significance of pixels that deviate from the fit by 
young and intermediate-aged stellar populations. Thus, there is too little information present for the best-fit solutions to encode sufficient amounts 
of old stars. 

A more probabilistic approach, whereby the 95\% upper limit is used to select late bloomers with a threshold of $\z5fract2_{95\%}<0.5$, yields
significantly more accurate LBFs, with significantly improved purities. Answer No.\ 1 is thus: {\it When there are late bloomers, 
we can count them most robustly using a selection of $\z5fract2_{95\%}<0.5$.\/}

\paragraph{LBF Measurement Accuracy}
While the measurements of late bloomer Fractions using the Maximum Likelihood selection are always too high, the selection by 95\% upper 
confidence limits appears not to overestimate the LBF by more than a few percent. The presence of dust appears to diminish the ability to classify 
$\sim$10\% of galaxies, for which the noise and resulting covariances/degeneracies between reddening and old stellar mass simply produces 
confidence intervals that are too broad for robust late bloomer identification, even if their $\z5fract2_{ML}$ values would have led to 
inclusion.   We would much rather lose late bloomers and {\it underestimate} the true LBF if it means strengthening the empirical evidence
for this populations that challenges ``established'' concepts and approaches such as the preservation of mass rank-ordering.
Answer No.\ 2 is thus: {\it Our measured late bloomer Fractions appear only to require small corrections.\/} In detail, we estimate corrections 
for the quiescent and star-forming fractions of galaxies separately, thus creating a weighted correction based on the QF in bins of stellar mass and 
redshift


\paragraph{LB Selection Purity}
In parentheses in Table \ref{tab:simlbf} we also list the ``purity'' of late bloomer samples selected using a given criterion; i.e, $\z5fract2_{ML}<0.5$
or $\z5fract2_{95\%}<0.5$.  In the models that most closely mimic the actual CSI sample---Models A--C---these numbers are $\sim$75\%;
i.e., $3/4$ of observational LB identifications should be physically meaningful. Answer No.\ 3 is thus: {\it Most of the late bloomers that
are selected truly grew more than half their mass in the last 2 Gyr.\/}  {\bfp And while we are confident that our selection of individual
late bloomers has this level of purity, the question of how many of our catalog entries in Section \ref{sec:catalog} will be verified by
subsequent follow-up, on an object-by-object basis, will follow that of strongly biased coin flips, with results that
depend sensitively to the purity with which the follow-up observations can measure the amount of old stellar mass in galaxies dominated by
young stellar populations. If follow-up techniques have similar purity to that probed by CSI in this paper, $\sim
75\%$ purity, then these follow-up observations will re-identify only $75\%\times 75\% + 25\%\times 25\%$, or $62.5\%$ of our late
bloomers as late bloomers. If follow-up techniques can provide an ability to identify late bloomers with $\simgt 95\%$ purity, then
$75\%\times 0.95\% + 25\%\times 05\%$, or $70\%$ of the individual galaxies we identified as late bloomers would be reidentified as such.}
In detail, and as discussed below, we also 
estimate purities for the quiescent and star-forming fractions of galaxies separately, and create a weighted purity based on the QF in bins of 
stellar mass and redshift.

\paragraph{Chief LB Contaminants}
Importantly, although the purity is never 100\%, in almost all cases, the majority of LB false-positives have intrinsic values $0.5\le\z5fract2\le 0.6$. 
Even for mock galaxies observed at the limit of $\z5fract2_{95\%}\sim 0.5$, $\simlt 75\%$ of these LBs grew more than 40\% of their mass in the 
last 2 Gyr. Answer No.\ 4 is thus: {\it Integrating over the broad range of possible intrinsic $\z5fract2$ values (which is unknown in a data set), 
between 50\% and 75\% of the contaminants have $\z5fract2<0.6$, with half of the remaining contaminants having grown at least a third of 
their stellar mass in the last 2 Gyr.}  In other words, almost all of these galaxies are {\it young\/}, even if not late bloomers by our conservative 
definition. Whether or not such systems are meaningfully distinct from ``true'' LBs is a topic we encourage other authors to explore.


\subsubsection{Corrections to Observed Late Bloomer Fraction by Type}

Table \ref{tab:simlbf_sf} dissects the mock observations further, given the modest dependence of LBF accuracy and LB purity on the
mix of star-formation activity in Models A--E. Table \ref{tab:simlbf_sf} lists the measured LBFs and purities for mock galaxies classified as 
star-forming and quiescent (again using their 200 Myr sSFRs). In principle, the statistics in Table \ref{tab:simlbf_sf} can be used to construct 
corrections to any measured LBFs if one takes into account the mix of quiescent and star-forming galaxies in a given sample.

For example, at intermediate redshifts in CSI, $\sim$70\% of galaxies at $\log\Mstel>10.8$ are quiescent.  For such galaxies, the models 
indicate we should observe an LBF of $\sim$0.05, with approximately zero systematic error.  For the $\sim $30\% of galaxies in 
that mass range that are star-forming, perhaps they are similar to the galaxies classified as star-forming in Model D, only that there are a 
few more of them (to lower the apparent QF from 87\% to 70\%). Table \ref{tab:simlbf_sf} suggests that we would observe an LBF of 
$\sim$0.39 for the  star-forming ensemble, even though the intrinsic LBF may be lower by $\sim 0.07$.

These estimates of systematic errors by galaxy or SED type let us construct systematic corrections to the CSI late bloomer fractions,
{\bfr $\Delta \hbox{LBF}$}, in each mass or redshift bin:
$\hbox{QF} \times \Delta \hbox{LBF}_Q  + (1-\hbox{QF}) \times \Delta \hbox{LBF}_{SF}$, the mean of the systematic offsets
between the measured and intrinsic LBF for quiescent and star-forming galaxies, weighted by the observed quiescent and star-forming fractions.

We can check if this approximation is valid by testing whether we can obtain, approximately, the right LBFs using a similar weighting. For galaxies 
at $\log M>10.8$, with $\hbox{QF}\sim 0.70$, one expects to measure an LBF of $0.70\times 0.06 + 0.30\times 0.39$, or $\sim 0.16$ ---roughly the 
fraction presented in Figure \ref{fig:lbfs} (though the intrinsic value may be more like $\sim$0.11 depending on the true mix of SFHs).  At higher 
redshifts, the SFMS moves up (e.g., Whitaker et al. 2014), suggesting that the star forming population will be more vigorous, perhaps more 
consistent with histories in Models A--C. If there is little-to-no dust attenuation in those galaxies, one might expect the LBF to approach 
$0.70\times 0.06 + 0.30\times 0.50$, $\sim 0.19$ ---consistent with the measurements in CSI's higher redshift slices (Figure \ref{fig:lbfMassTime}).

Similar calculations hold for intermediate ($10.4\le \log\Mstel\le 10.8$) and lower masses ($10.0\le\log\Mstel\le10.4$).  The observed QF in the 
former is $\sim$50\%, so one expects an observed LBF of $0.50\times 0.06 + 0.50\times 0.43$, or $\sim 0.25$ --- what we see. At lower masses, the 
QF is $\sim 25\%$, suggesting an LBF of $0.25\times 0.06 + 0.75\times 0.43$, or $\sim 0.34$ --- again, what we see in CSI. One can take these rough 
agreements with the observed LBFs either as indicating that the suites of Kelson (2014) SFHs are a good match to those of real galaxies,
or that the simulations can accurately be used to estimate (systematic) uncertainties and (systematic) corrections to the observed LBFs in CSI. 
We only require the latter.

Thus, our simple recipe for correcting CSI measurements of LBFs utilizes the fact that, according to Model D in Table \ref{tab:simlbf_sf}, the measured 
LBF is overestimated by about $+0.04$ for quiescent galaxies, and about $+0.07$ for starforming ones. Thus we might expect the CSI measurement 
of LBF at $\log M>10.8$ to be overestimated by $0.70\times 0.04 + 0.30\times 0.07$, or $\sim 0.05$. In the next two lower mass bins, the equivalent 
corrections are $0.50\times 0.04 + 0.50\times 0.07$, or $\sim 0.055$, and $0.25\times 0.04 + 0.75\times 0.07$, or $\sim 0.063$, respectively. 
We take these values as corrections to our LBFs for estimating plausible lower bounds.  We adopt a systematic uncertainty of $\pm 0.05$ in all of the 
LBFs quoted in this paper.


\subsubsection{Simulation Examples: Model C}

In the tables above, we see that the selection by $\z5fract2_{95\%}$ is significantly more conservative. While obtaining the correct LBF to within 
$\pm 0.05$ in the full samples of Models A, D, and E, it significantly underestimates the LBF in Models B and C.  There, more galaxies apparently 
scatter out than scatter in, with dust reddened young or intermediate SEDs admitting more old stellar mass than is intrinsically present. Here 
we explore such effects graphically to illustrate what kinds of objects scatter in and out of the LB classification.


\begin{figure*}[htb]\figurenum{A2}
\centerline{\includegraphics[width=1.00\hsize]{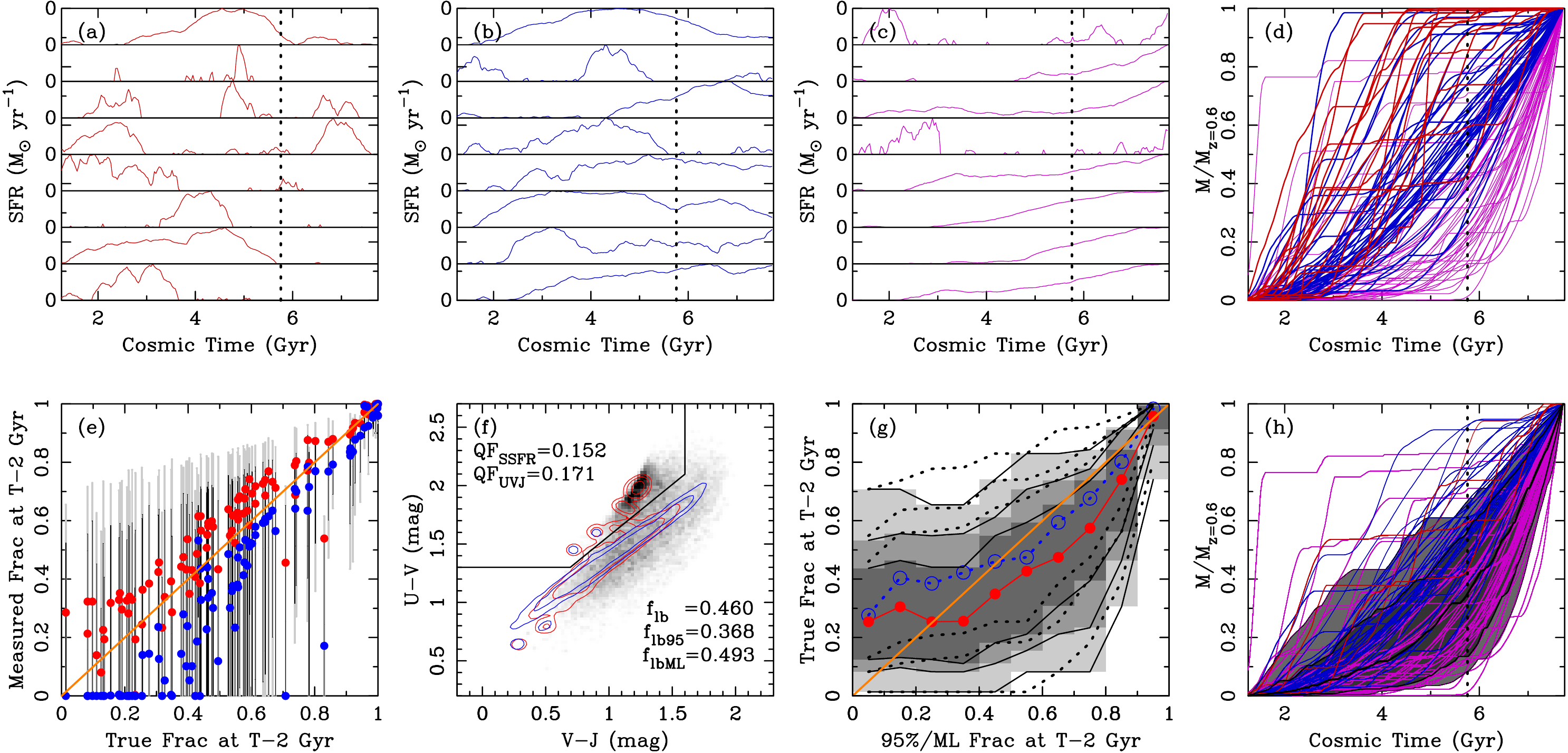}}
\caption{Summary of the Model C simulations of mock galaxies at $z=0.6$ observed at $S/N=40$ per pixel. These simulations include simple
screens of dust reddening in which the attenuation, $A_V$, is randomly selected uniformly up to an amount set by its ongoing
SFR, with the entire population of $A_V$ normalized to a maximum $A_V=2$ mag.
(a-c) Mock SFHs generated using the formalism of Kelson (2014) for use in simulating CSI's ability to accurately measure the amounts of
old stellar mass in systems. Red trajectories have low star formation activity at the epoch of observation, blue intermediate, and
violet high.
(d) Integrated growth trajectories for the 100 mock galaxies used in the simulations.
(e) Distributions of measured best-fit (maximum likelihood, ML) $\z5fract2$ for the 100 observations of each of 100 mock galaxies
observed. Dark gray vertical bars extend over the 16th-84th percentiles of the ML measurements for each
galaxy. The light gray extend over the 5th-95th percentiles. Blue circles show the median ML measurement for each mock galaxy,
skewing to lower and lower values as the amount of old mass present decreases. Red circles show the median of the 95\% upper
limits derived from each mock measurement. Noise in the data skews the distribution of ML measurements such that the typical
values are biased low.
(f) Distribution of restframe $UVJ$ diagram for the mock galaxies. The quiescent galaxy fractions are shown, measured using
$UVJ$ classification or classification by SSFR as measured by the last 200 Myr age bin. The quiescent fraction of this batch of
simulated galaxies is $\sim 16\%$, low compared to the $\sim 40\%$ quiescent fraction in CSI for galaxies at $\log M>10$. Here
the simulation also has a high intrinsic late bloomer fraction of 46\%, underestimated by our $\z5fract2_{95\%}$ selection
criterion. Blue contours trace galaxies selected thusly to be late bloomers. Red contours show the distribution of the remaining
galaxies. The grayscale in the background is the distribution of galaxies at $z\sim 0.6$ in CSI with stellar masses $\log M>10$.
The mock galaxies reproduce the general morphology of the ``quiescent clump'' and star forming sequence, suggesting
that simulations such as these, agnostic to assumptions about the forms of early SFHs, can provide a useful proxy for
getting at systematic issues in SED fitting and estimates of early mass fractions.
(g) Distribution of true $\z5fract2$ values given different $\z5fract2_{95\%}$ or $\z5fract2_{ML}$ thresholds when selecting late
growing mock galaxies. Blue circles trace the median true value of selected galaxies when using the ML $\z5fract2$ measurements.
Black dashed lines trace the 5th, 16th, 25th, 75th, 84th, and 95th percentiles. Red circles and solid black lines trace the
equivalent percentiles when selecting late bloomers using $\z5fract2_{95\%}$ instead of the maximum likelihood values. Selecting
late bloomers using $\z5fract2_{95\%}$ results in a purer identification of galaxies that grew the bulk of their stellar mass
late.
(h) Growth trajectories for the mock galaxies with observations that led to their selection as late bloomers. Shaded regions show
the 5th, 16th, 84th, and 95th percentiles of these growth trajectories, to highlight the range of mass fractions captured
by our LB selection criteria. An additional solid black line traces the 50th percentile.
\label{fig:simAv2simple}}
\end{figure*}


Figure \ref{fig:simAv2simple} quantitatively shows for Model C the measurements of mass in stars formed before $z=1$ relative to the 
mass measured at the epoch of observation ($z= 0.6$). While these plots present a simple analysis of $10^4$ mock observations at 
$S/N=40$ per pixel, the basic picture only degrades significantly once one reaches $S/N\approx 10$. Overall these mock observations show 
that we can place strong constraints on the fraction of old stellar mass {\it for individual galaxies\/}.

\paragraph{Panels (a--c)} show example SFHs for mock galaxies in the simulations. Red, blue, and violet code trajectories for $\log {\dot
\Mstel}/\Mstel<-10.5$, $-10.5\le \log {\dot \Mstel}/\Mstel \le -10$, and $\log {\dot \Mstel}/\Mstel>-10$, respectively, where 
$\Mstel=\int {\dot \Mstel}\,dt$.  Note: these sSFRs reflect SFR on a $\sim$30 Myr timescale because the stochastic SFHs have 200 timesteps.

\paragraph{Panel (d)} shows the 100 growth trajectories using the same color coding. The Kelson (2014) non-Markovian stochastic SFHs naturally 
produce starforming and quiescent galaxies at a ratio of $\sim$5:1 (in their final 30 Myr timestep), which also holds in the late bloomer population.
Hence, while most late bloomers are star-forming at \Tobs, {\it quite a few are not}. Having simulations with both star-forming and quiescent LBs
among the broad diversity of possible SFHs is critical for revealing what kind(s) of LBs we can reliably identify in CSI, or conversely
whether CSI can help determine the duty cycles and frequencies with which galaxies may exit and/or enter the quiescent zone of, for 
example, the {\it UVJ} diagram.

\paragraph{Panel (e)} shows the distribution of best-fit, $\z5fract2_{ML}$ estimates for each galaxy. Since each mock galaxy was observed 
100 times, we have a well defined distribution of constraints on the mass present by $z=1$ for each one.  Each vertical light (dark) gray bar 
extends over the 5$^{\rm th}$--95$^{\rm th}$ (16$^{\rm th}$--84$^{\rm th}$) percentiles of the \z5fract2$_{ML}$ distributions for each 
mock galaxy. Blue filled circles show the medians.  Below $\z5fract2_{\rm true}\simlt 0.5$ it becomes increasingly difficult to detect and 
measure the presence of old stars using our techniques unless one has $S/N\simgt 100$ per pixel. At the typical CSI $S/N$, galaxies with low 
intrinsic fractions of stellar mass older than 2 Gyr are increasingly driven towards best-fit measurements containing identically zero old stars, 
and with increasing frequency.

That said, red circles in Figure \ref{fig:simAv2simple}{\it e} mark the medians of the 95\% upper confidence limits on \z5fract2\ for 
each galaxy.  When the best-fit solutions are consistent with zero old stars, our methodology maintains broad confidence limits, accurately 
reflecting the increased uncertainty on the old stellar mass so as to capture at least the potential for old stars to be present.

We again conclude that the right approach in CSI is to use the 95\% $\z5fract2$ upper-limits to identify LBs. These problems are likely to be
generally applicable, hence classification by best-fit SFH properties should be performed with great caution.

\paragraph{Panel (f)} shows the distribution of ``observed'' {\it UVJ} colors for these mock galaxies in comparison to the {\it UVJ} colors 
of CSI galaxies at $z\sim 0.6$ and $\log\Mstel> 10$. Blue contours to trace the mock LBs as classified by $z5fract2_{95\%}<0.5$; red contours trace 
the rest.  The quiescent clump is consistent with CSI's because those galaxies have very low intrinsic $A_{V}$, matching the dust-free mock galaxies.  
However, the star-forming mock galaxies in Figure \ref{fig:simAv2simple} do not span the full range of CSI colors because this {\it does} require dust 
reddening (see Patel et al 2010).

\paragraph{Panel (g)} schematically traces the distributions of true $\z5fract2$ given different $\z5fract2_{95\%}$ and $\z5fract2_{ML}$ 
observational selection thresholds. Perfect inferences of SFHs, and thus perfect ability to identify late bloomers, would trace the orange, diagonal, 
unity line. Observational noise results in errors and uncertainties in the inferred SFHs, quantified here as an increasing dispersion and (upward) 
bias in true $\z5fract2$ values as the amount of old mass inferred from the SED fits decreases.

Blue circles trace the median true value of selected galaxies using $\z5fract2_{ML}$. Black dashed lines trace the 
5$^{\rm th}$, 16$^{\rm th}$, 25$^{\rm th}$, 75$^{\rm th}$, 84$^{\rm th}$, and 95$^{\rm th}$ percentiles. Selecting late growing galaxies 
by $\z5fract2_{ML}<0.5$ produces a sample where $\sim$75\% intrinsically have $\z5fract2\simlt 0.6$, and another 20\% of reach up to 
$\z5fract2\sim 0.85$.

The red circles and solid black lines trace the equivalent percentiles using $\z5fract2_{95\%}$, instead. Here $\sim$84\% of galaxies with 
measured $\z5fract2_{95\%}<0.5$ intrinsically have $\z5fract2\simlt 0.6$, and another $\sim$10\% reach $\z5fract2\sim 0.75$. Thus, selection 
of late bloomers by $\z5fract2_{95\%}$ results in a purer sample.

\paragraph{\it Panel (h)} uses selection by $\z5fract2_{95\%}$ to identify galaxies as (probable) LBs and shows their growth trajectories. The black 
lines trace the 5th, 16th, 50th, 84th, and 95th percentiles, to better illustrate the small fraction of the sample that are not true LBs. Identifying 
nontrivial samples of galaxies at $z\ll 1$ that have had such growth histories is a major challenge to notions that most normal and massive galaxies 
formed their stars and quenched early, leaving open the question of how such galaxies lingered for so long before experiencing rapid late-time growth. 
Furthermore, such galaxies break analytical approaches to ensemble galaxy evolution that do not explicitly account for the leap-frogging of 
subpopulations past others in their mass growth.


\begin{figure*}[tb]\figurenum{A3}
\centerline{\includegraphics[width=1.00\hsize]{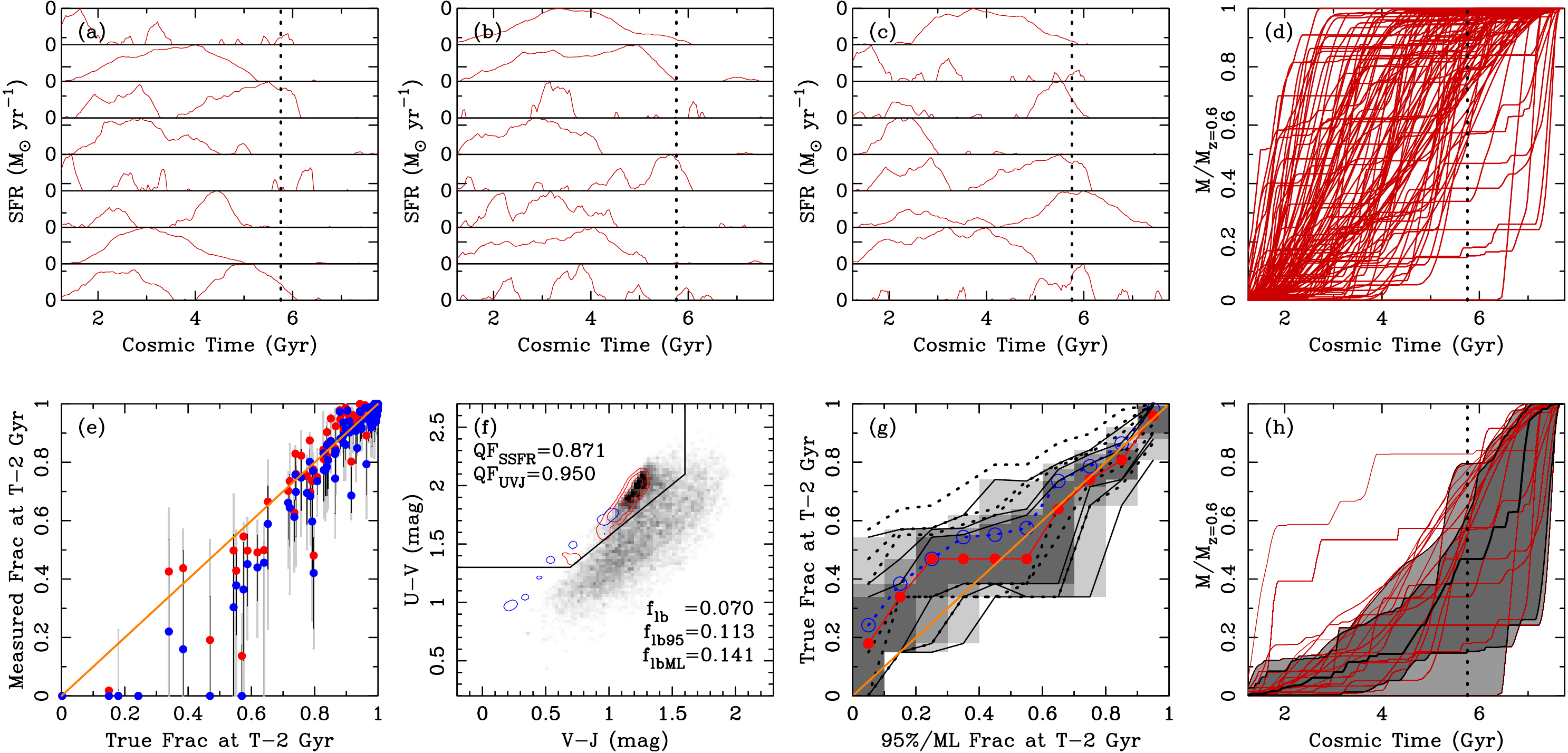}}
\caption{Summary of the Model D simulations. Panels as in Figure \ref{fig:simAv2simple}.
\label{fig:simOldsimple}}
\end{figure*}


\subsubsection{Simulation Examples: Model D}

Figure \ref{fig:simOldsimple} shows the same information but for Model D---mock galaxies with $\sfr(\Tobs)=0$. Overall these mock observations show 
that we can place strong constraints on the fraction of stellar mass already in place by $z=1$ for individual passive galaxies.

Even for mock galaxies where we have ``observed'' $\z5fract2_{95\%}\sim 0.5$, 84\% of such objects intrinsically have $\z5fract2<0.6$.

Identifying nontrivial samples of quiescent galaxies at $z\ll 1$ that have had such growth histories allows us to better probe the dynamism
hidden in the histories of galaxies that are largely viewed as ``red and dead.'' Measuring the late bloomer fractions in seemingly quiescent
populations also quantifies the extent to which galaxies enter the red sequence and stop growing new stars for the remainders of their lifetimes.


\subsubsection{{\bfp Simulation Examples: The Effects of Subsolar Metallicities and Lower Signal-To-Noise Ratios}}

\begin{figure*}[htb]\figurenum{A4}
\includegraphics[width=\hsize]{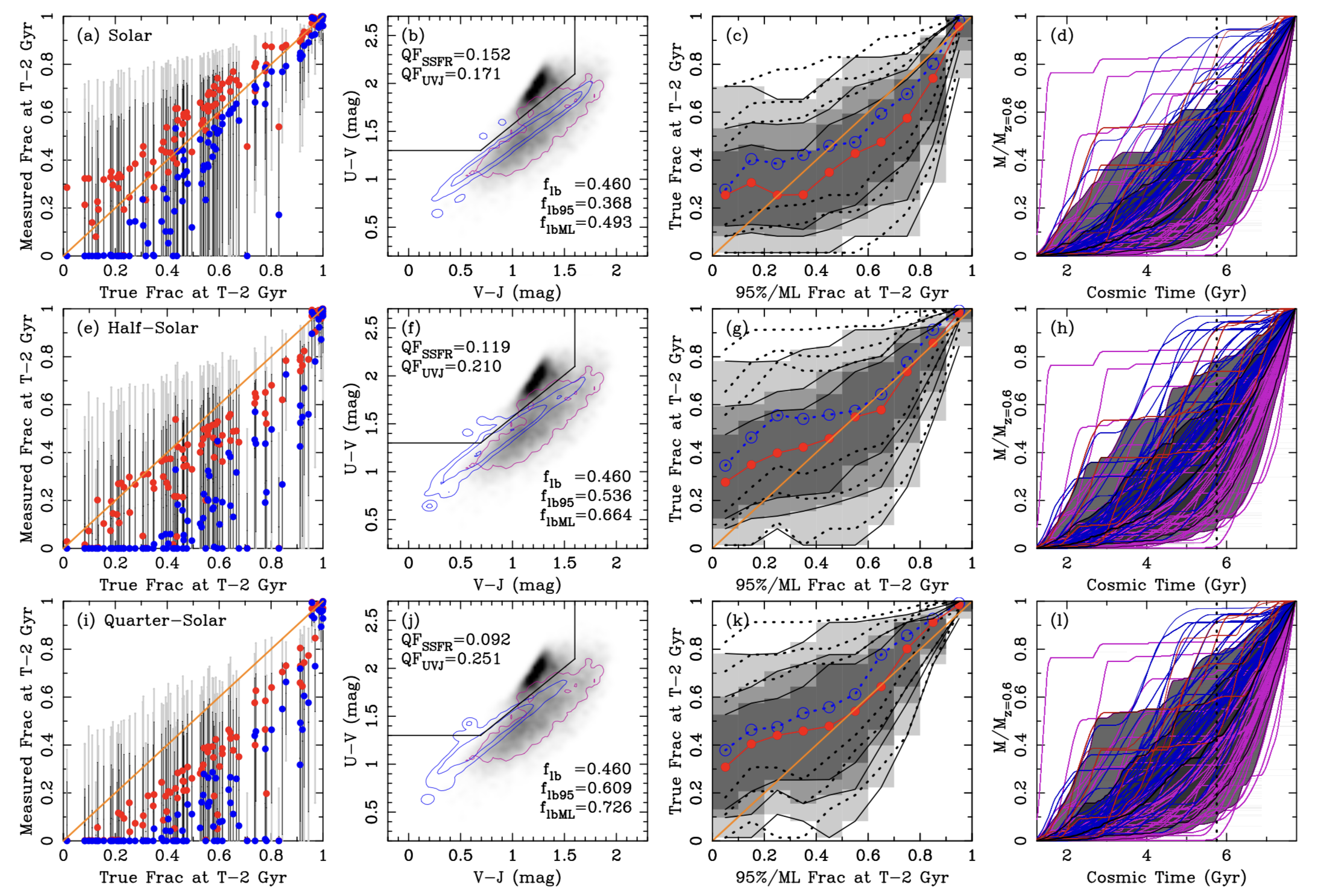}
\caption{{\bfp Summary of the effects of metallicity using Model C at $S/N=40$.
Each row is analogous to panels (e-h) of Figure \ref{fig:simAv2simple}, with
the exception of the UVJ diagrams which now only show (blue) the distribution of mock galaxies selected to be late-bloomers after
SED fitting, and (violet) the contour containing 75\% of the late-bloomers in CSI with stellar masses above $10^{10}$ M$_\odot$.
Top row shows the results for solar metallicity stellar populations.
Center row shows the results of our template fitting when the galaxies have half-solar metallicity stellar populations.
Bottom row shows the results of our template fitting when the galaxies have quarter-solar metallicity stellar populations.}
\label{fig:simMetals}}
\end{figure*}

{\bfp
When $S/N$ ratios are insufficient to simultaneously constrain both the ages and metallicities of stellar population, SED fitting
procedure(s) can produce results skewed by any input priors on stellar population metallicity---which in CSI are biased towards solar.
To test the sensitivity of CSI's ability to select late bloomers over a range of metallicities, we ran three variants of Model (C)--- with
stellar populations having solar metallicity, half-solar metallicity, and quarter-solar metallicity. Figure \ref{fig:simMetals} recreates
the final four panels of Figure \ref{fig:simAv2simple} for these simulations at the three metallicities, and at $S/N=40$. The UVJ diagrams
(b,f, and j) are, however, presented with violet contours that trace (75\% of) the late bloomers in CSI. For comparison, the blue
contours trace the distribution of mock galaxies that were selected by our fitting procedures to be late bloomers. Table
\ref{tab:simlbf_s2n} summarizes the key statistics from these simulations and for additional $S/N$ ratios than are displayed in these
figures.}

{\bfp
The key takeway from the simulations that vary metallicity and signal-to-noise ratio is that estimates of late bloomer fractions are only
mildly sensitive to the range of signal-to-noise ratios being considered in the samples used in this paper. In contrast, our methodology
for estimating a population's late bloomer fraction is sensitive to the metallicity(ies) of the underlying stellar populations\footnote{To
avoid this sensitivity, the $S/N$ ratios should be $\gg 40$ per pixel at the resolution of CSI.}.  Quantitatively, if the stellar
populations in {\it all\/} galaxies selected by CSI as late bloomers had quarter-solar metallicity, then our analysis of their
spectra makes $33\%-50\%$ more of them appear as late bloomers than are currently present. However,
given past work on the ages and metallicities of galaxies at these epochs (see, e.g., Gallazzi et al. 2014) or the distributions of UVJ
colors that are observed (see Panels b,f, and j). there is no evidence to suggest $[\hbox{Fe}/\hbox{H}] \approx -0.6$ dex is representative
of the metallicities for a majority of the late bloomers identified by CSI. These simulations indicate that the effect is $\sim 20\%$ when
the underlying galaxies identified as late bloomers have stellar populations with metallicities that are half-solar, i.e. bringing any
observed estimate of $25\%$ down to $20\%$---a systematic error within our already stated levels of uncertainty.}


\clearpage
\LongTables

\tabcolsep=0.11cm

\begin{deluxetable*}{c l l c c c c c c c c c }
\tablecaption{Properties of CSI Galaxies in CANDELS/UDS
\label{tab:inCANDELS0}}
\tablehead{
\colhead{Panel} &
\colhead{RA} &
\colhead{DEC} &
\colhead{$i$} &
\colhead{$S/N$} &
\colhead{$z$} &
\colhead{$\log M_{\hbox{*}}\tablenotemark{a,b}$} &
\multicolumn{4}{c}{Fractional Growth History} &
\colhead{$n_2$} \\
\colhead{\#} &
&
&
\colhead{(mag)} &
&
&
\colhead{$(M_\odot)$} &
\colhead{(2 Gyr)\tablenotemark{c}}&
\colhead{(1 Gyr)\tablenotemark{d}}&
\colhead{(0.5 Gyr)}&
\colhead{(0.2 Gyr)}&
\colhead{(Mpc$^{-3}$)} 
}
\startdata
  1 &    2:17:48.94 &   -5:13:12.4 & $ 21.94 $ & $ 17$ & $ 0.652_{-0.007}^{+0.018} $ & $  10.00_{-0.02}^{+0.04} $  & $   0.00_{-0.00}^{+0.00} $ & $   0.73_{-0.02}^{+0.04} $ & $   0.84_{-0.09}^{+0.01} $ & $   0.95_{-0.03}^{+0.01} $ &   5.04 \\ 
  2 &    2:17:57.09 &   -5:11:00.5 & $ 20.56 $ & $ 54$ & $ 0.526_{-0.021}^{+0.021} $ & $  10.37_{-0.01}^{+0.04} $  & $   0.00_{-0.00}^{+0.00} $ & $   0.74_{-0.54}^{+0.16} $ & $   0.87_{-0.02}^{+0.03} $ & $   0.87_{-0.02}^{+0.03} $ &   1.53 \\ 
  3 &    2:17:43.38 &   -5:12:33.0 & $ 20.97 $ & $ 35$ & $ 0.656_{-0.007}^{+0.012} $ & $  10.34_{-0.02}^{+0.01} $  & $   0.00_{-0.00}^{+0.00} $ & $   0.49_{-0.01}^{+0.02} $ & $   0.89_{-0.07}^{+0.02} $ & $   0.93_{-0.02}^{+0.00} $ &   6.02 \\ 
  4 &    2:16:55.44 &   -5:09:01.5 & $ 21.06 $ & $ 32$ & $ 0.604_{-0.006}^{+0.037} $ & $  10.58_{-0.02}^{+0.12} $  & $   0.00_{-0.00}^{+0.00} $ & $   0.00_{-0.00}^{+0.72} $ & $   0.98_{-0.29}^{+0.02} $ & $   1.00_{-0.00}^{+0.00} $ &   1.99 \\ 
  5 &    2:17:37.05 &   -5:12:26.7 & $ 21.94 $ & $ 18$ & $ 0.656_{-0.011}^{+0.024} $ & $  10.02_{-0.02}^{+0.03} $  & $   0.00_{-0.00}^{+0.13} $ & $   0.00_{-0.00}^{+0.25} $ & $   0.98_{-0.27}^{+0.01} $ & $   0.99_{-0.00}^{+0.01} $ &   6.17 \\ 
  6 &    2:16:58.33 &   -5:07:52.4 & $ 21.24 $ & $ 40$ & $ 0.632_{-0.021}^{+0.008} $ & $  10.57_{-0.02}^{+0.03} $  & $   0.00_{-0.00}^{+0.00} $ & $   0.21_{-0.08}^{+0.38} $ & $   0.97_{-0.07}^{+0.00} $ & $   0.97_{-0.01}^{+0.01} $ &   2.92 \\ 
  7 &    2:17:11.68 &   -5:07:42.7 & $ 22.18 $ & $ 15$ & $ 0.714_{-0.030}^{+0.001} $ & $  10.24_{-0.05}^{+0.02} $  & $   0.00_{-0.00}^{+0.00} $ & $   0.76_{-0.39}^{+0.02} $ & $   0.98_{-0.00}^{+0.00} $ & $   0.98_{-0.00}^{+0.00} $ &   1.47 \\ 
  8 &    2:17:54.10 &   -5:12:49.9 & $ 19.96 $ & $ 31$ & $ 0.572_{-0.011}^{+0.007} $ & $  10.69_{-0.03}^{+0.01} $  & $   0.00_{-0.00}^{+0.00} $ & $   0.00_{-0.00}^{+0.31} $ & $   0.99_{-0.23}^{+0.01} $ & $   1.00_{-0.00}^{+0.00} $ &   1.16 \\ 
  9 &    2:17:37.81 &   -5:14:23.3 & $ 21.62 $ & $ 15$ & $ 0.684_{-0.013}^{+0.018} $ & $  10.38_{-0.02}^{+0.02} $  & $   0.00_{-0.00}^{+0.28} $ & $   0.00_{-0.00}^{+0.52} $ & $   0.98_{-0.41}^{+0.01} $ & $   0.98_{-0.00}^{+0.01} $ &   2.64 \\ 
 10 &    2:16:59.33 &   -5:16:23.7 & $ 21.20 $ & $ 30$ & $ 0.668_{-0.018}^{+0.015} $ & $  10.13_{-0.02}^{+0.01} $  & $   0.00_{-0.00}^{+0.00} $ & $   0.00_{-0.00}^{+0.34} $ & $   0.75_{-0.25}^{+0.04} $ & $   0.75_{-0.04}^{+0.04} $ &   2.91 \\ 
 11 &    2:17:27.41 &   -5:15:20.8 & $ 20.34 $ & $ 36$ & $ 0.458_{-0.007}^{+0.006} $ & $  10.25_{-0.02}^{+0.01} $  & $   0.00_{-0.00}^{+0.01} $ & $   0.93_{-0.06}^{+0.05} $ & $   0.98_{-0.01}^{+0.00} $ & $   0.98_{-0.01}^{+0.00} $ &   1.52 \\ 
 12 &    2:17:00.65 &   -5:10:31.6 & $ 21.25 $ & $ 37$ & $ 0.634_{-0.013}^{+0.009} $ & $  10.01_{-0.04}^{+0.03} $  & $   0.00_{-0.00}^{+0.00} $ & $   0.31_{-0.09}^{+0.08} $ & $   0.79_{-0.07}^{+0.09} $ & $   0.93_{-0.01}^{+0.00} $ &   3.16 \\ 
 13 &    2:17:29.56 &   -5:15:33.3 & $ 21.58 $ & $ 25$ & $ 0.658_{-0.010}^{+0.019} $ & $  10.44_{-0.03}^{+0.01} $  & $   0.00_{-0.00}^{+0.21} $ & $   0.05_{-0.01}^{+0.32} $ & $   0.93_{-0.55}^{+0.04} $ & $   0.96_{-0.00}^{+0.00} $ &   2.31 \\ 
 14 &    2:17:47.17 &   -5:08:46.3 & $ 21.46 $ & $ 19$ & $ 0.600_{-0.012}^{+0.006} $ & $  10.19_{-0.02}^{+0.01} $  & $   0.00_{-0.00}^{+0.00} $ & $   0.06_{-0.06}^{+0.32} $ & $   0.76_{-0.35}^{+0.14} $ & $   0.99_{-0.00}^{+0.01} $ &   1.33 \\ 
 15 &    2:16:53.85 &   -5:16:15.6 & $ 20.98 $ & $ 57$ & $ 0.754_{-0.008}^{+0.015} $ & $  10.05_{-0.01}^{+0.08} $  & $   0.00_{-0.00}^{+0.00} $ & $   0.00_{-0.00}^{+0.23} $ & $   0.00_{-0.00}^{+0.26} $ & $   0.40_{-0.05}^{+0.15} $ &   1.36 \\ 
 16 &    2:16:56.46 &   -5:12:57.4 & $ 20.76 $ & $126$ & $ 0.506_{-0.009}^{+0.004} $ & $  10.43_{-0.02}^{+0.01} $  & $   0.00_{-0.00}^{+0.00} $ & $   0.00_{-0.00}^{+0.13} $ & $   1.00_{-0.00}^{+0.00} $ & $   1.00_{-0.00}^{+0.00} $ &   0.32 \\ 
 17 &    2:17:18.69 &   -5:14:00.9 & $ 21.46 $ & $ 20$ & $ 0.590_{-0.010}^{+0.064} $ & $  10.17_{-0.02}^{+0.10} $  & $   0.00_{-0.00}^{+0.49} $ & $   0.08_{-0.02}^{+0.74} $ & $   0.93_{-0.20}^{+0.01} $ & $   0.93_{-0.01}^{+0.03} $ &   2.45 \\ 
 18 &    2:17:05.49 &   -5:12:44.7 & $ 21.60 $ & $ 39$ & $ 0.608_{-0.007}^{+0.017} $ & $  10.25_{-0.01}^{+0.03} $  & $   0.00_{-0.00}^{+0.00} $ & $   0.70_{-0.01}^{+0.04} $ & $   0.95_{-0.06}^{+0.00} $ & $   0.96_{-0.02}^{+0.00} $ &   2.09 \\ 
 19 &    2:17:22.99 &   -5:13:23.4 & $ 21.94 $ & $ 19$ & $ 0.656_{-0.036}^{+0.021} $ & $  10.21_{-0.03}^{+0.03} $  & $   0.00_{-0.00}^{+0.22} $ & $   0.74_{-0.02}^{+0.07} $ & $   0.87_{-0.13}^{+0.07} $ & $   0.92_{-0.01}^{+0.03} $ &   3.36 \\ 
 28 &               &               &            & $ 21$ & $ 0.642_{-0.013}^{+0.020} $ & $  10.25_{-0.06}^{+0.04} $  & $   0.43_{-0.43}^{+0.13} $ & $   0.43_{-0.43}^{+0.13} $ & $   0.97_{-0.07}^{+0.00} $ & $   0.97_{-0.02}^{+0.01} $ &   3.63 \\ 
 20 &    2:17:11.41 &   -5:15:56.4 & $ 20.90 $ & $ 57$ & $ 0.676_{-0.017}^{+0.008} $ & $  10.47_{-0.04}^{+0.03} $  & $   0.00_{-0.00}^{+0.00} $ & $   0.24_{-0.24}^{+0.35} $ & $   0.79_{-0.09}^{+0.13} $ & $   0.96_{-0.00}^{+0.00} $ &   2.65 \\ 
 21 &    2:17:42.14 &   -5:13:56.1 & $ 22.55 $ & $ 17$ & $ 0.676_{-0.032}^{+0.010} $ & $  10.22_{-0.04}^{+0.03} $  & $   0.06_{-0.06}^{+0.18} $ & $   0.98_{-0.04}^{+0.02} $ & $   1.00_{-0.00}^{+0.00} $ & $   1.00_{-0.00}^{+0.00} $ &   4.97 \\ 
 22 &    2:16:57.42 &   -5:14:45.3 & $ 21.89 $ & $ 20$ & $ 0.670_{-0.020}^{+0.022} $ & $  10.19_{-0.05}^{+0.03} $  & $   0.21_{-0.21}^{+0.26} $ & $   0.78_{-0.09}^{+0.08} $ & $   0.90_{-0.18}^{+0.06} $ & $   0.95_{-0.09}^{+0.01} $ &   2.41 \\ 
 23 &    2:16:58.70 &   -5:10:28.2 & $ 21.38 $ & $ 46$ & $ 0.642_{-0.008}^{+0.004} $ & $  10.32_{-0.03}^{+0.04} $  & $   0.32_{-0.24}^{+0.16} $ & $   0.32_{-0.08}^{+0.17} $ & $   1.00_{-0.09}^{+0.00} $ & $   1.00_{-0.00}^{+0.00} $ &   2.66 \\ 
 24 &    2:17:04.27 &   -5:13:25.9 & $ 22.20 $ & $ 20$ & $ 0.704_{-0.014}^{+0.009} $ & $  10.31_{-0.02}^{+0.03} $  & $   0.34_{-0.34}^{+0.18} $ & $   0.34_{-0.15}^{+0.21} $ & $   0.81_{-0.21}^{+0.17} $ & $   0.99_{-0.01}^{+0.00} $ &   1.62 \\ 
 69 &               &               &            & $ 35$ & $ 0.682_{-0.008}^{+0.007} $ & $  10.47_{-0.12}^{+0.03} $  & $   0.90_{-0.05}^{+0.01} $ & $   0.90_{-0.04}^{+0.02} $ & $   1.00_{-0.04}^{+0.00} $ & $   1.00_{-0.00}^{+0.00} $ &   2.09 \\ 
 25 &    2:16:59.73 &   -5:10:35.1 & $ 22.31 $ & $ 21$ & $ 0.654_{-0.016}^{+0.015} $ & $  10.26_{-0.06}^{+0.06} $  & $   0.36_{-0.36}^{+0.27} $ & $   0.61_{-0.06}^{+0.16} $ & $   1.00_{-0.00}^{+0.00} $ & $   1.00_{-0.00}^{+0.00} $ &   2.80 \\ 
 26 &    2:16:59.16 &   -5:12:56.7 & $ 20.42 $ & $ 40$ & $ 0.578_{-0.011}^{+0.005} $ & $  10.44_{-0.06}^{+0.07} $  & $   0.36_{-0.36}^{+0.33} $ & $   0.36_{-0.36}^{+0.33} $ & $   0.99_{-0.09}^{+0.01} $ & $   0.99_{-0.01}^{+0.01} $ &   1.00 \\ 
 27 &    2:17:12.00 &   -5:09:12.9 & $ 21.91 $ & $ 23$ & $ 0.640_{-0.012}^{+0.008} $ & $  10.20_{-0.03}^{+0.05} $  & $   0.38_{-0.06}^{+0.27} $ & $   0.66_{-0.04}^{+0.08} $ & $   1.00_{-0.00}^{+0.00} $ & $   1.00_{-0.00}^{+0.00} $ &   2.75 \\ 
 29 &    2:17:38.68 &   -5:15:53.0 & $ 20.77 $ & $ 23$ & $ 0.604_{-0.017}^{+0.010} $ & $  10.85_{-0.04}^{+0.05} $  & $   0.50_{-0.21}^{+0.21} $ & $   0.50_{-0.21}^{+0.21} $ & $   0.88_{-0.13}^{+0.12} $ & $   1.00_{-0.00}^{+0.00} $ &   3.38 \\ 
 30 &    2:17:44.51 &   -5:09:10.1 & $ 21.39 $ & $ 16$ & $ 0.762_{-0.008}^{+0.032} $ & $  10.81_{-0.05}^{+0.03} $  & $   0.50_{-0.50}^{+0.22} $ & $   0.50_{-0.24}^{+0.22} $ & $   0.88_{-0.20}^{+0.12} $ & $   1.00_{-0.00}^{+0.00} $ &   0.67 \\ 
 31 &    2:17:43.30 &   -5:14:04.4 & $ 20.59 $ & $ 25$ & $ 0.640_{-0.029}^{+0.014} $ & $  10.42_{-0.08}^{+0.05} $  & $   0.50_{-0.50}^{+0.08} $ & $   0.50_{-0.10}^{+0.08} $ & $   0.96_{-0.07}^{+0.01} $ & $   0.96_{-0.01}^{+0.01} $ &   5.45 \\ 
 32 &    2:17:33.01 &   -5:13:14.0 & $ 22.01 $ & $ 34$ & $ 0.634_{-0.009}^{+0.007} $ & $  10.32_{-0.01}^{+0.05} $  & $   0.51_{-0.13}^{+0.14} $ & $   0.72_{-0.17}^{+0.08} $ & $   1.00_{-0.00}^{+0.00} $ & $   1.00_{-0.00}^{+0.00} $ &   6.50 \\ 
 33 &    2:16:58.27 &   -5:11:02.7 & $ 21.84 $ & $ 28$ & $ 0.658_{-0.009}^{+0.009} $ & $  10.45_{-0.07}^{+0.05} $  & $   0.53_{-0.24}^{+0.23} $ & $   0.53_{-0.23}^{+0.23} $ & $   1.00_{-0.19}^{+0.00} $ & $   1.00_{-0.00}^{+0.00} $ &   2.98 \\ 
 34 &    2:17:34.33 &   -5:12:44.4 & $ 21.91 $ & $ 26$ & $ 0.640_{-0.008}^{+0.011} $ & $  10.28_{-0.06}^{+0.01} $  & $   0.57_{-0.23}^{+0.02} $ & $   0.57_{-0.23}^{+0.02} $ & $   1.00_{-0.05}^{+0.00} $ & $   1.00_{-0.00}^{+0.00} $ &   6.52 \\ 
 35 &    2:17:03.39 &   -5:09:40.9 & $ 20.22 $ & $ 56$ & $ 0.504_{-0.007}^{+0.003} $ & $  10.72_{-0.05}^{+0.04} $  & $   0.60_{-0.07}^{+0.16} $ & $   0.80_{-0.05}^{+0.05} $ & $   1.00_{-0.00}^{+0.00} $ & $   1.00_{-0.00}^{+0.00} $ &   0.52 \\ 
 36 &    2:17:33.54 &   -5:15:33.9 & $ 21.70 $ & $ 39$ & $ 0.590_{-0.008}^{+0.010} $ & $  10.03_{-0.04}^{+0.09} $  & $   0.61_{-0.41}^{+0.11} $ & $   0.61_{-0.06}^{+0.11} $ & $   1.00_{-0.00}^{+0.00} $ & $   1.00_{-0.00}^{+0.00} $ &   3.00 \\ 
 37 &    2:17:25.70 &   -5:12:43.7 & $ 21.10 $ & $ 18$ & $ 0.674_{-0.023}^{+0.016} $ & $  10.63_{-0.10}^{+0.01} $  & $   0.62_{-0.41}^{+0.02} $ & $   0.69_{-0.21}^{+0.24} $ & $   0.99_{-0.00}^{+0.00} $ & $   0.99_{-0.00}^{+0.00} $ &   2.30 \\ 
 38 &    2:16:55.47 &   -5:11:10.3 & $ 22.20 $ & $ 18$ & $ 0.664_{-0.020}^{+0.017} $ & $  10.27_{-0.10}^{+0.02} $  & $   0.62_{-0.62}^{+0.14} $ & $   0.74_{-0.17}^{+0.04} $ & $   1.00_{-0.25}^{+0.00} $ & $   1.00_{-0.00}^{+0.00} $ &   3.38 \\ 
 39 &    2:17:29.41 &   -5:12:25.4 & $ 22.29 $ & $ 29$ & $ 0.650_{-0.014}^{+0.012} $ & $  10.19_{-0.06}^{+0.02} $  & $   0.63_{-0.23}^{+0.01} $ & $   0.63_{-0.22}^{+0.01} $ & $   1.00_{-0.00}^{+0.00} $ & $   1.00_{-0.00}^{+0.00} $ &   4.62 \\ 
 40 &    2:17:53.28 &   -5:09:27.8 & $ 21.15 $ & $ 19$ & $ 0.644_{-0.005}^{+0.016} $ & $  10.24_{-0.02}^{+0.07} $  & $   0.63_{-0.02}^{+0.12} $ & $   0.63_{-0.02}^{+0.12} $ & $   0.81_{-0.11}^{+0.06} $ & $   0.94_{-0.03}^{+0.01} $ &   2.05 \\ 
 41 &    2:17:31.26 &   -5:12:17.2 & $ 21.10 $ & $ 52$ & $ 0.600_{-0.007}^{+0.008} $ & $  10.56_{-0.07}^{+0.01} $  & $   0.65_{-0.18}^{+0.06} $ & $   0.65_{-0.18}^{+0.06} $ & $   1.00_{-0.00}^{+0.00} $ & $   1.00_{-0.00}^{+0.00} $ &   3.07 \\ 
 42 &    2:16:55.69 &   -5:12:48.2 & $ 20.99 $ & $ 48$ & $ 0.638_{-0.010}^{+0.007} $ & $  10.63_{-0.06}^{+0.02} $  & $   0.65_{-0.19}^{+0.03} $ & $   0.65_{-0.19}^{+0.03} $ & $   1.00_{-0.08}^{+0.00} $ & $   1.00_{-0.00}^{+0.00} $ &   3.01 \\ 
 43 &    2:17:24.12 &   -5:15:56.1 & $ 20.48 $ & $ 68$ & $ 0.440_{-0.006}^{+0.004} $ & $  10.28_{-0.01}^{+0.08} $  & $   0.65_{-0.08}^{+0.14} $ & $   0.92_{-0.03}^{+0.02} $ & $   0.93_{-0.01}^{+0.01} $ & $   0.93_{-0.01}^{+0.01} $ &   1.73 \\ 
 44 &    2:17:55.57 &   -5:11:54.7 & $ 22.14 $ & $ 16$ & $ 0.632_{-0.014}^{+0.012} $ & $  10.36_{-0.06}^{+0.01} $  & $   0.67_{-0.23}^{+0.07} $ & $   0.67_{-0.22}^{+0.07} $ & $   0.81_{-0.10}^{+0.18} $ & $   1.00_{-0.00}^{+0.00} $ &   2.11 \\ 
 45 &    2:17:33.67 &   -5:12:04.1 & $ 20.62 $ & $ 44$ & $ 0.698_{-0.006}^{+0.006} $ & $  10.95_{-0.06}^{+0.01} $  & $   0.69_{-0.13}^{+0.03} $ & $   0.69_{-0.13}^{+0.03} $ & $   1.00_{-0.08}^{+0.00} $ & $   1.00_{-0.00}^{+0.00} $ &   2.44 \\ 
 46 &    2:17:05.65 &   -5:12:55.0 & $ 21.39 $ & $ 40$ & $ 0.632_{-0.015}^{+0.012} $ & $  10.57_{-0.01}^{+0.01} $  & $   0.71_{-0.08}^{+0.07} $ & $   0.71_{-0.08}^{+0.07} $ & $   0.88_{-0.09}^{+0.07} $ & $   1.00_{-0.00}^{+0.00} $ &   2.53 \\ 
 47 &    2:17:21.06 &   -5:08:56.7 & $ 20.60 $ & $ 40$ & $ 0.604_{-0.008}^{+0.009} $ & $  10.71_{-0.07}^{+0.01} $  & $   0.71_{-0.10}^{+0.04} $ & $   0.71_{-0.10}^{+0.04} $ & $   1.00_{-0.00}^{+0.00} $ & $   1.00_{-0.00}^{+0.00} $ &   1.24 \\ 
 48 &    2:17:27.08 &   -5:09:50.5 & $ 21.21 $ & $ 41$ & $ 0.610_{-0.007}^{+0.006} $ & $  10.59_{-0.06}^{+0.02} $  & $   0.72_{-0.08}^{+0.03} $ & $   0.72_{-0.08}^{+0.03} $ & $   1.00_{-0.07}^{+0.00} $ & $   1.00_{-0.00}^{+0.00} $ &   1.53 \\ 
 59 &               &               &            & $ 43$ & $ 0.588_{-0.006}^{+0.006} $ & $  10.53_{-0.01}^{+0.11} $  & $   0.82_{-0.06}^{+0.10} $ & $   0.87_{-0.01}^{+0.04} $ & $   1.00_{-0.00}^{+0.00} $ & $   1.00_{-0.00}^{+0.00} $ &   1.31 \\ 
 49 &    2:17:38.00 &   -5:13:09.9 & $ 21.40 $ & $ 49$ & $ 0.642_{-0.008}^{+0.008} $ & $  10.57_{-0.07}^{+0.02} $  & $   0.73_{-0.11}^{+0.05} $ & $   0.73_{-0.11}^{+0.05} $ & $   1.00_{-0.04}^{+0.00} $ & $   1.00_{-0.00}^{+0.00} $ &   6.78 \\ 
 50 &    2:17:33.78 &   -5:14:01.9 & $ 20.61 $ & $ 38$ & $ 0.662_{-0.010}^{+0.005} $ & $  10.97_{-0.06}^{+0.01} $  & $   0.73_{-0.12}^{+0.02} $ & $   0.73_{-0.12}^{+0.02} $ & $   0.90_{-0.10}^{+0.10} $ & $   1.00_{-0.00}^{+0.00} $ &   4.96 \\ 
 51 &    2:17:36.61 &   -5:13:34.2 & $ 21.64 $ & $ 19$ & $ 0.648_{-0.004}^{+0.010} $ & $  10.72_{-0.04}^{+0.06} $  & $   0.74_{-0.05}^{+0.18} $ & $   0.74_{-0.05}^{+0.18} $ & $   1.00_{-0.07}^{+0.00} $ & $   1.00_{-0.00}^{+0.00} $ &   6.10 \\ 
 52 &    2:17:25.59 &   -5:13:31.4 & $ 20.39 $ & $ 45$ & $ 0.630_{-0.008}^{+0.010} $ & $  10.96_{-0.10}^{+0.05} $  & $   0.75_{-0.15}^{+0.10} $ & $   0.75_{-0.15}^{+0.10} $ & $   0.99_{-0.05}^{+0.01} $ & $   1.00_{-0.00}^{+0.00} $ &   4.95 \\ 
 53 &    2:17:14.46 &   -5:15:46.0 & $ 20.56 $ & $ 67$ & $ 0.686_{-0.005}^{+0.003} $ & $  11.01_{-0.01}^{+0.03} $  & $   0.77_{-0.00}^{+0.05} $ & $   0.77_{-0.00}^{+0.05} $ & $   1.00_{-0.07}^{+0.00} $ & $   1.00_{-0.00}^{+0.00} $ &   1.71 \\ 
 54 &    2:17:08.28 &   -5:16:11.6 & $ 21.46 $ & $ 36$ & $ 0.682_{-0.013}^{+0.010} $ & $  10.65_{-0.07}^{+0.02} $  & $   0.78_{-0.11}^{+0.05} $ & $   0.78_{-0.09}^{+0.05} $ & $   1.00_{-0.01}^{+0.00} $ & $   1.00_{-0.00}^{+0.00} $ &   2.39 \\ 
 55 &    2:16:54.22 &   -5:14:56.4 & $ 19.90 $ & $154$ & $ 0.624_{-0.005}^{+0.002} $ & $  11.24_{-0.01}^{+0.01} $  & $   0.79_{-0.00}^{+0.01} $ & $   0.79_{-0.00}^{+0.01} $ & $   0.87_{-0.02}^{+0.01} $ & $   1.00_{-0.00}^{+0.00} $ &   2.24 \\ 
 56 &    2:17:09.69 &   -5:08:53.8 & $ 21.44 $ & $ 38$ & $ 0.634_{-0.014}^{+0.011} $ & $  10.54_{-0.07}^{+0.01} $  & $   0.79_{-0.08}^{+0.07} $ & $   0.89_{-0.05}^{+0.01} $ & $   1.00_{-0.00}^{+0.00} $ & $   1.00_{-0.00}^{+0.00} $ &   2.80 \\ 
 57 &    2:17:43.33 &   -5:13:30.6 & $ 21.21 $ & $ 23$ & $ 0.640_{-0.005}^{+0.010} $ & $  10.80_{-0.06}^{+0.03} $  & $   0.79_{-0.09}^{+0.10} $ & $   0.79_{-0.09}^{+0.10} $ & $   1.00_{-0.09}^{+0.00} $ & $   1.00_{-0.00}^{+0.00} $ &   5.31 \\ 
 62 &               &               &            & $ 32$ & $ 0.630_{-0.006}^{+0.012} $ & $  10.80_{-0.10}^{+0.04} $  & $   0.83_{-0.09}^{+0.08} $ & $   0.94_{-0.19}^{+0.00} $ & $   1.00_{-0.00}^{+0.00} $ & $   1.00_{-0.00}^{+0.00} $ &   5.41 \\ 
 58 &    2:17:28.93 &   -5:13:17.9 & $ 21.28 $ & $ 30$ & $ 0.604_{-0.010}^{+0.009} $ & $  10.69_{-0.05}^{+0.01} $  & $   0.81_{-0.09}^{+0.03} $ & $   0.81_{-0.09}^{+0.03} $ & $   1.00_{-0.05}^{+0.00} $ & $   1.00_{-0.00}^{+0.00} $ &   4.57 \\ 
 60 &    2:17:11.60 &   -5:09:01.0 & $ 20.59 $ & $ 59$ & $ 0.646_{-0.007}^{+0.002} $ & $  10.97_{-0.08}^{+0.01} $  & $   0.82_{-0.07}^{+0.00} $ & $   0.84_{-0.09}^{+0.00} $ & $   1.00_{-0.00}^{+0.00} $ & $   1.00_{-0.00}^{+0.00} $ &   2.86 \\ 
 67 &               &               &            & $ 54$ & $ 0.650_{-0.007}^{+0.003} $ & $  11.03_{-0.09}^{+0.01} $  & $   0.90_{-0.05}^{+0.00} $ & $   0.90_{-0.05}^{+0.00} $ & $   1.00_{-0.00}^{+0.00} $ & $   1.00_{-0.00}^{+0.00} $ &   3.03 \\ 
 61 &    2:17:38.95 &   -5:13:05.2 & $ 21.11 $ & $ 31$ & $ 0.652_{-0.007}^{+0.005} $ & $  10.78_{-0.07}^{+0.01} $  & $   0.83_{-0.08}^{+0.06} $ & $   0.83_{-0.08}^{+0.08} $ & $   1.00_{-0.11}^{+0.00} $ & $   1.00_{-0.00}^{+0.00} $ &   6.37 \\ 
 63 &    2:17:37.48 &   -5:14:31.1 & $ 20.70 $ & $ 72$ & $ 0.624_{-0.008}^{+0.006} $ & $  10.86_{-0.01}^{+0.08} $  & $   0.83_{-0.10}^{+0.04} $ & $   0.85_{-0.06}^{+0.09} $ & $   1.00_{-0.00}^{+0.00} $ & $   1.00_{-0.00}^{+0.00} $ &   5.27 \\ 
 64 &    2:17:31.04 &   -5:12:37.5 & $ 21.57 $ & $ 50$ & $ 0.636_{-0.006}^{+0.006} $ & $  10.58_{-0.07}^{+0.03} $  & $   0.84_{-0.07}^{+0.06} $ & $   0.84_{-0.07}^{+0.06} $ & $   1.00_{-0.08}^{+0.00} $ & $   1.00_{-0.00}^{+0.00} $ &   5.75 \\ 
 65 &    2:17:36.68 &   -5:13:41.2 & $ 22.59 $ & $ 25$ & $ 0.642_{-0.008}^{+0.007} $ & $  10.04_{-0.01}^{+0.11} $  & $   0.85_{-0.07}^{+0.05} $ & $   0.85_{-0.05}^{+0.06} $ & $   1.00_{-0.00}^{+0.00} $ & $   1.00_{-0.00}^{+0.00} $ &   6.39 \\ 
 66 &    2:17:31.86 &   -5:09:57.7 & $ 20.14 $ & $ 64$ & $ 0.498_{-0.006}^{+0.004} $ & $  10.94_{-0.09}^{+0.01} $  & $   0.87_{-0.06}^{+0.01} $ & $   0.87_{-0.06}^{+0.02} $ & $   1.00_{-0.00}^{+0.00} $ & $   1.00_{-0.00}^{+0.00} $ &   1.63 \\ 
 68 &    2:17:34.17 &   -5:13:39.3 & $ 19.80 $ & $132$ & $ 0.438_{-0.003}^{+0.002} $ & $  10.93_{-0.07}^{+0.01} $  & $   0.90_{-0.02}^{+0.07} $ & $   1.00_{-0.00}^{+0.00} $ & $   1.00_{-0.00}^{+0.00} $ & $   1.00_{-0.00}^{+0.00} $ &   0.84 \\ 
 70 &    2:17:38.16 &   -5:13:19.1 & $ 20.80 $ & $ 56$ & $ 0.634_{-0.006}^{+0.003} $ & $  10.93_{-0.08}^{+0.01} $  & $   0.92_{-0.06}^{+0.04} $ & $   0.97_{-0.02}^{+0.01} $ & $   1.00_{-0.01}^{+0.00} $ & $   1.00_{-0.00}^{+0.00} $ &   6.39 \\ 
 71 &    2:17:45.71 &   -5:13:27.7 & $ 21.05 $ & $ 22$ & $ 0.774_{-0.007}^{+0.007} $ & $  10.92_{-0.08}^{+0.02} $  & $   0.92_{-0.02}^{+0.01} $ & $   0.92_{-0.02}^{+0.01} $ & $   0.92_{-0.02}^{+0.01} $ & $   0.99_{-0.01}^{+0.00} $ &   1.10 \\ 
 72 &    2:17:14.97 &   -5:12:29.3 & $ 21.87 $ & $ 30$ & $ 0.760_{-0.008}^{+0.007} $ & $  10.45_{-0.09}^{+0.01} $  & $   0.93_{-0.04}^{+0.00} $ & $   0.93_{-0.03}^{+0.00} $ & $   0.93_{-0.03}^{+0.00} $ & $   0.98_{-0.01}^{+0.00} $ &   1.08 \\ 
 73 &    2:17:06.22 &   -5:13:17.8 & $ 20.23 $ & $ 75$ & $ 0.622_{-0.005}^{+0.003} $ & $  11.19_{-0.01}^{+0.01} $  & $   0.94_{-0.01}^{+0.00} $ & $   0.94_{-0.01}^{+0.00} $ & $   0.99_{-0.01}^{+0.01} $ & $   1.00_{-0.00}^{+0.00} $ &   2.32 \\ 
 74 &    2:17:29.02 &   -5:15:49.1 & $ 21.57 $ & $ 27$ & $ 0.772_{-0.012}^{+0.008} $ & $  10.67_{-0.07}^{+0.03} $  & $   0.96_{-0.01}^{+0.01} $ & $   0.96_{-0.01}^{+0.01} $ & $   0.96_{-0.01}^{+0.01} $ & $   0.96_{-0.01}^{+0.01} $ &   1.40 \\ 
 75 &    2:17:44.55 &   -5:15:22.3 & $ 19.37 $ & $138$ & $ 0.490_{-0.004}^{+0.001} $ & $  11.11_{-0.01}^{+0.02} $  & $   0.96_{-0.01}^{+0.04} $ & $   1.00_{-0.00}^{+0.00} $ & $   1.00_{-0.00}^{+0.00} $ & $   1.00_{-0.00}^{+0.00} $ &   1.32 \\ 
 79 &               &               &            & $ 86$ & $ 0.488_{-0.006}^{+0.004} $ & $  11.28_{-0.01}^{+0.01} $  & $   0.98_{-0.01}^{+0.00} $ & $   0.98_{-0.01}^{+0.00} $ & $   1.00_{-0.00}^{+0.00} $ & $   1.00_{-0.00}^{+0.00} $ &   1.31 \\ 
 76 &    2:17:36.32 &   -5:11:01.1 & $ 21.40 $ & $ 32$ & $ 0.492_{-0.007}^{+0.007} $ & $  10.48_{-0.01}^{+0.01} $  & $   0.96_{-0.02}^{+0.00} $ & $   0.96_{-0.02}^{+0.00} $ & $   1.00_{-0.00}^{+0.00} $ & $   1.00_{-0.00}^{+0.00} $ &   1.57 \\ 
 77 &    2:17:29.91 &   -5:08:44.1 & $ 21.23 $ & $ 27$ & $ 0.490_{-0.010}^{+0.011} $ & $  10.50_{-0.12}^{+0.01} $  & $   0.96_{-0.04}^{+0.03} $ & $   0.96_{-0.02}^{+0.03} $ & $   1.00_{-0.03}^{+0.00} $ & $   1.00_{-0.00}^{+0.00} $ &   1.33 \\ 
 78 &    2:18:01.19 &   -5:08:22.4 & $ 21.07 $ & $ 53$ & $ 0.564_{-0.007}^{+0.005} $ & $  10.75_{-0.01}^{+0.01} $  & $   0.97_{-0.01}^{+0.00} $ & $   0.97_{-0.01}^{+0.00} $ & $   1.00_{-0.02}^{+0.00} $ & $   1.00_{-0.00}^{+0.00} $ &   1.80 \\ 
 80 &    2:17:45.53 &   -5:10:06.9 & $ 21.76 $ & $ 16$ & $ 0.518_{-0.010}^{+0.010} $ & $  10.33_{-0.15}^{+0.03} $  & $   0.99_{-0.02}^{+0.01} $ & $   0.99_{-0.02}^{+0.01} $ & $   0.99_{-0.01}^{+0.00} $ & $   1.00_{-0.00}^{+0.00} $ &   1.42 \\ 
\enddata
\tablecomments{
Confidence intervals reflect formal/random errors only.
}
\tablenotetext{a}{Formal errors on stellar masses are typically 0.03-0.05 dex.}
\tablenotetext{b}{Systematic uncertainties can be up to 0.1-0.3 dex,
based on our simulations of CSI data using synthetic star formation histories.}
\tablenotetext{c}{$z5fract2$}
\tablenotetext{d}{$z5fract$}
\end{deluxetable*}


\begin{deluxetable*}{l l l c c c}[t]
\tablecaption{True and Recovered LBFs and Purity\tablenotemark{a} of Selection of Simulated Data
\label{tab:simlbf}}
\tablehead{
\multicolumn{1}{l}{Model} &
\colhead{QF} & \colhead{LBF} & \colhead{LBF$_{ML}(P_{ML})$} & \colhead{LBF$_{95\%}(P_{95\%})$}
}
\startdata
(A) $A_V=0$                                       & 0.15 & {\bf 0.46} & $0.58(0.70)$ & ${\bf 0.43}(0.76)$ \\
(B) $\hbox{max}A_V=1$                             & 0.15 & {\bf 0.46} & $0.49(0.69)$ & ${\bf 0.35}(0.76)$ \\
(C) $\hbox{max}(A_V)\propto\log{\dot M}_{z=0.6}$  & 0.15 & {\bf 0.46} & $0.49(0.69)$ & ${\bf 0.37}(0.75)$ \\
(D) ${\dot M}_{z=0.6}=0$\tablenotemark{b}, $A_V=0$& 0.87 & {\bf 0.07} & $0.15(0.44)$ & ${\bf 0.11}(0.55)$ \\
(E) $3:1$ mix of B $+$ Dead                       & 0.17 & {\bf 0.27} & $0.41(0.46)$ & ${\bf 0.28}(0.51)$
\enddata
\tablenotetext{a}{Fraction of observations selected to be late bloomers for which the mock galaxy intrinsically had $\z5fract2<0.5$. The majority
of false positives have $\z5fract2<0.6$.}
\tablenotetext{b}{Quiescent SFH defined as ${\dot M}\equiv 0$ in the final 30 Myr timestep.}
\end{deluxetable*}


\begin{deluxetable*}{c c c c c c c c c c}
\tablecaption{Late Bloomer Fractions and Purity\tablenotemark{a} of Selection in Simulated Data by Star Forming Activity
\label{tab:simlbf_sf}}
\tablehead{
&\multicolumn{3}{c}{Star Forming\tablenotemark{b}} &&
\multicolumn{3}{c}{Quiescent\tablenotemark{b}}\\
\cline{2-4}\cline{6-8}\\
\colhead{Model} &
\colhead{LBF} & \colhead{LBF$_{ML}$} & \colhead{LBF$_{95\%}$} &&
\colhead{LBF} & \colhead{LBF$_{ML}$} & \colhead{LBF$_{95\%}$}
}
\startdata
(A) & {\bf 0.53} & $0.67 (0.70)$ & ${\bf 0.50} (0.77)$ && {\bf 0.06} & $0.07 (0.42)$ & ${\bf 0.05} (0.44)$  \\
(B) & {\bf 0.53} & $0.55 (0.71)$ & ${\bf 0.40} (0.77)$ && {\bf 0.06} & $0.11 (0.31)$ & ${\bf 0.06} (0.40)$  \\
(C) & {\bf 0.53} & $0.57 (0.70)$ & ${\bf 0.43} (0.76)$ && {\bf 0.06} & $0.07 (0.41)$ & ${\bf 0.04} (0.39)$  \\
(D) & {\bf 0.32} & $0.47 (0.64)$ & ${\bf 0.39} (0.74)$ && {\bf 0.02} & $0.09 (0.24)$ & ${\bf 0.06} (0.31)$  \\
(E) & {\bf 0.32} & $0.46 (0.48)$ & ${\bf 0.32} (0.52)$ && {\bf 0.00} & $0.11 (0.00)$ & ${\bf 0.06} (0.00)$  
\enddata
\tablenotetext{a}{Fraction of observations selected to be late bloomers that indeed had $\z5fract2<0.5$. The majority
of false positives have $\z5fract2<0.6$.}
\tablenotetext{b}{Distinction between quiescent or star-forming based on the measurement of the 200 Myr age bin in
the mock SED fitting. Imperfect correspondence between observed SSFR on 200 Myr and intrinsic 30 Myr SSFR leads Model D
to contain some mock galaxies that could thus be classified empirically as ``star forming.''}
\end{deluxetable*}


\begin{deluxetable*}{c c l c c c}[h]
\tablecaption{The Effects of $S/N$ Ratio and Uncertain Metallicity on True and Recovered LBFs and Purity\tablenotemark{a} with Simulated Data
using Model (C)
\label{tab:simlbf_s2n}}
\tablehead{
\colhead{$S/N$ Ratio} &
\colhead{$[\hbox{Fe}/\hbox{H}]$} & \colhead{LBF} & \colhead{LBF$_{ML}(P_{ML})$} & \colhead{LBF$_{95\%}(P_{95\%})$}
}
\startdata
40 &  0.0 & {\bf 0.46} & $0.49(0.69)$ & ${\bf 0.37}(0.75)$ \\
40 & -0.3 & {\bf 0.46} & $0.66(0.60)$ & ${\bf 0.54}(0.67)$ \\
40 & -0.6 & {\bf 0.46} & $0.73(0.59)$ & ${\bf 0.61}(0.63)$ \\
\noalign{\vskip 6pt}
20 &  0.0 & {\bf 0.46} & $0.50(0.65)$ & ${\bf 0.35}(0.69)$ \\
20 & -0.3 & {\bf 0.46} & $0.69(0.56)$ & ${\bf 0.54}(0.61)$ \\
20 & -0.6 & {\bf 0.46} & $0.79(0.52)$ & ${\bf 0.65}(0.55)$ \\
\noalign{\vskip 6pt}
10 &  0.0 & {\bf 0.46} & $0.54(0.61)$ & ${\bf 0.34}(0.66)$ \\
10 & -0.3 & {\bf 0.46} & $0.69(0.53)$ & ${\bf 0.51}(0.59)$ \\
10 & -0.6 & {\bf 0.46} & $0.84(0.47)$ & ${\bf 0.67}(0.49)$ 
\enddata
\tablenotetext{a}{Fraction of observations selected to be late bloomers for which the mock galaxy intrinsically had $\z5fract2<0.5$. The majority
of false positives have $\z5fract2<0.6$.}
\end{deluxetable*}

\clearpage

\end{document}